\documentclass{article}

\usepackage[a4paper,bindingoffset=0.2in, top=1.65 in, %
           left= 1.65 in,right=1.65 in, bottom= 1.75 in,%
           footskip=.25in]{geometry}
\usepackage{blindtext}

\usepackage{graphicx}  % standard LaTeX graphics tool;
\usepackage{amsmath}   % For getting proper math eqns
\usepackage[compress]{cite}
\usepackage{amssymb}   % Used for getting various symbols eg. gtrsim, lesssim etc.
\usepackage{bm} % For bold math (esp of lower greek letters)
\usepackage{graphicx}
\usepackage{subcaption}
\usepackage{mathrsfs}
\usepackage{mathtools,amsmath}   % For getting proper math eqns
\usepackage{amssymb}   % Used for getting various symbols eg. gtrsim, lesssim etc.
\usepackage{bm} % For bold math (esp of lower greek letters)
\usepackage{dcolumn}% Align table columns on decimal point
\usepackage{color}
\usepackage{mathrsfs}
\usepackage{amsfonts}
\usepackage{varioref}
\usepackage{braket}
\usepackage{bm}

\usepackage{varioref}
\usepackage{braket}

\usepackage[active]{srcltx}
\def\eq#1{{Eq.~(\ref{#1})}}

\def\fig#1{{Fig.~\ref{#1}}}

\def\frab#1#2{\left(\frac{#1}{#2}\right)} 

\def\ket#1{|#1\rangle}                    %%%%   ket 
\def\bk#1#2#3{{\langle #1|#2|#3\rangle}}  %%%%   bracket
\newcommand{\be}{\begin{equation}}
\newcommand{\ee}{\end{equation}}
\newcommand{\bea}{\begin{eqnarray}}
\newcommand{\eea}{\end{eqnarray}}

\newcommand{\diff}{\mathrm{d}}

\title{Quantum Correlators in Friedmann Spacetimes: The omnipresent de Sitter and the invariant vacuum noise}
\author{\small{Kinjalk Lochan\footnote{kinjalk@iisermohali.ac.in}$~^{1}$,  Karthik Rajeev\footnote{karthik@iucaa.in}$~^{2}$, Amit Vikram\footnote{amitvikram523@gmail.com}$~^{3}$, T. Padmanabhan\footnote{paddy@iucaa.in}$~^{2}$}\\
\\
$~^{1}${\small{Department of Physical Sciences, IISER Mohali, Sector 81, Manauli 140306, India}}\\
$~^{2}${\small{IUCAA, Post Bag 4, Ganeshkhind, Pune University Campus, Pune 411 007, India}}\\
$~^{3}${\small{Department of Physics,  Indian Institute of Technology Madras,  Chennai 600036,  India}}}
\date{ }
\begin{document}
\maketitle
\begin{abstract}
We discuss several aspects of quantum field theory of a scalar field in a Friedmann universe. (i) We begin by showing that it is possible to map the dynamics of a scalar field with a given mass, in a given Friedmann background to another scalar field of a different mass in another Friedmann universe. In particular one can map the dynamics of (1) a massless scalar field in a universe with power-law expansion to (2) a massive scalar field in 
the de Sitter spacetime. This allows us to understand several features of either system in a simple manner and clarifies several issues related to the massless limit.  (ii) We relate the Euclidean Green's function for the de Sitter spacetime to the solution of a hypothetical electrostatic problem in D=5 and obtain, in a very simple manner, a useful integral representation for the Green's function. This integral representation is helpful in the study of several relevant limits, and in recovering some key results which are --- though known earlier --- not adequately appreciated. One of these results is the fact that, in \textit{any} Friedmann universe, sourced by a negative pressure fluid, the  Wightman function for a massless scalar field is  divergent. This shows that the divergence of Wightman function for the massless field in the de Sitter spacetime is just a special, limiting, case of this general phenomenon. (iii) We provide a generally covariant  procedure for defining the power spectrum of vacuum 
fluctuations in terms of the different Killing vectors present in the spacetime. This allows one to study the interplay of the 
choice of 
vacuum state and the nature of the power spectrum in different co-ordinate systems, in the de 
Sitter 
universe, in a unified manner. (iv) As a specific application of this formalism, we discuss the power spectra of vacuum fluctuations in the static (and Painlev\'{e}) vacuum states in the de Sitter spacetime and compare them with the corresponding power 
spectrum in the Bunch-Davies vacuum. We demonstrate how these power spectra are related to each other in a manner similar to  the power spectra detected by the inertial and Rindler observers in flat spacetime. This also gives rise to a notion of an invariant vacuum noise in the corresponding spacetimes which is observer independent. (v) In addition,
several conceptual and technical issues regarding quantum fields in general cosmological spacetimes are clarified as a part of this study.

\end{abstract}

\newpage

\tableofcontents
%\listoffigures*

\newpage

\section{Introduction and Summary}
Quantum fields in Friedmann universes  have been investigated extensively in the past leading to fairly vast amount of literature (see for instance,\cite{Parker:1968mv, Parker:1969au, Parker:1971pt, Mottola:1984ar, Fulling:1989nb, Parker:1999td, Hollands:2014eia}). In addition to enriching our theoretical understanding of quantum field theory, such  studies also seem to be relevant to identify the seeds of structure formation as the quantum fluctuations in the early universe \cite{Brandenberger:1984cz,Parker:2009uva, gravitation, Akhmedov:2013vka}. This problem, as well as the backreaction of quantum fluctuations, have been the subject of numerous investigations (e.g., \cite{Vilenkin:1982wt,Markkanen:2017edu}). Moreover, such investigations, especially in the context of a de Sitter universe, have highlighted several theoretical issues which are rather special to this context \cite{Miao:2010vs, Krotov:2010ma, Polyakov:2012uc, Akhmedov:2012dn, Anderson:2013ila, Wang:2015eaa, Kahya:2011sy, Mora:2012kr,
Mora:2012zi, Mora:2012zh}.

In this work, we revisit the study of a minimally coupled scalar field $\phi(x)$ of mass $m$ (which could be zero or non-zero) in a Friedmann universe with a power-law expansion $a^2(\eta)\propto \eta^{-2q}$ in terms of the conformal time $\eta$, with $q=1$ representing the de Sitter universe. Though this subject has a literature running to several hundreds of papers (we provide a handful sample which  a reader can approach for a quick survey, viz., \cite{ Ford:1984hs, Allen:1985ux, Antoniadis:1985pj, Allen:1987tz, Polarski:1991ek, Kirsten:1993ug, Ratra:1984yq, Dolgov:1994ra, Takook:2000qn, Tolley:2001gg, Garbrecht:2006df, Page:2012fn,  Tanaka:2013caa, Woodard:2004ut, Miao:2010vs, Akhmedov:2013vka, Wetterich:2015gya, Wetterich:2015iea}), we find that fresh insights and new results are still possible.
We summarize   these below in order to guide the reader through this, rather lengthy, paper.

We begin, in Sec. \ref{sec:coor}, with a brief description of a few co-ordinate systems which are useful in the study of Friedmann universes in general and de Sitter spacetime in particular. The de Sitter universe has a time translational invariance which is \textit{not} manifest in the standard Friedmann co-ordinates (in which $a(t)\propto \exp Ht$) or in conformal Friedmann co-ordinates (in which $a(\eta)\propto \eta^{-1}$). This explicit time dependence of the metric prevents defining vacuum states by choosing modes which evolve as $\exp{(-i\omega t)}$ or as $\exp{(-i\omega \eta)}$ (except approximately or asymptotically). On the other hand the same de Sitter spacetime can be expressed in terms of Painlev\'{e}-type co-ordinates so that the metric%
\footnote{\textit{Notation:} We use the signature $(-,+,+,+)$ and  natural units with $c=1, \hbar =1$.  Latin letters $i, j$ etc. range over spacetime indices and the Greek letters $\alpha, \beta$ etc. range over the spatial indices. We will write $x$ for $x^i$, suppressing the index, when no confusion is likely to arise.} 
is actually \textit{stationary} (i.e $g_{ab}$ is independent of the cosmic or conformal time co-ordinate). In this co-ordinate system one \textit{does} have modes evolving as $\exp{(-i\omega t)}$ at \textit{all times} allowing one to define a vacuum state \textit{with respect to the cosmic time} $t$. A further transformation reduces the metric from the stationary to the static form with a new co-ordinate $\tau$ which is timelike in a region of spacetime. We briefly describe these constructions and emphasize the fact the static and Painlev\'{e} vacua are the same and both can be defined with respect to modes which evolve as $\exp{(-i\omega t)}$. (While the static spherically symmetric co-ordinates for de Sitter is well-known in the literature, the Painlev\'{e} co-ordinates, which retains the cosmic, geodesic, time co-ordinate $t$ has not attracted much attention.) 

Another co-ordinate system for the de Sitter universe which we describe is the one in which the geodesic distance between two events $\ell(x_2,x_1)$ --- or a simple function of the same, like $Z(x_2,x_1)=\cos(H\ell)$ (when the two events are spacelike) --- itself is used as one of the co-ordinates. This co-ordinate system turns out to be particularly useful to discuss two-point functions $G(x_2,x_1)$ which are de Sitter invariant. Such de Sitter invariant two-point functions depend on the pair of co-ordinates only through $\ell(x_2,x_1)$. When we use the geodesic distance as one of the co-ordinates, the differential equation obeyed by $G(\ell)$ depends only on one of the co-ordinates and it is easy to find and analyze the resulting ``static'' solutions. We use these properties to simplify the technical issues throughout the paper.

We next turn our attention (in Sec.\ref{sec:PSThoughKilling}) to the study of the power spectra of vacuum fluctuations in  different contexts. Since the power spectra are most useful when defined in the Fourier space, we introduce a generally covariant procedure for defining them using the Killing vectors present in the universe. Each Killing vector corresponds to a particular translation symmetry in the spacetime. When this translation invariance is reflected in the two-point function, there is a natural way of defining the corresponding power spectra by using the integral curves to the Killing vector field and the Killing parameter associated with these curves (see \cite{Romania:2012tb} for an alternate approach; the formal role of Killing vectors in  the structure of various correlation functions was explored, for e.g, in \cite{Marcori:2016oyn}). This procedure allows us to study the vacuum fluctuation spectra in several different contexts and for different vacuum states. In particular, we study the 
spectra in the case of Bunch-Davies vacuum, as well as the Painlev\'{e}/static vacuum, and discuss their physical interpretation.

We obtain, in Sec. \ref{sec:hiding}, 
 an easily proved --- but extremely useful --- result which allows us to relate the dynamics of a system $[a(\eta), \phi(x), m]$ made of a scalar field $\phi(x)$ of mass $m$
in a background universe with expansion factor $a(\eta)$ to another system $[b(\eta), \psi(x), M]$ in terms of a well-defined function. This mapping, in turn, allows us to relate the dynamics of a massless scalar field in a power-law Friedmann universe (with $a(\eta)\propto \eta^{-q}$)
to a massive scalar field in the de Sitter spacetime with a mass given by $M^2\propto (2+q)(1-q)=(9/4)-\nu^2$ where $\nu\equiv q+(1/2)$. This immediately tells you --- without any extensive calculation --- that the mass turns tachyonic and hence instabilities are expected for $|\nu|>(3/2)$. Further, the mapping allows us to study the dynamics of (a) massive fields in de Sitter and (b) massless fields in power-law Friedmann universes in a unified manner and understand the special features of either system by looking at the other one. 

One key application of this approach is the following: It is a well-known, ancient, result in this subject that the massless scalar field in the de Sitter spacetime exhibits several peculiar features e.g. divergent infrared behaviour. In the literature, these are usually thought of as a consequence of such a system not having a de Sitter invariant vacuum state \cite{Ford:1984hs, Allen:1985ux, Antoniadis:1985pj, Allen:1987tz, Polarski:1991ek, Kirsten:1993ug, Woodard:2004ut, TsamisWoodard, Pujolas:2004uj, Miao:2010vs}. There have been all sorts of attempts to handle this divergence (like e.g.,  \cite{Ratra:1984yq, Dolgov:1994ra, Takook:2000qn, Tolley:2001gg, Garbrecht:2006df, Marolf:2010zp, Page:2012fn, Tanaka:2013caa, Park:2015kua}). We will see that this is only part of the story. We recover the well known result \cite{FordParker} that the similar infrared divergences exist for massless fields in any spacetime sourced by matter with negative pressure; that is, whenever the equation of state parameter 
$w\equiv(p/\rho)$ is negative (see  \cite{FordParker, Janssen:2008dp, Janssen:2008dw, Janssen:2008px, Higuchi:2017sgj}). The de Sitter spacetime --- and the pathologies of a massless field in that spacetime --- is just a particular case of this general result when $w=-1$. In all Friedmann universes with $-1<w<0$ the massless scalar field will exhibit pathologies even though these spacetimes have no special invariance properties like the de Sitter spacetime. We describe these features in detail and from several perspectives in this work.

Another application of these results is in the approach to the massless field in de Sitter, viz., $m=0, q=1$,  in two different ways. We could have thought of $m=0, q=1$ as arising from (i) the limit $m\to0$ with $q=1$ (massless limit in de Sitter) or as (ii) the limit $q\to 1$ with $m=0$ (de Sitter limit of massless theory). Two-point functions, like e.g., the Wightman function,  do not exist in this limit irrespective of how we take it. However, if we treat the approaches in (i) and (ii) as two different ways of regularizing the limit (with the small parameters being $m/H$ and $(q-1)$ respectively), then the final results depend on the regularization scheme. One approach leads to a secular growth term for regularized $\langle\phi^2\rangle$ while the other does not. We explore several features of this and related results using our mapping.

There are also several technical results which are new in this work. For example, we show that (Sec \ref{sec:5ded}) the Euclidean Green's function for the de Sitter spacetime can be obtained very easily by mapping the problem to one of $D=5$ electrostatics. This leads to a simple integral representation for the Green's function (which, of course, is algebraically equivalent to the Gauss hypergeometric function) that is easy to analyze and understand. It also clarifies the issues involved in the analytic continuation to Lorentzian spacetime. We also provide a careful discussion of the $H\to0$ limit of de Sitter (and similar limits for power-law cosmologies) when the spacetime becomes flat. Our results, e.g., mode functions, Green's functions etc. are expected to go over to the flat spacetime expressions in this limits. However, this limit, as we show, is often technically nontrivial. In some cases (like e.g., in the case of the Feynman propagator) this limit even leads to fresh insights about the flat 
spacetime 
QFT (see e.g., Appendix \ref{app:derB1})! Several other derivations in this work also  contain new and useful techniques.

\section{Conceptual and mathematical background}

We begin by summarizing several conceptual and mathematical aspects in this section. While some of these results are well known, others are not. Even as regards some of the  better known results, our emphasis will be different from the conventional one in several cases. (So it will be useful for you to rapidly go through the subsections below, even if you are familiar with the literature in the subject!). Most of the discussion in this section can be directly generalized to a $D+1$ dimensional spacetime but we will confine ourselves to $3+1$ dimensions for simplicity.

\subsection{Coordinate systems}\label{sec:coor}

Let us start by listing the properties of several co-ordinate systems used to describe a Friedmann universe in general and a de Sitter spacetime in particular. All these co-ordinate systems, except probably the geodesic co-ordinates (discussed in Sec. \ref{sec:co-geo}) have appeared in the literature before. While the co-ordinate system used most frequently in the literature is the Friedmann co-ordinates in Sec. \ref{sec:co-frw} we will, however, make extensive use of all the four co-ordinate systems discussed below. 

\subsubsection{Friedmann co-ordinates}\label{sec:co-frw}

The co-ordinate system which makes the \textit{spatial} symmetries of the Friedmann spacetime manifest is the Friedmann co-ordinate system given by either of the two forms of the line element:
\begin{equation}
 ds^2 = - dt^2 + a^2(t) |d\bm{x}|^2 = a^2 (\eta) [-d\eta^2 + |d\bm{x}|^2].
 \label{tp1}
\end{equation} 
The co-ordinate $t$ has a direct physical meaning and measures the time shown by geodesic, freely falling, co-moving clocks in this spacetime and the spatial co-ordinate $x^\alpha$ makes the homogeneity and isotropy of the spatial co-ordinates apparent. These co-ordinates also has the interpretation that observers with $\bm{x}$ = constant are geodesic observers. We will call the $(t,\bm{x})$ system the \textit{cosmic co-ordinates}.
The conformal time $\eta$ is related to the cosmic time $t$ through $dt=a(\eta) d\eta$ and is often convenient for mathematical manipulations \textit{even though it does not have a direct physical meaning}, unlike the geodesic cosmic time $t$. We will call the $(\eta,\bm{x})$ system the conformal Friedmann co-ordinates or simply \textit{Friedmann co-ordinates}. 

For the most part of the paper, we will concentrate on Friedmann universes with a power law expansion with $a(t) \propto t^p$ corresponding to $a(\eta) \propto \eta^{-q}$ where $q=p/(p-1)$. In the limit of $p\to \infty$, corresponding to $q\to 1$, we get the de Sitter expansion. When we need to take this limit, it is often convenient to shift the origin of the time co-ordinate and use the expressions,
\begin{equation}
a(t) = \left( 1 + \frac{Ht}{p}\right)^p = \left( - \frac{H\eta}{q}\right)^{-q}; \quad \eta = - \frac{1}{H} \frac{p}{p-1}\left(1+\frac{Ht}{p}\right)^{1-p};\quad q=\frac{p}{p-1},
\label{expfactor}
\end{equation} 
where $H$ is a constant parameter introduced for dimensional reasons and $q$ and $p$ are dimensionless indices. These expressions have clear limits when $p\to \infty$, $q\to 1$ with $a(t) = \exp(Ht) = (-H\eta)^{-1}$ which is the de Sitter limit. They also make clear that  the algebraic behaviour of these relations differ significantly when $p>1$ (accelerating universes) compared to $p<1$ (decelerating universes). For example, $\eta \to +\infty$ when $t\to +\infty$ if $p<1$. But $\eta\to 0^-$ when $t\to \infty$ in the case of de Sitter expansion. 

When the limiting form for $a(t)$ is not explicitly required, we will continue to use the simpler forms with $a(t)\propto t^p \propto \eta^{-q}$ with $q=p/(p-1)$. These expansion indices $q$ and $p$ are related to the (constant) equation of state parameter $w\equiv P/\rho$, of an ideal fluid, which can act as the source to the power law expansion. We find that 
\begin{equation}
 p = \frac{2}{3(1+w)}; \qquad q = - \frac{2}{1+3w} .
 \label{tp3}
\end{equation} 
The power law expansion is also characterized by the condition that they have constant acceleration/deceleration parameters; that is, for a universe with power law expansion the quantity $\dot H/H^2 \equiv \epsilon$ is a constant given by $\epsilon = -1/p $. Such a parameterization is often used with a small (approximately)  constant $\epsilon$ to describe an approximately de Sitter evolution of the universe.

When described in terms of the cosmic time, the power-law expansions with $a(t) \propto t^p$ are clearly distinguished from the de Sitter expansion with $a(t) \propto \exp(Ht)$. But
when described in terms of the conformal time, $a(\eta) = \eta^{-q}$ does not seem to distinguish the de Sitter expansion with $q=1$ from any other power law. 
This appearance is, of course, illusory and the correct way to distinguish de Sitter expansion from the power-law expansion is from the \textit{extra symmetry} which arises when $a \propto \exp(Ht) \propto \eta^{-1}$. \textit{In spite of the apparent dependence of the metric on the time co-ordinate ($t$ or $\eta$), such a universe is in steady state and has no intrinsic time dependence.} Under a finite translation of the cosmic time, $t\to t+T$, along with the rescaling of spatial co-ordinates by $x^\alpha \to x^\alpha \exp(-HT)$ the metric remains invariant. In terms of the conformal time $\eta$, this symmetry manifests as a rescaling of \textit{all} the co-ordinates: i.e, the line interval remains invariant under $\eta \to \mu\eta, x^\alpha \to \mu x^\alpha$. This selects out the power law $a(\eta)\propto \eta^{-q}$ with $q=1$ --- which corresponds to the de Sitter expansion --- as special. One can easily verify such an extra symmetry (viz. time translation invariance in terms of cosmic time or the 
rescaling 
invariance in terms of conformal time) does not exist for any other power law. 

We will often require, in our future discussion, the expression for the geodesic distance $\ell(x_2,x_1)$ between two events in the de Sitter spacetime. For example, when $H\ell<1$, this can be expressed in terms of a quantity $Z(x_2,x_1)$ as $H\ell(x_2,x_1) \equiv \cos^{-1} Z(x_2,x_1)$ where
\begin{equation}
 Z(x_2,x_1) = \frac{1}{2\eta_1 \eta_2} \, \left( \eta_1^2 +\eta_2^2 - |\bm{x_1}-\bm{x}_2|^2\right),
 \label{zfrw}
\end{equation} 
in Friedmann co-ordinates. (A more general, geometric definition is given later on.)

\subsubsection{Painlev\'{e} co-ordinates} \label{sec:co-pan}

\begin{figure*}
\begin{center}
\includegraphics[scale=0.25]{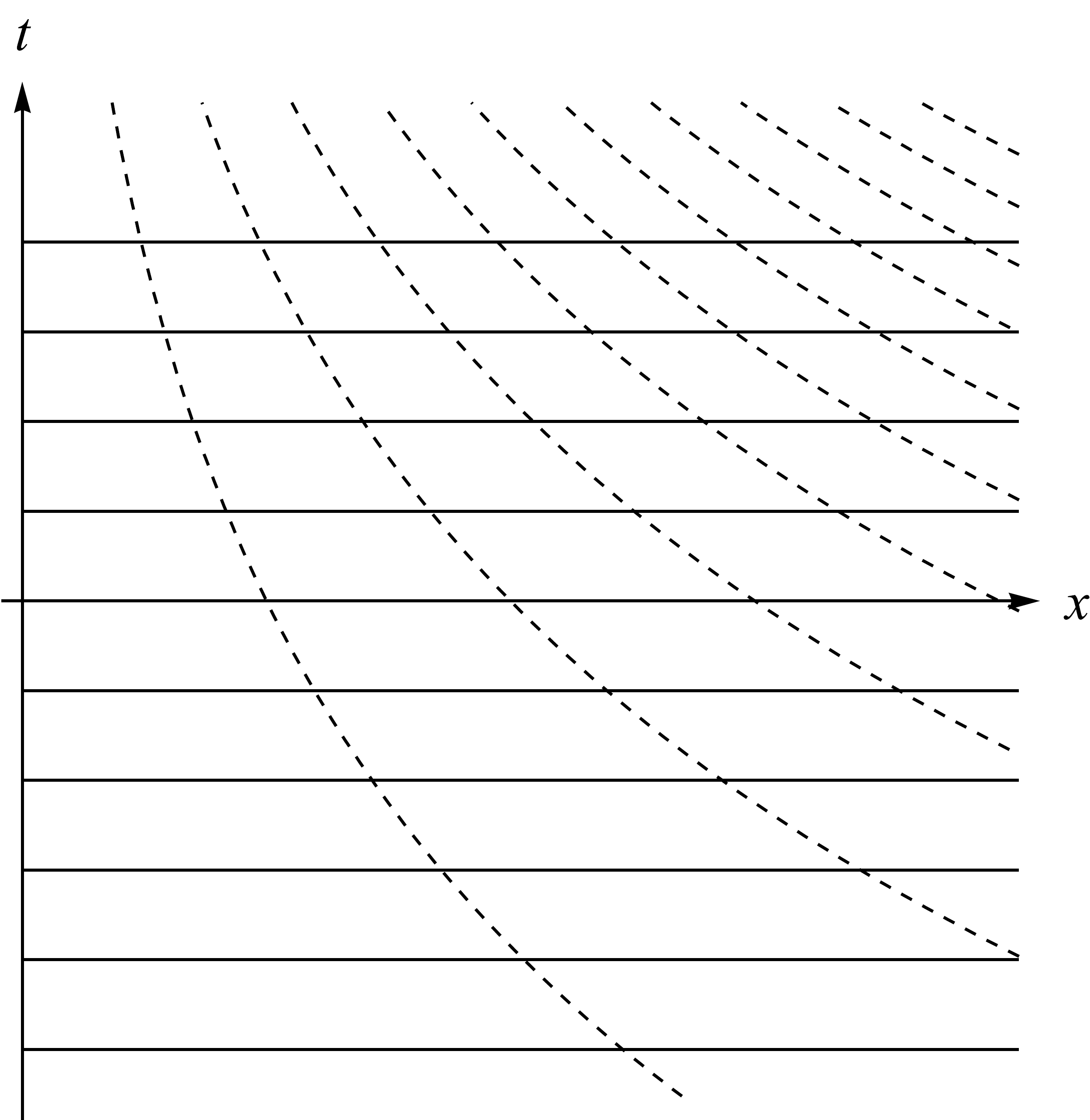}
\caption{Contours of curves corresponding to constant values of Painlev\'{e} co-ordinates  (defined in \eq{A50}) are  plotted in the co-moving $(t, x)$ co-ordinates. The solid lines correspond to curves of constant Painlev\'{e} time (which coincides with the co-moving time) and the dashed curves correspond to curves of constant Painlev\'{e} radial co-ordinate which is related to the co-moving co-ordinate $x$ as $r = a(t)x$. }\label{fig_01}
\end{center}
\end{figure*}

It is sometimes convenient to introduce a set of co-ordinates in which the expansion of the universe is made to vanish in terms of the spatial co-ordinates \cite{Padmanabhan:2016lul}. This is  done by using the (proper)  spatial co-ordinate $\bm{r} = a(t) \bm{x}$ instead of the original comoving co-ordinates $\bm{x}$. We retain the time co-ordinate to be the cosmic time $t$ with the physical meaning that this is the time registered by geodesic clocks. The metric, for an arbitrary Friedmann universe, now becomes 
\begin{equation}
ds^2 = -(1 -H^2(t) r^2)d{t}^2 - 2 H(t) r d{t} d r + dr^2 + r^2 d \Omega^2,
 \label{A50}
\end{equation} 
where $d\Omega^2$ is the metric on the unit 2-sphere. The constant $-t$ surfaces are now spatially flat, static Euclidean space \textit{with no sign of any cosmic expansion}. The observers located at $\bm{r} =$ constant,  are (in general) non-geodesic, accelerated observers. This is in contrast with the situation in the Friedmann co-ordinate system, in which $\bm{x}=$ constant represents geodesic curves. The time co-ordinate, however, continues to represent the time registered by geodesic clocks which now move along trajectories with $\bm{r}(t) = \bm{r}_0 a(t)$, where $\bm{r}_0$ is a constant vector. 
The only exception to these general comments is provided by the observer located at the \textit{origin} of the co-ordinate system $\bm{r} =0$ who will be a geodesic observer, because the spatial origin $\bm{r} =0$ maps to $\bm{x} =0$ in comoving co-ordinates which, of course, is a geodesic. Thus the time co-ordinate $t$ can also be interpreted as the time shown by a clock located at the origin of the Painlev\'{e} co-ordinate system (see \fig{fig_01}). The spatial homogeneity of the spacetime is not manifest in these co-ordinates which is the  price one has to pay to neutralize the effects of cosmic expansion.

Everything we said so far is applicable to \textit{all} Friedmann spacetimes and the metric in \eq{A50} depends on time through the function $H(t)$. The de Sitter spacetime is again special because $H(t)$ in \eq{A50} now becomes a constant thereby making the metric stationary. The time translation invariance under $t\to t+T$ of the de Sitter universe is \textit{now manifest in this stationary co-ordinate system which was not the case in the Friedmann co-ordinate system}. 

Because the metric coefficients are independent of time, the solutions to the wave equation $(\square -m^2)\phi(x)=0$ can now be expressed as the superposition of fundamental modes of the form $f_\omega(\bm{r})\exp(\pm i\omega t)$. While dealing with a quantum field, this allows the definition of a vacuum state using modes which are positive frequency with respect to \textit{cosmic time}. Note that this is \textit{not} possible when we describe the de Sitter spacetime in Friedmann co-ordinates because of the time dependence of the metric, arising from $a(t)=\exp{(Ht)}$ factor. 
We will make use of this property later on in our discussions. 

\subsubsection{Spherical co-ordinates}\label{sec:co-sph}

Since homogeneity and isotropy  of spatial cross sections necessarily imply spherical symmetry, it is also possible to describe \textit{any} Friedmann spacetime in a spherically symmetric form. This can be done by introducing  a new time co-ordinate  $\tau$ (in addition to the spatial co-ordinate $\bm{r} = a(t) \bm{x}$ which we have already introduced in the last section) with  
\begin{equation}
 \tau \equiv F(\sigma); \qquad \sigma \equiv \left( \int^r x\, dx + \int^\tau \frac{ \mathrm{d} t}{a(t)\dot a(t)}\right);\quad r=a(t)x,
\end{equation} 
where $F(\sigma) $ is an arbitrary function of the variable $\sigma$. It is easy to verify that this will lead to a metric given by
\begin{equation}
ds^2 = - e^\nu \, d\tau^2 + e^\lambda \, dr^2 + r^2 d\Omega^2; \quad e^\nu = \frac{a^2 \, \dot a^2 }{1 - r^2 H^2}\frab{dF}{d\sigma}^{-2}; \quad e^\lambda = \frac{1}{1- r^2 H^2}.
 \label{B11}
\end{equation} 
Since the spatial co-ordinates used in this metric are the same as those used in the Painlev\'{e} co-ordinates (see Sec. \ref{sec:co-pan})  all the comments related to spatial co-ordinates continue to apply. In particular, $\bm{r}$= constant observers are non-geodesic observers except for  the special observer located at the origin, $\bm{r}=0$, who is a geodesic observer.  The $\tau$ co-ordinate,  no longer measures the geodesic clock time except for a clock located at the origin. 

\begin{figure*}
\begin{center}
\includegraphics[scale=0.25]{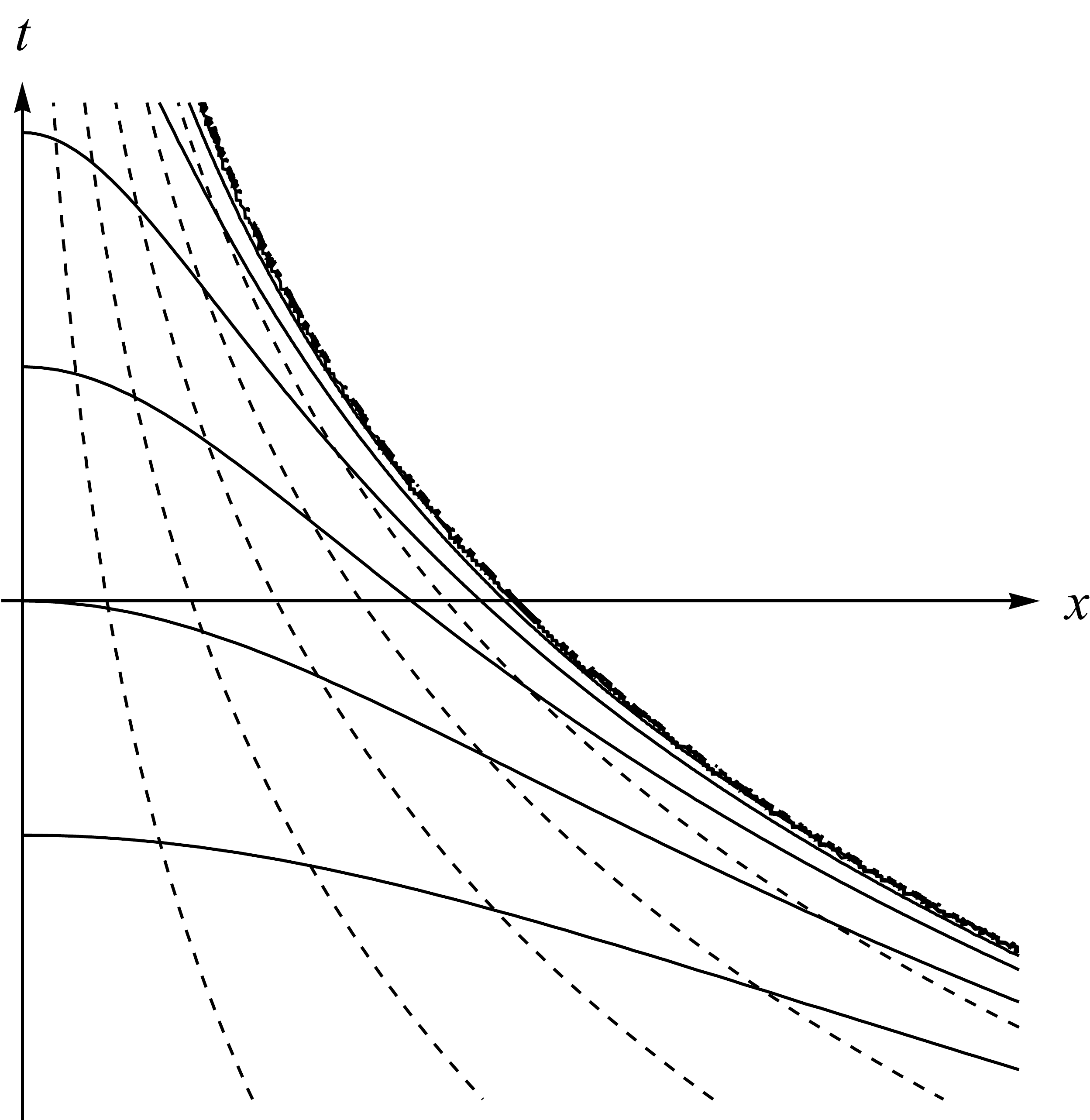}
\caption{Contours of curves corresponding to constant values of static co-ordinates (as defined in \eq{A16}) are plotted in the co-moving $(t, x)$ co-ordinates. The solid curves correspond to curves of constant static co-ordinate time $\tau$ (as given in \eq{A49}) and the dashed curves correspond to curves of constant radial co-ordinate which is related to the co-moving co-ordinate $x$ as $r = a(t)x$. The dark wavy line corresponds to the surface $H r(t,x)=1$, the static co-ordinate patch.}\label{fig_02}
\end{center}
\end{figure*}

In particular, in the case of de Sitter spacetime with $a(t) \propto \exp(Ht)$, the choice   $F(\sigma)=-(1/2H)\ln \sigma$ reduces the metric to the form
\begin{equation}
  ds^2 = - (1-H^2r^2)\, d\tau^2 + \frac{dr^2}{1-H^2r^2} + r^2\, d\Omega^2 .
 \label{A16}
\end{equation}
The metric is now static (rather than stationary which was the case in the Painlev\'{e} co-ordinates) and is invariant under the time translation $\tau\to \tau +$ constant.
In this case, the relation between $\tau$ and the geodesic time (used in Friedmann and Painlev\'{e} co-ordinates) is given by
\begin{equation}
 t = \tau+ \frac{1}{2H } \, \log ( 1 - H^2 r^2),
 \label{A49}
\end{equation} 
(see \fig{fig_02}).
Clearly translation in cosmic time $t$ corresponds to the translation in $\tau$ so that the symmetry is manifest.

Moreover, 
because the metric coefficients in \eq{A16} are independent of time, the solutions to the wave equation $(\square -m^2)\phi(x)=0$ can again be expressed as the superposition of fundamental modes of the form $g_\omega(\bm{r})\exp(\pm i\omega \tau)$ thereby allowing us to define a vacuum state using modes which are positive frequency with respect to $\tau$. (We will call this state the \textit{cosmic vacuum} or the \textit{static vacuum}.) But, from \eq{A49}, it is clear that
the positive frequency mode with respect to $\tau$, of the form
  $e^{-i\omega \tau}g_\omega(\bm{r})$ will translate to a positive frequency solution $e^{-i\omega t} f_\omega(\bm{r})$ with respect to the cosmic time $t$ with:
  \begin{equation}
f_\omega(\bm{r})=g_\omega(\bm{r})(1-H^2r^2)^{i\omega/2H},
\label{painstatic}
\end{equation} 
   under the co-ordinate transformation in \eq{A49}. 
 Therefore the positive frequency modes in static, spherically symmetric co-ordinate system actually correspond to those which are positive frequency with respect to the cosmic time $t$ and \textit{the static vacuum can be reinterpreted as the one corresponds to positive frequency modes
with respect to the cosmic time $t$}. The only issue we need to be careful about is the fact that
 $\tau$ itself retains its time-like character only for $r<H^{-1}$ in this co-ordinate system due to the existence of a horizon at $r=H^{-1}$. We will have occasion to use these results later on.

 The existence of Friedmann co-ordinates (discussed in Sec. \ref{sec:co-frw}) as well as the static co-ordinates described by the metric in \eq{B11} shows that geodesic observers can actually be associated with two distinct co-ordinate systems. Consider, for example,  a geodesic observer whose  world-line is described by $\bm{x}=$ constant in the Friedmann co-ordinates. Because of spatial homogeneity of the spacetime, we can  always choose this world-line to be $\bm{x} =0$ by a suitable choice of the origin. The clock carried by this geodesic observer will show the flow of the cosmic time $t$. One can now introduce a static spherically symmetric co-ordinate system around this observer, again describing her world-line as 
 $\bm{r}=0$ in the static co-ordinate system (which, of course, corresponds to $\bm{x} =0$ in the Friedmann co-ordinate system) maintaining the geodesic nature. We see from \eq{A49} that
  at $\bm{r}=0$ we have  $\tau=t$; so proper time measured by the geodesic clock carried by the observer continues to track the  cosmic time. In other words,  a given geodesic observer can place herself at the origin of the spatial co-ordinate system either in the Friedmann co-ordinates or in the spherically symmetric co-ordinate system and describe the spacetime around her using either of the co-ordinate patches. In particular, the geodesic observers in de Sitter spacetime can use either the Friedmann co-ordinates with the metric having the form in \eq{tp1} or the static spherically symmetric co-ordinate system with the metric having the form in \eq{A16}. We will come back to this feature later on in our analysis.

  For future reference, we give the form of the geodesic distance between two events in this spherical co-ordinate system. As in the case of \eq{zfrw}, the geodesic distance $\ell(x_2,x_1)$ can again be written as $H\ell(x_2,x_1)= \cos^{-1}Z(x_2,x_1)$ where 
  \begin{equation}
   Z(x_2,x_1)=H^2( \bm{ r}_1 \cdot \bm{r}_2)  +\sqrt{1-H^2r_1^2}\sqrt{1-H^2r_2^2}\cosh\left[H(\tau_2-\tau_1)\right],
   \label{zstatic}
  \end{equation} 
  in this co-ordinate system. (The dot product $\bm{ r}_1 \cdot \bm{r}_2$ is just a notation for $\delta_{\alpha\beta}r_1^\alpha r_2^\beta$ in terms of Cartesian components.)  Note that the geodesic distance depends on $\tau_1, \tau_2$ only through $(\tau_2 - \tau_1)$ because of the time translational invariance. 

\subsubsection{Geodesic co-ordinates}\label{sec:co-geo}

Finally, we will describe a co-ordinate system which is somewhat special to de Sitter spacetime \cite{Bros:1995js, PerezNadal:2009hr, Page:2012fn, Akhmedov:2013vka}. As is well known, the 4-dimensional de Sitter manifold can be thought of as a hyperboloid embedded in a 5-dimensional flat Minkowski  spacetime\footnote{In fact any Friedmann spacetime can be embedded in a 5-dimensional flat Minkowski spacetime and de Sitter embedding is just a special case of this general result.}
with Cartesian co-ordinates $X^A$ where $A = 0 - 4$. 
Let $\ell(x,x_0)$ be the geodesic distance between two events in the de Sitter spacetime. Using the embedding properties it is straightforward to introduce a co-ordinate system in which the geodesic distance $\ell$ itself is one of the co-ordinates (see Appendix \ref{app:dera} for the derivation). For example, if the two events are separated by a space-like distance, then such a co-ordinate system will describe the de Sitter spacetime in terms of the line element 
\begin{equation}
ds_H^2=-\frac{\sin^2(H\ell)}{H^2}d\tau^2+d\ell^2+\frac{\sin^2(H\ell)}{H^2}\cosh^2\tau d\Omega_2^2.
 \label{C11}
\end{equation} 
The explicit co-ordinate transformation from the conformal Friedmann co-ordinates ($\eta, r, \theta, \phi$) to the geodesic co-ordinates $(\tau, \ell, \theta, \phi)$ is given by
 \begin{equation}
 \cosh\tau=-\frac{2\eta_0 r}{\sqrt{4\eta^2\eta_0^2-(\eta^2+\eta_0^2-r^2)^2}};\qquad
 \cos(H\ell)=\frac{\eta^2+\eta_0^2-r^2}{2\eta\eta_0}; \qquad
 \label{cot1}
 \end{equation} 
where  $\eta_0$ is a constant. 
\begin{figure*}
\begin{center}

\includegraphics[scale=0.25]{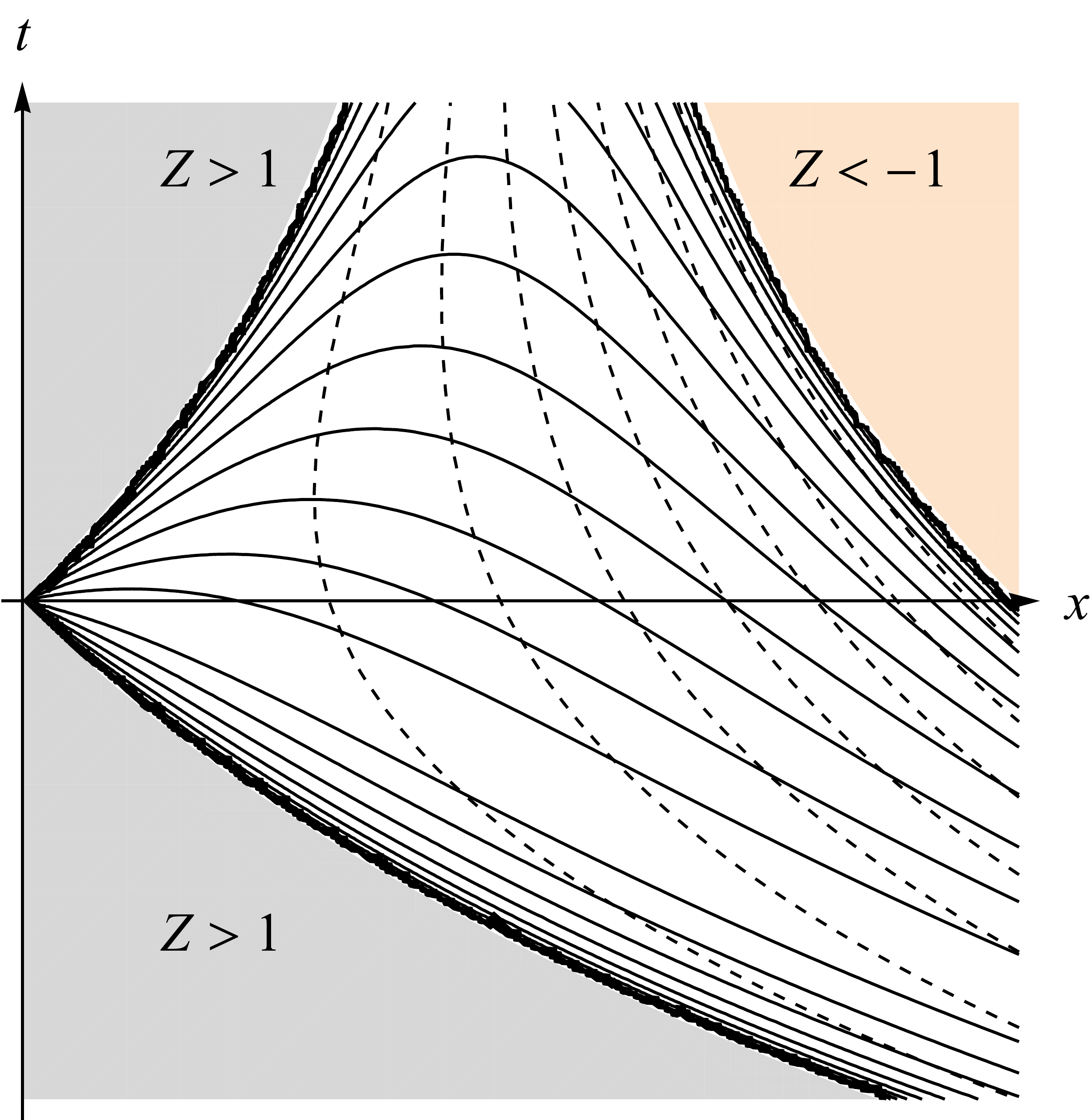}
\caption{Contours of curves corresponding to constant values of the geodesic co-ordinates (as defined in \eq{C11}) are plotted in the co-moving $(t, x)$ co-ordinates. The solid curves correspond to curves of constant timelike co-ordinate $\tau$ and the dashed curves correspond to curves of constant spacelike co-ordinate $l$. The co-ordinates $(\tau,l)$ are related to the co-moving co-ordinates through conformal time $\eta=-e^{-Ht}/H$ and \eq{cot1}. The shaded regions are not covered by this co-ordinate system because: (i) the shaded region corresponding to $Z>1$ are timelike separated from the origin $O$ of the comoving co-ordinates and (ii) no points in the shaded region marked $Z<-1$ can be connected to $O$ by a spacelike \textit{geodesic} segment (see the discussion around \eq{geodist}).}\label{fig_03}

\end{center}
\end{figure*}

This co-ordinate system  (see \fig{fig_03}) has an interesting limit when $H\to 0$ and the spacetime becomes flat. We see that it reduces to the form:
\begin{equation}
\lim_{H\rightarrow0} ds_H^2=-\ell^2d\tau^2+d\ell^2+\ell^2\cosh^2\tau d\Omega_2^2,
\label{sphrin}
\end{equation} 
which is indeed flat spacetime, but  expressed in what is  called the spherical Rindler co-ordinate system. The spherical Rindler co-ordinates are obtained from the standard spherical polar co-ordinates of flat spacetime ($t,r,\theta,\phi$) by the transformation $r= \ell \cosh \tau, t = \ell \sinh \tau$ in the region $r^2 > t^2$. Obviously $\ell^2 = r^2 - t^2$ is the square of the geodesic distance from the origin to the event ($t,r,\theta,\phi$). What we  have in \eq{C11} is just a generalization of this co-ordinate system, but now  based on the geodesic distance in the de Sitter spacetime. 
For the sake of completeness  we mention that, when $l$ corresponds to a time-like separation, we get a slightly modified version of \eq{C11}.
\begin{eqnarray}
ds^2=-dl^2+\frac{\sinh^2(Hl)}{H^2}\left(d\tau^2+\sinh^2(\tau)d\Omega_2^2\right).
\end{eqnarray}
The $H\rightarrow 0$ limit of this, as expected, gives the corresponding version of spherical Rindler in the time-like wedge:
\begin{eqnarray}
ds^2=-dl^2+l^2d\tau^2+l^2\sinh^2\tau~d\Omega^2_2.
\end{eqnarray}

The line element in \eq{C11} can be further simplified by introducing the co-ordinate $Z = \cos(H\ell)$ where $\ell(x,x_0)$ is now the geodesic distance between the events $(\eta_0,\bm{x}_0)$ and $(\eta,\bm{x})$ with some fixed values for $x_0$. 
The line element in \eq{C11} now becomes 
 \begin{equation}
H^2\, ds_H^2=-(1-Z^2)\,d\tau^2+\frac{dZ^2}{(1-Z^2)}+(1-Z^2)\,\cosh^2\tau \ d\Omega_2^2.
  \label{C27}
 \end{equation} 
From \eq{zfrw} we know that 
$
Z = (1/2)(1/\eta \eta_0)( \eta^2 +\eta_0^2 - |\bm{x}-\bm{x}_0|^2).
$
This equation, along with the first equation in \eq{cot1}, gives the direct transformation from the ($\eta, r, \theta, \phi$) co-ordinates to the  co-ordinates $(\tau, Z, \theta, \phi)$. 
 
 The co-ordinate $Z$ has a simple geometrical meaning in terms of the embedding space. 
Let $X^A =(X^0,X^1,X^2,X^3, X^4)$ and $X^A_0= (0,H^{-1}, 0,0,0)$ be two Cartesian 5-vectors in the embedding space. We  can then easily verify that  the de Sitter invariant,  dimensionless Cartesian dot product  between these two vectors is just $Z$; i.e., $Z= - H^2 \eta_{(5) A B} X^A X^B_0$. Therefore, the general definition for $Z$ is given by the 5D Lorentzian inner product,
$
Z(x_1,x_2)\equiv -H^2\eta_{(5)A B}X_1^{A}X_2^{B},
$
where, $X_{1,2}^{A}$ corresponds to the 5D Cartesian co-ordinates of the point, say, $P_1$ and $P_2$ in the $dS_4$ hyperboloid embedded in the 5D Minkowski space. When the points $P_1$ and $P_2$ \textit{can be connected} by a geodesic of length-squared $l^2$, then $Z(x_1,x_2)$ takes the different forms
\begin{eqnarray}
Z(x_1,x_2)=\begin{cases} \cos(Hl)~~; \textrm{$P_1$ and $P_2$ are spacelike seperated}\\
\cosh(Hl_t)~~;\textrm{$P_1$ and $P_2$ are timelike seperated}\\
0~~; \textrm{$P_1$ and $P_2$ are lightlike seperated}
\end{cases}
\label{geodist}
\end{eqnarray}
where, we have defined $l_t^2=-l^2$ for the timelike separated events. When $Z<-1$ (a special case of spacelike separated), even though there exist many spacelike \textit{curves} connecting $P_1$ and $P_2$, there are no spacelike \textit{geodesics} connecting them \cite{Schmidt:1993ga}. This has the consequence that for $Z<-1$, there is no analogue of \eq{geodist}. The expressions in \eq{geodist} can be used to obtain the geodesic co-ordinate charts in different regions.

The importance of this co-ordinate system (which does not seem to have been realized in the literature) arises from the fact that it allows one to deal with de Sitter invariant solutions to wave equations in a simple manner. For example,  consider any two point function $G_{\rm dS}(x, x')$ for a scalar field of mass $m$ which satisfies the equation $(\Box - m^2) G_{\rm dS} =0$. If $G_{\rm dS}$ is  de Sitter invariant, it will depend only on $x$ and $x'$ only through $\ell(x,x')$ or, equivalently, only on $Z(x,x')$ so that $G_{\rm dS}(x, x') = G_{\rm dS}[Z(x,x')]$. In the  equation $(\Box - m^2) G_{\rm dS} =0$ we can easily evaluate the $\square$ operator in the co-ordinate system with the metric in \eq{C27}, retaining only the $Z$ dependence. (This means that we are looking for static, ``radially'' dependent solutions to the Klein-Gordon operator in this co-ordinate system.)  We will get:
 \begin{equation}
  (Z^2-1)\frac{\mathrm{d}^2 G_{\mathrm{dS}}}{\mathrm{d} Z^2} + 4Z\frac{\mathrm{d} G_{\mathrm{dS}}}{\mathrm{d} Z} + \frac{m^2}{H^2}G_{\mathrm{dS}} = 0,
  \label{D156}
 \end{equation} 
 when $G_{\rm dS}$ depends only on $Z$.
 In terms of $\ell$ the same equation reduces to
 \begin{equation}
 \left[\frac{d^2}{d\ell^2}+3H\cot(H\ell)\frac{d}{d\ell}+m^2\right]  G_{\rm dS}=0.
 \end{equation} 
 As we shall see later, this approach leads to an interesting way of determining de Sitter invariant two point functions and analyzing their properties. 
  
  Incidentally, in the limit of $H\to 0$ de Sitter spacetime reduces to flat Minkowski spacetime (but in the spherical Rindler co-ordinates) and the equation for the two point function reduces to 
  \begin{equation}
 \left[\frac{d^2}{d\ell^2}+\frac{3}{\ell}\frac{d}{d\ell}+m^2\right] G_{\rm dS}=0.
 \label{C30}
  \end{equation}
  It can be easily verified that the acceptable solutions to this equation, given in terms of $K_1(ml)$ reproduces the correct two point functions of the flat spacetime Lorentz invariant field theory. 

\subsection{Quantum Correlators and power spectra}   \label{sec:PSThoughKilling}

\subsubsection{Quantum correlators}

The quantum fluctuations of a field living in a curved spacetime can be described by the correlation functions of the field in any given state.
Therefore, these correlators provide a unique way to analyze the background spacetime and the symmetries it comes with \cite{Fukuma:2013mx, Bros:1995js, Wetterich:2015gya, Frob:2017gyj,Brooker:2017kij}.

The simplest correlator, which is relevant for the free field theories,  is the two point correlation function in a, suitably defined, vacuum state $\ket{0}$,  called the Wightman function.  For a scalar field, this is defined by 
\begin{equation}
G(x,y) \equiv \bk{0}{\phi(x)\phi(y)}{0}.
\label{defgxy}
\end{equation}
Many other two point functions like, for example, the Feynman propagator $iG_F(x,y) \equiv \bk{0}{T\{\phi(x)\phi(y)\}}{0}$ or the commutator function $G^{\rm c}(x,y) \equiv \bk{0}{[\phi(x),\phi(y)]}{0}$ etc. can be expressed in terms of the Wightman function in a fairly straightforward manner. Hence we shall concentrate on the Wightman function as a key measure of quantum fluctuations in a curved spacetime. 

From the definition in \eq{defgxy} it is clear that $G(x,y)$ transforms as a biscalar in $x$ and $y$ when the co-ordinate system is changed, \textit{if we keep the vacuum state  the same}. Obviously the two point function depends on the choice of the vacuum state $\ket{0}$ and --- as is well known --- this choice is far from unique (or even physically well-defined) in an arbitrary curved spacetime. Very often, the choice of the co-ordinate system could itself suggest a natural vacuum state adapted to that particular co-ordinate system. For example, when the flat spacetime is described in the inertial co-ordinates, it is natural to use mode functions which are positive frequency solutions with respect to the inertial time and use it to define the inertial vacuum $\ket{0,\rm In}$. This will, in turn, define the Wightman function for the inertial vacuum state as $G_{\rm In} = \bk{0,\rm In}{\phi(x)\phi(y)}{0,\rm In}$. This function $G_{\rm In}(x,y)$ can, of course, be expressed in any other co-ordinate system 
including, say, 
the 
Rindler co-ordinate system. But when we use Rindler co-ordinate system one may find it natural or convenient to choose mode functions which are positive frequency with respect to the Rindler time co-ordinate thereby defining another vacuum state, viz., the Rindler vacuum $\ket{0,\rm Rin}$. The corresponding Rindler-Wightman function $G_{\rm Rin}(x,y) \equiv \bk{0,\rm Rin}{\phi(x)\phi(y)}{0,\rm Rin}$ is, of course, quite different from $G_{\rm In}(x,y)$ and they are not related by a co-ordinate transformation because the vacuum states are different. We will have occasion to use similar constructs for different vacuum states in Friedmann spacetimes later on.

\subsubsection{Power spectra from Killing vectors}\label{sec:psgen}

Given the fact that $G(x,y)$ describes the fluctuations of a quantum field, it is natural to inquire about the power spectrum of these fluctuations. Power spectra, conventionally, are represented in a, suitably defined, Fourier space and are useful when some natural co-ordinate choice induces some symmetries on $G(x,y)$. Since the symmetries of the spacetime are described by Killing vector fields, it is possible to provide a natural, covariant, definition of power spectrum associated with any Killing vector field along the following lines:

 Let $\xi^a(x)$ be a Killing vector field which exists in some region of the spacetime and let $\mathcal{C}(\lambda)$ be an integral curve of this Killing vector field satisfying the equation $dx^a/d\lambda = \xi^a(x)$ where the Killing parameter $\lambda$ is assumed to run over the entire real line. We will assume that a congruence of such integral curves, corresponding to a given $\xi^a$, exists in some region of spacetime. We can now introduce $\lambda$ itself as one of the co-ordinates in this region and we will denote the rest of the (`transverse') co-ordinates by $x^a_\perp$. Consider now the Wightman function between two events $x_1$ and $x_2$, located on a given integral curve, with $x_1=(\lambda_1, x^a_\perp)$ and $x_2= (\lambda_2, x^a_\perp)$. (Since the events are on the integral curve, their co-ordinates will  only differ  in the $\lambda$-co-ordinate value and they will have the same transverse co-ordinates $x^a_\perp$.)
 Clearly, because the Killing vector generates a translational symmetry along $\lambda$-co-ordinate, and --- \textit{if} we choose a vacuum state that respects this symmetry --- the Wightman function will only depend on $\lambda\equiv \lambda_1-\lambda_2$ with the structure $G(x,y) = G(\lambda,x^a_\perp)$. One can now define a power spectrum from the Fourier transform of the two point function with respect to the Killing parameter, $\lambda$, which is
 one of the co-ordinates in this co-ordinate system. That is we define:
 \begin{equation}
  P_+(\omega;x^a_\perp) \equiv \int_{-\infty}^\infty \frac{ \mathrm{d} \lambda}{2 \pi}\exp(i\omega \lambda)\, G(x,y;\lambda) = \int_{-\infty}^\infty \frac{ \mathrm{d} \lambda}{2 \pi} \exp(i\omega \lambda)\, G(\lambda,x^a_\perp).
  \label{pkilling}
 \end{equation} 
 This is the primitive definition of power spectrum; usually we will multiply it with some measure based on physical considerations to give suitable dimensions but this is just kinematics.
 
 As an aside, we will mention an important subtlety as regards this definition, postponing its detailed discussion to a later section. Note that we could also have defined the power spectrum with $\exp(-i\omega \lambda)$ instead of with $\exp(+i\omega \lambda)$, thereby obtaining:
  \begin{equation}
  P_-(\omega;x^a_\perp) \equiv \int_{-\infty}^\infty \frac{ \mathrm{d} \lambda}{2 \pi}\exp(-i\omega \lambda)\, G(x,y) = \int_{-\infty}^\infty \frac{ \mathrm{d} \lambda}{2 \pi} \exp(-i\omega \lambda)\, G(\lambda,x^a_\perp).
  =P(-\omega).
  \label{pkilling2}
 \end{equation} 
 If the two-point-function has support only for positive or negative frequencies, then one of these two definitions will be more natural than the other. But, in general, $G(\lambda)$ will be a complex function and its Fourier transform with respect to $\lambda$ will have support for both positive and negative $\omega$. Then, the interpretation of power spectrum will depend on whether we use $P_+(\omega)$ or $P_-(\omega)$. As an elementary --- but important --- example, consider the situation when, one of them, say, $P_-(\omega)=\omega n(\omega)=\omega[e^{\beta\omega}-1]^{-1}$ is Planckian. Then $P_+(\omega)=P_-(-\omega)=\omega[1+n(\omega)]$. This difference between $n(\omega)$ and $1+n(\omega)$ corresponds to the existence of spontaneous emission in the interactions. We need to keep this aspect in mind while interpreting the power spectra. 
 As we shall see, this issue is relevant only when we define Fourier transforms with respect to timelike Killing trajectories;  in the spacelike  case,  the Wightman function usually depends on the spacelike separation in a symmetric fashion and this issue does not arise.

 If the spacetime has more than one Killing vector field, then it is possible to introduce a Fourier transform with respect to each one of them and define the corresponding power spectrum. A simple example is provided in the case of the 
Friedmann spacetime, in which the spatial homogeneity provides 3 Killing vector fields corresponding to spatial translations. This symmetry is manifest when we use the   conformal Friedmann co-ordinates in which  the two point function will have a structure $G(x_1,x_2) = G(\eta_1,\eta_2; |\bm{x}_2 -\bm{x}_1|) =  G(\eta_1,\eta_2; |\bm{x}|)$  where $\bm{x} \equiv \bm{x}_1 - \bm{x}_2$. 
The corresponding power spectrum arises most naturally in terms of the Fourier transform with respect to $\bm{x}$ after setting $\eta_1=\eta_2 = \eta$. That is,
\begin{equation}
 P(k,\eta) = \int \frac{ \mathrm{d}^3{\bf x}}{(2 \pi)^3} e^{i\bm{k\cdot {\bf x}}} \, G(\eta,\eta; \bm{x}).
 \label{pketa}
\end{equation} 
This power spectrum will depend on the magnitude of $\bm{k}$ (due to rotational invariance) and on the conformal time $\eta$. But note that since $G(\eta,\eta; \bm{x})= \bk{0}{\phi(\eta,\bm{x}_1)\phi(\eta,\bm{x}_2)}{0}$ it will crucially depend on the choice of the vacuum state $\ket{0}$.

For a generic Friedmann spacetime, this is the only 
 natural definition of the power spectrum. But other interesting possibilities for defining the power spectrum exist in the context of de Sitter spacetime which has an intrinsic time translational invariance. Both in the static co-ordinate system as well as in the Painlev\'{e} co-ordinate system, the metric in \eq{A50} (with $H(t)=$ constant) and \eq{A16} exhibit translational symmetry with respect to cosmic time $t$ and the static time $\tau$ corresponding to the Killing vector field with components $\xi^a=(1,\bm{0})$ in these co-ordinates. 
 One can now repeat the analysis leading to \eq{pkilling} using this Killing vector field.%%%%%%%%%% 
 \footnote{The Killing vector with components $\xi^a=\delta^a_0$ in Painlev\'{e} co-ordinates will correspond to a vector with components $\xi^a_{\rm Friedmann} =(1,-H x)$ in the Friedmann co-ordinates. Similarly, the Killing vector with components $\xi^a=\delta^a_0$ in static co-ordinates will correspond to a vector with components
 $\xi_{\rm Friedmann}^a=(\partial_{\tau}t,\partial_{\tau}x)=(1,-H x/(1-H^2e^{2Ht}x^2))$ in the Friedmann co-ordinates.}
 The adapted co-ordinate system is then just the Painlev\'{e} or static co-ordinates and the Killing parameter $\lambda$ will coincide with $t$ or $\tau$.
 If we choose a vacuum state which respects the time translational symmetry, then the corresponding Wightman function will have the structure $G(t_2-t_1; \bm{r}_2,\bm{r}_1)$ in the Painlev\'{e} co-ordinates and similarly have the form $G(\tau_2-\tau_1; \bm{r}_2,\bm{r}_1)$ in the static spherically symmetric co-ordinates. This will happen, for example, if the vacuum state is defined using mode functions which are positive frequency with respect to $t$ which --- as we noted earlier --- is the same as the mode functions being positive frequency with respect to $\tau$. Writing $\tau = \tau_2-\tau_1$ and taking $\bm{r}_1=\bm{r}_2=\bm{r}$,  we again have a natural power spectrum defined through the equation 
\begin{equation}
 P(\omega,\bm{r}) = \int_{-\infty}^\infty \frac{ \mathrm{d} \tau}{2\pi} e^{i\omega \tau} \, G(\tau; \bm{r}).
 \label{pomegar}
\end{equation}

  The situation which will concern us in the later sections, will correspond to one in which the chosen vacuum state respects the geometrical symmetry of the underlying spacetime. In the case of de Sitter spacetime and flat spacetime (which arises in the limit $H\to 0$ of the de Sitter spacetime), the relevant geometrical symmetry is de Sitter invariance and the Lorentz invariance respectively.  When the vacuum state respects these symmetries, the Wightman function $G(x_1, x_2)$  will depend on the co-ordinates only through the geodesic distance $\ell(x_1,x_2)$; that is, $G(x_1, x_2)= G[\ell(x_1,x_2)]$. The Killing symmetries of the spacetime now manifest in terms of the dependence of $\ell(x_1,x_2)$ on the co-ordinates. In particular, if the two events are situated along the integral curve of a Killing trajectory with $x_1=x(\lambda_1), x_2 = x(\lambda_2)$ then $\ell$ will have the structure $\ell(x_1,x_2) = \ell(\lambda,x_\perp)$ where $\lambda \equiv \lambda_1 - \lambda_2$. The Fourier transform with 
respect 
to 
$\lambda$ --- which determines the power spectrum --- now depends essentially on the dependence of $\ell$ on $\lambda$. 
  
  Let us illustrate these abstract ideas in terms of two concrete examples. Consider first the inertial vacuum state in flat spacetime which respects Lorentz invariance so that the relevant two point function depends on the geodesic distance $\ell(x_1,x_2)$ between the two events (apart from an imaginary $(t-t')\log{\ell}$, which renders the Wightman function complex, leading to the commutator structure of the field, something we will come back to later). In the standard inertial co-ordinate system the existence of  a Killing vector corresponding to translations in Minkowski  co-ordinates  implies that $\ell $ will have the form $\ell(T,\bm{X}) $ where $T\equiv T_2-T_1$ and $\bm{X} \equiv \bm{X}_2 - \bm{X}_1$. The power spectrum corresponding to translations in Minkowski time co-ordinate can now be defined as
  \begin{equation}
   P(\Omega, \bm{0}) = \int_{-\infty}^\infty \frac{ \mathrm{d}T}{2\pi}e^{i\Omega T} G[\ell(T,\bm{0})].
   \label{psin}
  \end{equation} 
  
  On the other hand, we also have a Killing vector in flat spacetime corresponding to the Lorentz boosts\footnote{This vector corresponds to the translational symmetry in Rindler time and has components $\xi^a_{boost}=\delta^a_0$ in the Rindler frame which leads to the components $\xi^a_{boost}=N (x^1,x^0,0,0)$ in the inertial frame} which is time-like, for example, in the right and left Rindler wedges, $|\bm{X}|>|T|$. In the Rindler co-ordinate system,  the Lorentz boost symmetry manifests itself as translational symmetry in Rindler time co-ordinate.  The geodesic distance, expressed in terms of Rindler co-ordinate in the right wedge will have the form $\ell = \ell(\tau, \bm{x}_2, \bm{x}_1)$ where $\tau \equiv \tau_2 - \tau_1 $ is the Rindler time difference between the events with spatial Rindler co-ordinates $\bm{x}_2 $ and $\bm{x}_1$. We can now define the power spectrum by taking $\bm{x}_2=\bm{x}_1=\bm{x}$ and Fourier transforming $G$ with respect to $\tau$. This will give:\footnote{The $P(\omega, \bm{x}
)$ 
will 
be 
a Planckian with a suitably red-shifted Rindler temperature. These Fourier transforms with respect to a time co-ordinate occur in the response of Unruh-DeWitt detectors because these detectors --- though thought of as `particle' detectors --- actually  respond to the quantum fluctuations  by their very construction. We prefer to keep the discussion more general, allowing power spectra to be defined either by spatial Fourier transform or by temporal Fourier transform, depending on the context.}
  \begin{equation}
   P(\omega, \bm{x}) = \int_{-\infty}^\infty \frac{ \mathrm{d}\tau}{2\pi} e^{i\omega \tau} G[\ell(\tau,\bm{x})].
   \label{psrin}
  \end{equation} 
  We stress that the vacuum state has \textit{not} been changed when we go from \eq{psin} to \eq{psrin} and we have only transformed the Wightman function treating it as a biscalar on the co-ordinates. One could have also computed a different Wightman function corresponding to, say, the Rindler vacuum state, and evaluated its power spectrum with respect to Rindler time co-ordinate which, of course, would have led to a different result. More importantly this Wightman function constructed from the Rindler vacuum will not be a function of $\ell(x_2,x_1)$ alone.
  
  We will see later that the situation  is conceptually similar --- but algebraically more complicated --- in the case of de Sitter spacetime. 
  The existence of spatial or temporal Fourier transforms  allows us to define three natural power spectra in the context of de Sitter spacetime. We shall briefly mention them here, postponing their detailed discussion to later sections: 
 
 (a) To begin with, one can choose the Friedmann co-ordinates and define a vacuum state by some physical criterion and compute the Wightman function.  In the literature, one often uses a quantum state called Bunch-Davies vacuum $\ket{0,\rm BD}$ for this exercise, which respects the de Sitter invariance. Therefore,  the Wightman function
 $G_{\rm BD}(\eta,\bm{x}) \equiv \bk{0, \rm BD}{\phi(\eta,\bm{x}_2)\phi(\eta,\bm{x}_1)}{0,\rm BD}$ actually depends only on the geodesic distance between the two events for the \textit{massive} scalar field. (There are some subtleties in the case of the massless field which we will discuss later on.)
 We can then evaluate the $P(k,\eta)$ as the spatial Fourier transform of $G_{\rm BD}(\eta,\bm{x}) \equiv \bk{0, \rm BD}{\phi(\eta,\bm{x}_2)\phi(\eta,\bm{x}_1)}{0,\rm BD}$. (This definition is used extensively in the study of inflationary perturbations.) 
 
 (b) One can instead decide to use the static co-ordinates and a vacuum state $\ket{0, \rm ss}$ defined through positive frequency modes with respect to $\tau$, leading to the Wightman function $G_{\rm ss}(\tau,\bm{r}) \equiv \bk{0, \rm ss}{\phi(\tau_2,\bm{r})\phi(\tau_1,\bm{r})}{0, \rm ss}$. 
 (This vacuum state $\ket{0, \rm ss}$ --- in contrast to $\ket{0,\rm BD}$ --- is \textit{not} de Sitter invariant and hence we cannot express $G_{\rm ss}$ as a function of the geodesic distance alone.)
 We can, however, use the definition in \eq{pomegar} to define the corresponding power spectrum. In particular, an observer at the spatial origin will define the power spectrum to be $P(\omega, \bm{0})$ by Fourier transforming 
 $G_{\rm ss}(\tau,\bm{0})$ with respect to $\tau$.
 
 (c) The  two choices mentioned above are rather natural. It is also possible to define yet another power spectrum. Notice that an observer at the origin of the static co-ordinate system is a geodesic observer. It is therefore possible to take the Wightman function $G_{\rm BD}$ defined using the Bunch-Davies vacuum, transform it as a biscalar to the static co-ordinate system and evaluate the power spectrum by Fourier transforming with respect to $\tau$ with, say, at $\bm{r}_1 = \bm{r}_2 =0$ . In other words, one can define \textit{two} different power spectra for the Bunch-Davies vacuum by Fourier transforming either with respect to the spatial co-ordinates or with respect to the static time co-ordinate at the spatial origin. 
 
 In general, we do not expect the power spectra defined by these three procedures (a), (b) and (c) to have any simple relation with each other. However, we will find that it is actually possible to relate them to each other and provide a physical interpretation for the power spectrum.  This will be one of tasks we will address in the later sections.

 It is worth emphasizing the role played by de Sitter invariance (or its absence) in these constructions.  
Whenever we can choose a de Sitter invariant vacuum state the Wightman function will only depend on the de Sitter geodesic distance. If we express such a Wightman function in the Friedmann co-ordinates, spatial homogeneity and isotropy implies that it will have the form $G= G[\ell(\eta_1,\eta_2; |\bm{x}|)]$ where $\bm{x} \equiv\bm{x}_1 - \bm{x}_2  $. One can now define a power spectrum  by Fourier 
   transforming this expression with respect to $\bm{x}$ and setting $\eta_1 = \eta_2 = \eta$. This is what is usually done in the literature, especially in the context of inflationary models and corresponds to item (a) in the previous paragraph.
   
 As we mentioned it is indeed possible to define another power spectrum for the \textit{same} de Sitter invariant vacuum state. When we use the Painlev\'{e} or spherically symmetric co-ordinate systems, the line interval is invariant under corresponding time translations. This implies that when the same Wightman function is expressed in, say, spherically symmetric co-ordinate system, it will have the structure $G = G[\ell(\tau,\bm{x}_1, \bm{x}_2)]$, where $\tau = \tau_2 - \tau_1$. (This is clear from the functional form of $Z$ in \eq{zstatic}.)
   We can Fourier transform this expression with respect to $\tau$ and define another power spectrum.  In doing this, we are retaining the same de Sitter invariant vacuum state and are merely transforming the Wightman function as a biscalar in the co-ordinates. 
   (This is analogous to expressing the inertial vacuum of flat spacetime in two different co-ordinate systems and computing two different power spectra). This corresponds to item (c) in the earlier discussion.
   
Finally, one can also compute the Wightman function in a vacuum state adapted to the spherically symmetric co-ordinate system, viz the static vacuum defined through  positive frequency modes with respect to $\tau$ (This is  analogous to the Rindler vacuum.)
  This Wightman function, however, will not be a function of $\ell$ alone (since the static vacuum is not de Sitter invariant) but will depend only on $\tau\equiv\tau_2-\tau_1$ because of the static nature of the metric. Using this feature, we can  define yet another power spectrum
  by Fourier transforming this Wightman function with respect to $\tau$. 
  This power spectrum will, of course, be quite different from the previous ones. We shall discuss these features in detail in later sections.

\section{Same actions lead to same physics:  The de Sitter spacetime hiding in  power law expansion}\label{sec:hiding}

  We will begin our discussion by proving an equivalence between different Friedmann models, as far as the dynamics of a massive scalar field is concerned. It turns out that the dynamics of a scalar field $\phi$ with mass $m$, living in a Friedmann universe with expansion factor $a(\eta)$, is identical to the dynamics of another scalar field $\psi$ with a mass $M$, living in another Friedmann universe with an expansion factor $b(\eta)$. This equivalence, in particular, allows the mapping of the dynamics of (1) a massless scalar field in a  Friedmann universe with a power law expansion to that of (2) a massive scalar field in a de Sitter universe. We will first prove the equivalence, which is relatively straightforward, and then describe its consequences. 
  
  The action for a scalar field $\phi$ with mass $m$ in a Friedmann universe, described by the expansion factor $a(\eta)$ in conformal Friedmann co-ordinates, is given by
  \begin{equation}
 {\cal A}= \frac{1}{2} \int  \mathrm{d}\eta\,  \mathrm{d}^3 \bm{x}\, a^2 \left[ \dot \phi^2 - |\nabla \phi|^2 - m^2 a^2 \phi^2\right],
   \label{tp5}
  \end{equation} 
  where an overdot indicates the time derivative with respect to $\eta$. 
  Let us introduce a function $\mathcal{F}(\eta)$ and make a field redefinition from $\phi(x)$ to $\psi(x) \equiv \phi(x)/\mathcal{F}(\eta)$. The action in \eq{tp5} can now be rewritten in terms of the new field $\psi$. Expanding out $\dot{\phi}^2$, we will get  terms involving $\dot\psi^2, \psi^2 $ and a cross term containing $\psi\dot \psi$.  By doing an integration by parts and ignoring the boundary term in the action, the cross term involving $\psi\dot\psi$ can be expressed as a term containing $\psi^2$. This allows us, after some algebraic simplifications, (see Appendix \ref{app:derb}) to express the action in the form 
  \begin{equation}
  {\cal A} = \frac{1}{2} \int  \mathrm{d}\eta\,  \mathrm{d}^3 \bm{x}\, b^2 \left[ \dot \psi^2 - |\nabla \psi|^2 - M^2 b^2 \psi^2\right],
   \label{tp6}
  \end{equation}
  where $M^2$ is a constant, $b^2 = a^2 \mathcal{F}^2$ and  $\mathcal{F}$  is chosen to satisfy  the differential equation 
  \begin{equation}
 \frac{\ddot{\mathcal{F}}}{\mathcal{F}} + \frac{2\dot a}{a} \, \frac{\dot{\mathcal{F}}}{\mathcal{F}} + a^2 m^2 = a^2 \mathcal{F}^2 M^2.
   \label{tp7}
  \end{equation} 
   This action in \eq{tp6} represents a scalar field $\psi$ of mass $M$ in a universe with expansion factor $b(\eta)=a(\eta)\mathcal{F}(\eta) $ with $\mathcal{F}(\eta)$ determined as a solution to \eq{tp7}. Given a scalar field with mass $m$ in an Friedmann universe with expansion factor $a(\eta)$, we can solve \eq{tp7}, determine $\mathcal{F}(\eta)$ and thus transform from the system $[a(\eta), \phi(x), m]$ to the system $[b(\eta), \psi(x), M]$ \footnote{This duality can also be trivially extended to interacting theories as well. An interaction term of the kind  $\lambda_{(n)} \phi^n$ in the Lagrangian of $[a(\eta), \phi(x), m]$ set is mapped to $\tilde{\lambda}_{(n)} \psi^n$ in $[b(\eta), \psi(x), M]$, with  $\tilde{\lambda}_{(n)} = \lambda_{(n)} \mathcal{F}^{n-4}$. Such a duality allows us to handle interacting theories in cosmological backgrounds through a new approach. In this paper though, we will be concerned with free fields only.}. Clearly the physics of both these fields will be identical; this 
fact is useful in several conceptual and mathematical contexts.
  
  Our specific interest will be  in the context of a massless scalar field ($m=0$) in a Friedmann universe with a power-law expansion with $a(\eta) \propto (H\eta)^{-q}$ for some constant parameter $H$. In this case, 
  \eq{tp7} has the solution $\mathcal{F}\propto(H\eta)^{-k}$ with $k=1-q$
  (in suitable dimensionless units).  The mass of the rescaled scalar field is given by $M^2 = H^2 (2+q) (1-q)$. What is interesting for our purpose is that the new expansion factor is given by 
  \begin{equation}
 b = a\mathcal{F} \propto(H\eta)^{-(k+q)} \propto(H\eta)^{-1},
 \label{effM1}
  \end{equation}
  which is just the de Sitter spacetime in conformal Friedmann co-ordinates. In other words, a massless scalar field in a Friedmann universe with power law expansion $a(\eta) = a_0(H\eta)^{-q}$, has the same physics as a massive scalar field with mass parameter $M^2 = H^2 (2+q) (1-q)$ in a de Sitter spacetime. 
  
  As an aside, we mention that these ideas can be extended to include a, non-minimal, curvature coupling  term  of the kind $-\xi R \phi^2$ in the action. We can again provide a mapping between the set of quantities $[a(\eta), \phi(x), m, \xi]$ and $[b(\eta), \psi(x), M, \xi]$, through  a straightforward generalization of \eq{tp7} (See Appendix \ref{app:derb} for details). A massless field in the power law universe will then get mapped to to a field with mass $M$, where
    \begin{equation}
   M^2 = (1-6\xi)(2+q) (1-q) \nonumber
  \end{equation} 
   in the units of $H^2$, in the de Sitter spacetime. From this relation, we  see immediately that, in the case of the conformally invariant coupling $\xi =1/6$, we get  a massless theory in the de Sitter spacetime  as well, which is to be expected. However, the range of stability of the theory now  depends on the value of $6\xi$ as well.  We hope to pursue this and related issues in a subsequent work. In this paper, we will continue to deal with minimally coupled scalar fields, i.e., with $\xi =0$.

Returning to the minimal coupling, we note that the expression for the effective mass   $M^2 = (2+q) (1-q)$ can be written more symmetrically and usefully by introducing a parameter 
  \begin{equation}
\nu\equiv q+\frac{1}{2}=\frac{3}{2}\frac{(w-1)}{1+3w},
\label{nufirstdef}
  \end{equation} 
  in place of $q$ so that the effective mass becomes:
  \begin{equation}
   M^2 = (2+q) (1-q)=\left(\frac{3}{2}+\nu\right)\left(\frac{3}{2}-\nu\right)=\frac{18 w(w+1)}{(1+3w)^2}.
   \label{instability}
  \end{equation} 
 
  Clearly, this function remains positive only for $-3/2<\nu<3/2$ and the mass will turn tachyonic for $\nu$ outside this range. Such a theory will be pathological. We can therefore conclude, \textit{without any detailed analysis}, that the massless scalar field theory in a power law universe will exist only if $-3/2<\nu<3/2$.
  In terms of the equation of state parameter this corresponds to the condition $w>0$ (when we exclude the phantom regime with $w<-1$ \cite{Onemli:2004mb}). In other words \textit{we expect a massless scalar field to exhibit a pathology in any power law universe with a source having negative pressure,} including, of course, the de Sitter universe which is just a \textit{special} case. We will see later by explicit analysis that the theory does not exist for $w<0$.
  
  It is, of course, possible to verify this result in terms of the field equations satisfied by the respective scalar fields in the two spacetimes. In the conformal Friedmann co-ordinates, we can choose the fundamental solution to  the scalar field equation to have the form $f_k(\eta)\exp(i\bm{k\cdot x})$ with the general solution obtained by superposing these solutions for different $\bm k$.  The dynamics is contained in the mode functions $f_k(\eta)$. In the case of a  power law universe, these mode functions $f_{\mathbf{k}}(\eta)$ satisfy the equation 
  \begin{equation}
   \eta^2\frac{d^2 f_{\mathbf{k}}}{d\eta^2}-2q\eta\frac{df_{\mathbf{k}}}{d\eta}+k^2\eta^2f_{\mathbf{k}}=0,
  \end{equation} 
  where $q=p/(p-1)$ where $a(\eta) \propto (H\eta)^{-q} \propto t^p$ for some constant $H$. (In the de Sitter limit, corresponding to $p\to \infty $ we take $q=1$.)  
 It then follows that the differential equation satisfied by the rescaled function $h_{\mathbf{k}}\equiv (H\eta)^{1-q}f_{\mathbf{k}}$ (corresponding to a field redefinition) is given by
\begin{eqnarray}\label{scaled_var}
	\eta^2\frac{d^2 h_{\mathbf{k}}}{d\eta^2}-2\eta\frac{dh_{\mathbf{k}}}{d\eta}+(k^2\eta^2-q^2-q+2)h_{\mathbf{k}}=0.
\end{eqnarray}
Comparing \eq{scaled_var} with the differential equation corresponding to the mode function of a scalar field of mass(squared) $M^2$ in de Sitter, namely,
\begin{eqnarray}
		(H\eta)^2\frac{d^2 h_{\mathbf{k}}}{d\eta^2}-2(H^2\eta)\frac{dh_{\mathbf{k}}}{d\eta}+\left[k^2(H\eta)^2+M^2\right]h_{\mathbf{k}}=0,
		\label{effM2}
\end{eqnarray}
we identify that $M^2=H^2(q+2)(1-q)$.
This reproduces the previous result obtained in terms of the action principle with the correct dimensional constants \footnote{In the subsequent discussions we will not explicitly write the parameter $H$ in expressions where its explicit appearance is not necessary. These expressions are to be understood with proper $H$ scalings.}. 
  
This equivalence is extremely useful and allows us to discuss the physics of two separate situations at one go. These two situations corresponds to (i) massive scalar field in a de Sitter universe and (ii) massless scalar field in a universe with power law expansion.  It is also possible to make use of the existence of a hidden de Sitter expansion to address some other interesting issues. For example, it is not easy to define a natural vacuum state for a massless scalar field living in a power law universe. But if we map it to a massive scalar field in a de Sitter universe, we can make use of the de Sitter invariant vacuum states available for massive fields in de Sitter.  

This equivalence also provides a mapping between quantum correlators and -in particular, the Wightman functions. From the scaling  $\phi = \mathcal{F}\psi$ with $\mathcal{F}= \eta^{q-1}$, we immediately see that the Wightman function $G_\phi$ for $\phi$ can be expressed in the form $G_\phi (x_2,x_1) = (\eta_2\eta_1)^{q-1} G_\psi (x_2,x_1)$. But if we define the vacuum state for the $\phi$ field using the hidden de Sitter invariance, then $G_\psi$ will have the form $G_\psi [\ell(x_2,x_1)]$ where $\ell(x_2,x_1)$ is the geodesic distance in the de Sitter spacetime. So we have the result
  \begin{equation}
   G_\phi (x_2,x_1) = (\eta_2\eta_1)^{q-1} G_\psi [\ell (x_2,x_1)],
   \label{ggrelation}
  \end{equation} 
  connecting the Wightman function of a massless scalar field in a power law universe to that of a massive scalar field in a de Sitter universe in a preferred vacuum state. We shall make use of this equivalence extensively in our analysis. Most of the time we will concentrate on massive fields in de Sitter spacetime but these results can be translated for massless fields in power law universes if we choose the de Sitter invariant vacuum for these fields. 
  
Finally, we mention that \eq{tp7} is applicable even for a  general Friedmann universe in which $a(\eta) $ is \textit{not} a power law.  In fact, the entire analysis can be generalized for  scalar fields in  \textit{arbitrary} curved spacetimes along the following lines. One can show that (see Appendix \ref{app:derb}) the dynamics of a scalar field $\phi$ with mass $m$ in a spacetime with metric $g_{ab}$ is the same as the dynamics of another scalar field $\psi = e^{\Lambda}\phi$ with mass $M$ in a spacetime with metric $q_{ab} = e^{-2\Lambda} g_{ab}$ where $\Lambda(x)$ is the solution to the differential equation $\Box \Lambda + (\partial \Lambda)^2 + m^2 e^{2\Lambda} = M^2$. (The derivative operations are carried out with $q_{ab}$ in this equation.) We hope to analyze the more general cases in a future work.

\section{Mode functions and their limiting forms}

We will now start discussing several aspects of quantum field theory of a massive scalar field in a Friedmann universe, concentrating on two cases mentioned above: (1) Case A corresponds to a massive scalar field in a de Sitter universe with the massless field treated as a limiting case. 
(2) Case B corresponds to a massless scalar field in a Friedmann universe with power law expansion. The results obtained in Sec. \ref{sec:hiding} tell us that these two cases can be mapped to each other by a suitable redefinition of the field. Taking advantage of this fact, we will discuss the results, most of the time, for Case A and merely quote the special features for case B. 
In particular, the \textit{massless} scalar field in a de Sitter universe is supposed to exhibit several peculiar features, all of which are usually attributed in the literature, to the fact that there are no de Sitter invariant vacuum states for such a field. As we will see, this is not the real cause of trouble and the peculiar features which occur for a massless field in de Sitter also occurs in the context of power law Friedmann spacetimes.

As we saw in Sec. \ref{sec:coor}, it is possible to choose several physically relevant co-ordinates systems to describe the Friedmann universe in general and the de Sitter universe in particular. The natural solutions to the wave equations can be chosen to preserve the symmetries exhibited by these co-ordinates systems. Further the choice of solutions also have implications for the choice of the vacuum state. We will review the solutions in two co-ordinate systems: (i) conformal Friedmann co-ordinates (see Sec.\ref{sec:co-frw}) for both de Sitter and power law cosmologies and (ii) static, spherically symmetric co-ordinates (see Sec.\ref{sec:co-sph})  for the de Sitter case.

\subsection{Conformal Friedmann co-ordinates and the Bunch-Davies vacuum}

We are interested in a massive scalar field $\phi$ obeying the Klein-Gordon equation in a de Sitter spacetime ($a_{\rm dS}(t) = e^{Ht}$; $t\in \mathbb{R}$) and a massless scalar field in a power-law expanding spacetime ($a_{p}(t) = (1+Ht/p)^{p}$)  where $t$ is the cosmic time co-ordinate and $a(t)$ is the expansion factor. With this choice of a power law metric, $\lim_{p\rightarrow\infty} a_p(t) = a_{\rm dS}(t)$, and we obtain the de Sitter spacetime as a limiting case.
The conformal time co-ordinate is defined by $\mathrm{d}\eta =   \mathrm{d} t/a(t)$, with the integration constant chosen so that in the de Sitter and power law cases, we have:
\begin{equation}
\eta_{\rm dS} = -H^{-1}e^{-Ht}; \quad \eta_{p} = -\frac{p}{H(p-1)}a_p^{-\frac{p-1}{p}}(t),
\end{equation} 
With this choice $\eta$ has the same range in both cases:  $\eta \in (-\infty, 0)$ with $\eta\to -\infty$ corresponding to $t\to -\infty$ and $\eta\to 0$ to $t \to \infty$, for $p > 1$. We will now discuss the form of the mode functions.

\subsubsection {Massive scalar field in  de Sitter}

In any Friedmann universe, described in conformal Friedmann co-ordinates, we can take the mode functions to be
\begin{equation}
u_{\mathbf{k}}(\eta, \bm{x}) =  f_{\mathbf{k}}(\eta)\left[\frac{1}{(2\pi)^{\frac{3}{2}}}\exp(i\bm{k}\cdot\bm{x})\right],
\end{equation}
so that all the dynamics is contained in $f_{\mathbf{k}}(\eta)$ which satisfies the equation:
\begin{eqnarray}
\eta^2\frac{d^2 f_{\mathbf{k}}}{d\eta^2}-2\eta\frac{df_{\mathbf{k}}}{d\eta}+(k^2\eta^2+m^2)f_{\mathbf{k}}=0.
\end{eqnarray}
There are two independent solutions to this equation which can be taken to be proportional to $(-\eta)^{3/2}\mathrm{H}_\nu^{(1)}(-k\eta)$ and
$(-\eta)^{3/2}\mathrm{H}_\nu^{(2)}(-k\eta)$ where
\begin{equation}
\nu \equiv \sqrt{\frac{9}{4}-\frac{m^2}{H^2}}.
\label{nudS}
\end{equation} 
Any particular linear combination as the choice for $f_{\mathbf{k}}(\eta)=(-\eta)^{3/2}[A\mathrm{H}_\nu^{(1)}+B\mathrm{H}_\nu^{(2)}]$ will lead to a corresponding definition of `vacuum' state for the quantum field theory. Conventionally, one sets $B=0$ and chooses $f_{\mathbf{k}}$ to be proportional to $(-\eta)^{3/2}\mathrm{H}_\nu^{(1)}$ (We will comment on the reasons for this choice later on). In quantum field theory the commutation rules for the field uniquely fixes the overall normalization of the solution, except for a constant phase. This leads to the following expression for the mode function:
\begin{equation}
f_{\mathbf{k}}(\eta) = \frac{\sqrt{\pi}H}{2} e^{i\theta}e^{\frac{i\nu\pi}{2}} (-\eta)^{\frac{3}{2}} \mathrm{H}_\nu^{(1)}(-k\eta),
\label{fkdSmassive}
\end{equation}
where $\theta$ is a constant phase. These mode functions define a vacuum state called the Bunch-Davies vacuum.

Conventionally one sets $\theta=0$ which is acceptable for most purposes. But the solution with $\theta=0$ will \textit{not} have the correct limit when $H\to0$. In this limit we expect the positive frequency  mode functions to become proportional to $\exp(-i\omega_k t+i\bm{k}\cdot\bm{x})$ with $\omega_k=\sqrt{k^2+m^2}$. To ensure this, it is necessary to make a specific, non-zero,  choice for $\theta$. This analysis is algebraically nontrivial so we shall just mention the key points, delegating the details to Appendix \ref{app:derc}. (We include this discussion since we have not seen this aspect addressed explicitly in the previous literature.)

Let us study  the behaviour of the  mode function in \eq{fkdSmassive} as $H\rightarrow 0$. In this limit,
the parameter $\nu$ approaches infinity along the positive imaginary axis with
$
\nu \approx im/H\equiv i\mu.
$
The conformal time is also affected by this limit and we must  consider both an $\mathcal{O}(H^{-1})$ (the dominant term) and $\mathcal{O}(H^0)$ part to arrive at any conclusions about the dependence of the modes on the cosmological time $t$. We can re-write
the relevant limiting form of $f_{\mathbf{k}}(\eta)$, now expressed as a function of t, as:
\begin{equation}
\label{fasymp0}
f_{\mathbf{k}}(t) \approx \frac{\sqrt{\pi} H}{2} e^{i\theta}e^{-\frac{\mu\pi}{2}}\left(\frac{1-Ht}{H}\right)^{\frac{3}{2}} \mathrm{H}_{i\mu}^{(1)}(\mu z),
\end{equation}
where $z \equiv (k/m)(1-Ht)$, a positive real number that is kept finite as $H\rightarrow 0$. 
We are therefore essentially interested in the asymptotic form of $\mathrm{H}_{i\mu}^{(1)}(\mu z)$ for fixed $z$ (the ${\cal O}(H)$ correction to $z$ is small, allowing such a treatment) as $\mu \rightarrow \infty$. It turns out that such asymptotic forms for large \textit{order} of the Hankel function are not easy to find; fortunately, we located a previous work \cite{dunster} which has the relevant result. The leading behaviour is given by:
\begin{equation}
\mathrm{H}_{i\mu}^{(1)}(\mu z) \simeq \left(\frac{2}{\pi\mu}\right)^{\frac{1}{2}} e^{\frac{\pi\mu}{2}} e^{-\frac{i\pi}{4}} (1+z^2)^{-\frac{1}{4}} e^{i\mu\xi(z)}\qquad (\mu \to \infty),
\end{equation}
where \begin{equation}
\xi(z) = (1+z^2)^{\frac{1}{2}} + \ln\left(\frac{z}{1+(1+z^2)^{\frac{1}{2}}}\right).
\end{equation}
The rest of the analysis is relatively straightforward and one can show that our mode function has the following limiting form when $H\to0$:
\begin{equation}
f_{\mathbf{k}}(t) \to e^{i\theta}\left[e^{-\frac{i\pi}{4}+\frac{i}{H}\left(\omega_k + m\ln\left(\frac{k}{\omega_k + m}\right)\right)}\right] \frac{e^{-i\omega_k t}}{\sqrt{2\omega_k}}\qquad (H \to 0).
\end{equation}
So $f_{\mathbf{k}}(t)$ does go over to positive frequency Minkowski mode $\exp(-i\omega_k t)$ when $H\to0$ which is (partial) justification for the choice of modes in de Sitter. But to get it right we need to  choose the phase $\theta$ to be:
\begin{equation}
\theta=\frac{\pi}{4}-\frac{1}{H}\left(\omega_k + m\ln\left(\frac{k}{\omega_k + m}\right)\right).
\end{equation} 
We will, however, continue to work with the mode function in \eq{fkdSmassive} with $\theta=0$ (as is usually done in the literature) when the phase is irrelevant.

We mentioned earlier that de Sitter spacetime has a hidden time translational invariance under the transformation
$t \to t+\tau, \eta \to \eta e^{-H\tau} \text{and }  \bm{x} \to \bm{x}e^{-H\tau}$. In Fourier space, it appropriate to supplement these with the rescaling
 ($\bm{k} \to \bm{k}e^{H\tau}$). It is clear from the form of $f_{\mathbf{k}}$ in \eq{fkdSmassive} that the mode $u_{\bm{k}}=f_{\mathbf{k}}\exp(i\bm{k}\cdot\bm{x})$
has the functional form:
$
u_{\bm{k}}(\eta, x) = k^{-\frac{3}{2}}U(\bm{k}\cdot\bm{x}, k\eta).$
The combinations $\bm{k}\cdot\bm{x}$ and $k\eta$ are invariant under our rescaling. The modes then transform as:
\begin{equation}
u_{\bm{k}}(\eta, \bm{x}) \to u_{\bm{k}e^{H\tau}}(\eta e^{-H\tau}, \bm{x}e^{-H\tau}) = e^{-\frac{3}{2}H\tau}u_{\bm{k}}(\eta, \bm{x}).
\end{equation}
So the modes themselves are \textit{not} invariant under this transformation. However, quantities such as the two-point functions, given by expressions involving integrals over  $  \mathrm{d}^3 k\ u_{\bm{k}}(x)u_{\bm{k}}^\ast(x')$ are invariant under this transformation, because $ \mathrm{d}^3 k \rightarrow e^{3H\tau} \mathrm{d}^3 k$. Thus the two-point functions inherit the hidden time translational invariance of the de Sitter spacetime, as they should.

\subsubsection{Massless scalar field in power-law universe}

\begin{figure*}
\begin{center}
\includegraphics[scale=0.35]{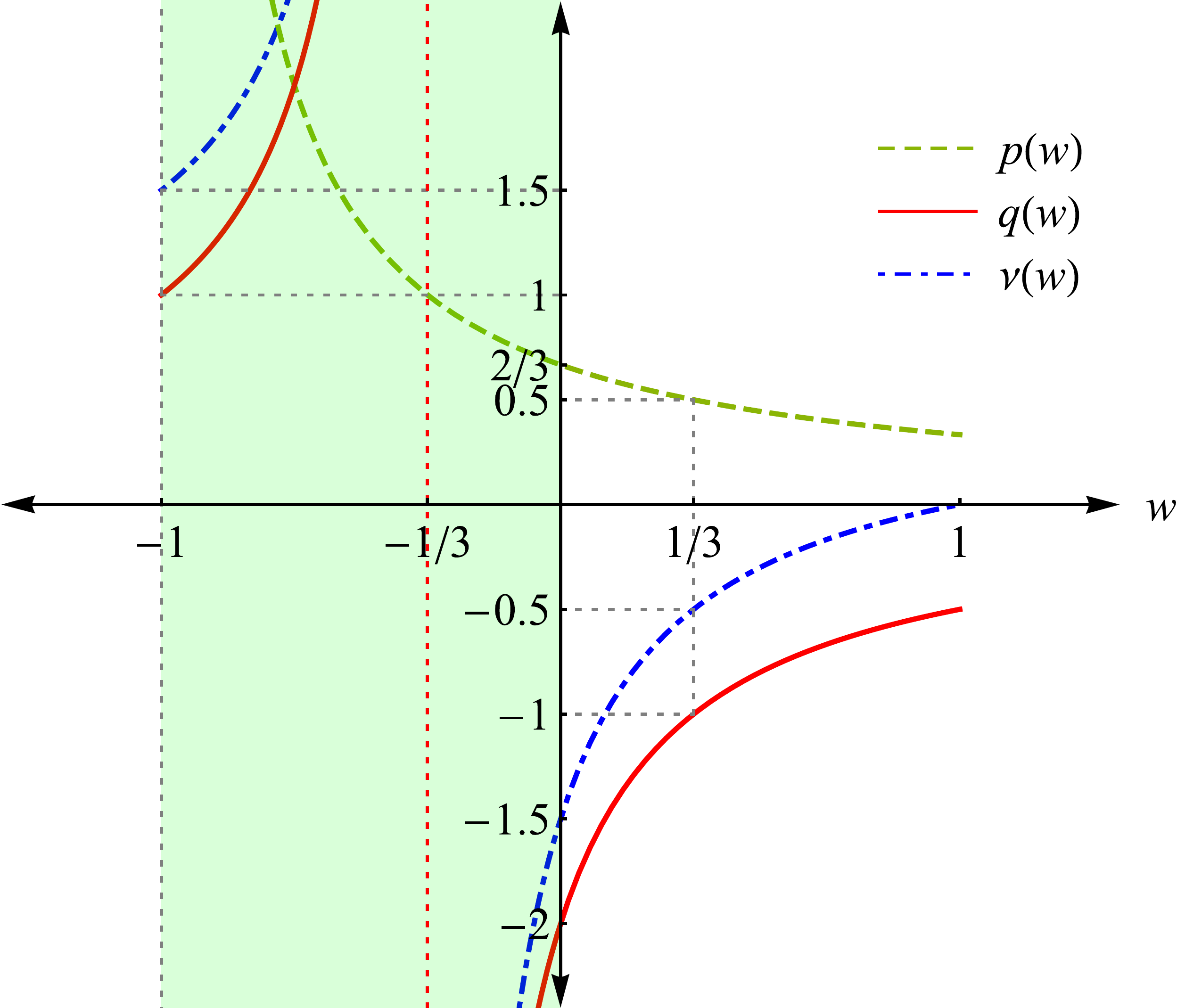}
\caption{Plots for (i) $p(w)=2/(3+3w)$ (ii) $q(w)=-2/(1+3w)$ and (iii) $\nu(w)=3(w-1)/2(1+3w)$. The parameters $q$ and $p$ are defined through the expansion factor as $a(t)=(1+Ht/p)^p=(-H\eta/q )^{-q}$ where, $\eta$ is the conformal time and $t$ is the comoving time and $\nu$  is just $q+(1/2)$. We have taken $w\in[-1,1]$, which corresponds to the physical range of the equation of states. The values of $p$, $\nu$ and $q$ for three special cases of the parameter $w$, viz., (i) $w=0$, the dust (ii) $w=1/3$, the radiation and (iii) $w=-1$, the de Sitter --- that are of main interest in cosmology are also marked. The region $-3/2<\nu<3/2$, where the field theory correlators are finite, which is the region outside the shaded part, maps to the range $0<w<\infty$ when we exclude the unphysical (`phantom') range of $w<-1$. So the field theory correlators are well-defined only if the equation of state parameter is positive semi-definite.}\label{fig_04}
\end{center}
\end{figure*}

We saw earlier that  the dynamics of a massless scalar field in a power-law universe (with $a \propto (H\eta)^{-q} \propto t^p $ with
$q=p/(p-1)$) can be translated to that of a massive field in de Sitter with a field redefinition by a factor $F\propto(H\eta)^{-k}$ where $k=1-q$.
This is indeed what happens when we solve the equation
\begin{equation}
   (H\eta)^2\frac{d^2 f_{\mathbf{k}}}{d\eta^2}-2q(H^2\eta)\frac{df_{\mathbf{k}}}{d\eta}+k^2(H\eta)^2f_{\mathbf{k}}=0,
\end{equation}
for the mode function. The properly normalized positive frequency solution can now be taken as:
\begin{equation}
u_{\bm{k}}(\eta, \bm{x}) = \frac{\sqrt{\pi}}{2}\left(\frac{H}{q}\right)^{q}e^{\frac{i\nu\pi}{2}}(-\eta)^{\nu}\mathrm{H}_\nu^{(1)}(-k\eta) \frac{e^{i\bm{k}\cdot\bm{x}}}{(2\pi)^{\frac{3}{2}}},
\label{fkpowlaw}
\end{equation}
with $\nu$ being the parameter introduced earlier in \eq{nufirstdef} which gets related to the exponent of expansion like many other variables of the theory (see \fig{fig_04}) :
\begin{equation}
\nu =\frac{1}{2}+q= \frac{1}{2}+\frac{p}{p-1} = \frac{3}{2}-(1-q)= \frac{3}{2}-k =\frac{3}{2}\frac{w-1}{1+3w}.
\label{nuuseful}
\end{equation} 
The factor $\eta^\nu=\eta^{3/2}\eta^{-k}$ is the product of the factor $\eta^{3/2}$ which occurs in the de Sitter mode functions (see \eq{fkdSmassive}) and the scaling factor $F\propto\eta^{-k}$ involved in the field redefinition. If we equate the $\nu$ in \eq{nuuseful} (relevant for a massless field in power law universe) with the $\nu$ in \eq{nudS} (relevant for a massive field in de Sitter universe), we can define an effective mass $M$. A simple calculation shows that $M^2=(q+2)(1-q)$ which is the same result we obtained earlier using the scaling arguments (see the discussion around \eq{effM1} and \eq{effM2}).
In fact, defining $\beta \equiv(H/q)^{q}$ and introducing the notation $q = \text{min}(1, \text{Re}\ \nu-\frac{1}{2})$ --- which translates to 
$q=p/(p-1)$ for the power law case with $q=1$ for the de Sitter --- the mode functions for \textit{both} the power law and the exponential expansion can  be written in the same form:
\begin{equation}
u_{\bm{k}}(\eta, \bm{x}) = \frac{\sqrt{\pi} \beta}{2} e^{\frac{i\pi\nu}{2}}(-\eta)^{q+\frac{1}{2}} \mathrm{H}_\nu^{(1)}(-k\eta).
\label{oneforboth}
\end{equation}
Based on our analysis in Sec. \ref{sec:hiding} we expect the theory to exhibit pathologies when $|\nu|>3/2$ when the effective mass turns pathological. As we shall see later, this pathology occurs at the level of two-point functions and the mode functions remain well-defined.

Finally, let us briefly review the choice of positive frequency modes, which in turn, decides the vacuum state. To justify the choice of the modes in \eq{fkpowlaw} as the appropriate positive frequency modes one can proceed in two different ways. First, from the results of Sec. \ref{sec:hiding} and the functional form of \eq{fkpowlaw} we know that these modes arise from the rescaling of positive frequency modes in the case of de Sitter, which justifies the choice. Second, we can again take the limit of flat spacetime 
$a(t) \rightarrow 1$ limit of the expansion factor and verify that the modes in \eq{fkpowlaw} have the correct limiting form. This is fairly straightforward if we think of the flat spacetime limit as arising due to $p\to 0$. In this limit, $q\to0$ and $\nu \to {1/2}$. In this case we have known form for the Hankel function:
\begin{equation}
\mathrm{H}_{\frac{1}{2}}^{(1)}(-k\eta) = -i\sqrt{\frac{2}{\pi}}\frac{e^{-ik\eta}}{\sqrt{-k\eta}},
\end{equation}
along with the result  $\eta \to t$. A simple calculation (see Appendix \ref{app:derc}) now shows that:
\begin{equation}
u_{\bm{k}}(t,\bm{x}) \to \frac{1}{(2\pi)^{\frac{3}{2}} \sqrt{2k}} e^{i\bm{k}\cdot\bm{x}-ikt},
\end{equation}
thereby again justifying the choice of positive frequency modes.\footnote{This procedure corresponds to taking the $p\to0$ limit in \eq{expfactor} keeping $H$ constant. One can also obtain the flat spacetime limit from power-law Friedmann universe in different manner, by taking the $H\to0$ limit in \eq{expfactor}, keeping $p$ constant. This case --- though it leads to the same conclusion --- is slightly more subtle,
and is discussed in Appendix \ref{app:derc}}

\subsubsection{Massless scalar field in de Sitter}

As far as mode functions are concerned, the massless limit of the massive scalar field exhibits no pathologies. 
This limit corresponds to $\nu = {3/2}$ and arises when we take $m \rightarrow 0$ in the de Sitter case or if we take  $p \rightarrow \infty, q\to1$ in the power law expansion. Both mode functions take an identical form  in these limits, giving:
\begin{equation}
u_{\bm{k}}(\eta, \bm{x}) = \frac{\sqrt{\pi}H}{2} e^{\frac{3\pi i}{4}}(-\eta)^{\frac{3}{2}}\mathrm{H}_{\frac{3}{2}}^{(1)}(-k\eta) \frac{e^{i\bm{k}\cdot\bm{x}}}{(2\pi)^{\frac{3}{2}}}
=\frac{H}{\sqrt{2}(2\pi)^{\frac{3}{2}}}e^{\frac{3\pi i}{4}}\left(\frac{k\eta-i}{k^{\frac{3}{2}}}\right)e^{i\bm{k}\cdot\bm{x} - ik\eta}.
\label{masslessdSmodes}
\end{equation}
The second relation follows from the fact that 
the Hankel function of order ${3/2}$ can expressed in terms of elementary functions:
\begin{equation}
H_{\frac{3}{2}}^{(1)}(-k\eta) = \sqrt{\frac{2}{\pi}}(-k\eta)^{-\frac{3}{2}}(k\eta - i)e^{-ik\eta}.
\end{equation}
We will use this form extensively later on.

In our discussions so far, we have used the Bunch-Davies vacuum.
While this  is a vacuum state preferred in the literature when one uses the conformal Friedmann co-ordinates, there are certain subtleties regarding this choice which needs to be emphasized. To begin with, the Bunch-Davies modes are \textit{not} pure positive frequency modes with respect to the conformal time $\eta$ in $(1+3)$ dimension.\footnote{The corresponding modes do evolve as $\exp(-i\omega\eta)$  in $(1+1)$ dimension due to conformal invariance. But we will not be concerned with $(1+1)$ dimensional case in this paper.} They do not have the dependence $\exp(-i\omega\eta)$ \textit{even in the asymptotic past} when $\eta \to -\infty$; in this limit, the mode functions behave like $[1/a(\eta)] \exp(-ik\eta)$ for the massless modes. The extra factor $[1/a(\eta)]$  prevents the pure sinusoidal behaviour even in the asymptotic past. 

The usual trick to circumvent this difficulty is to introduce a field redefinition and work with the field $v(x) \equiv a(\eta) \phi(x)$ where $\phi(x)$ is the original scalar field. This new field  --- closely related to what is called Mukhanov-Sasaki variable in inflationary literature \cite{gravitation, Parker:2009uva, Baumann:2009ds}   --- does have the dependence $\exp(-i k\eta)$ in the asymptotic past. One then quantizes the field $v(x)$ and adopts the resulting Hilbert space structure for the original scalar field $\phi(x)$ as well. 

More importantly, the Bunch-Davies modes are  complicated functions of the cosmic time $t$ containing an, understandable, exponential redshift factor. In the massless case, for example, the positive frequency Bunch-Davies modes in \eq{masslessdSmodes}, expressed in terms of the cosmic time, has the form:
\begin{equation}
f_k (t) = \frac{1}{\sqrt{2k}}\exp\left[-\frac{ik}{H}\left(1- e^{-Ht}\right)\right]\left(\frac{iH}{k} + e^{-Ht}\right).
\end{equation} 
On the other hand,  the de Sitter spacetime \textit{does} have an implicit translational invariance with respect to the cosmic time. So it makes sense to inquire about the positive and negative frequency components of $f_k(t)$ with respect to the cosmic time. These modes, $f_k(t)$, are indeed 
 a superposition of positive and negative frequency waves with respect to the cosmic time $t$. Writing 
\begin{equation}
 f_k (t) = \int_0^{\infty}\,\frac{d\gamma}{2\pi} \left(\alpha_\gamma\,e^{-i\gamma t} + \beta_\gamma\,e^{i\gamma t} \right),
\label{mix}
\end{equation} 
one can determine by inverse Fourier transform the coefficients $\alpha_\gamma$ and $\beta_\gamma$.  A straightforward calculation (see, for e.g., \cite{Singh:2013pxf}) now gives the result
\begin{equation}
|\alpha_\gamma|^2 =\frac{H^2}{2k^3\gamma}\frac{\beta e^{\beta \gamma}}{e^{\beta \gamma}-1}\left(1+\frac{\gamma^2}{H^2}\right); \quad
|\beta_\gamma|^2 = \frac{H^2}{2k^3\gamma}\frac{\beta}{e^{\beta \gamma}-1}\left(1+\frac{\gamma^2}{H^2}\right); \quad
\beta^{-1} = \frac{H}{2\pi}.
\label{thermal2}
\end{equation} 
We see that there is a thermal factor with temperature $H/2\pi$ modified by a kinematic factor $(1+\gamma^2/H^2)$. While the Planck spectrum is modified by this factor, the ratio of the coefficients, 
\begin{equation}
 \frac{|\beta_\gamma|^2}{|\alpha_\gamma|^2}=\exp (-\beta\gamma),
\end{equation} 
remains to be  the standard Boltzmann factor.
(Since $\alpha_\gamma$ and $\beta_\gamma$ are not Bogoliubov coefficients, the condition $|\alpha|^2 - |\beta|^2 =1 $ need not hold.)
We will see similar factors arising later on, in the context of power spectra of the vacuum noise.

\subsection{Static co-ordinates and the cosmic vacuum}\label{sec:staticmodes}

The most natural way to define a vacuum state is in terms of mode functions which are positive frequency with respect to a time co-ordinate.
The metric in Friedmann co-ordinates exhibit a time dependence, thereby preventing solutions to the wave equations which evolve as $\exp(-i\omega t)$, say. For a general Friedmann universe, and even for a universe with a power law expansion, there is no way around this situation. The geometry \textit{is} time dependent and we have to live with that fact. The vacuum state has to defined by some other criterion (like the ones we talked about in the last section) because of this intrinsic time dependence.

The situation, however, is different in the case of de Sitter expansion. The de Sitter universe is inherently time translation invariant and the apparent time dependence in $ a(t)=\exp Ht =\eta^{-1}$ is spurious. So, in this case, we must be able to choose mode functions which are indeed positive frequency with respect to the cosmic time and evolve as $\exp(-i\omega t)$. This choice, in turn, will define an appropriate vacuum state which may be called \textit{cosmic vacuum} since it is defined with respect to the cosmic time. As we saw in Secs. \ref{sec:co-pan} and \ref{sec:co-sph}, the static nature of the metric is manifest in Painlev\'{e} and spherical co-ordinates. The cosmic vacuum, defined by modes which evolve as $\exp(-i\omega t)$ in the Painlev\'{e} co-ordinates is the same as the static vacuum defined by modes which evolve as $\exp(-i\omega \tau)$ in the spherical co-ordinates because these positive frequency solutions map to each other; see \eq{painstatic}. Given the rather natural way in which 
such 
a vacuum 
state arises, it is important to look at the mode functions in the static and Painlev\'{e} co-ordinates.

Let us begin with the de Sitter spacetime described by the static line element in \eq{A16}. We take the 
mode functions to be of the  form
\bea
v_{\omega}(x)= e^{-i\omega \tau} \Phi_{\omega l m}(r)Y_{lm}(\theta,\phi),
\eea
separating out the time dependence and angular dependences. The resulting radial equation has two independent solutions
which are given by
\be
\Phi_{\omega l m}^{(1)}(r) = r^{l} (1-H^2 r^2)^{\frac{-i \omega}{2H}} {}_2F_1\left(\frac{3}{4} +\frac{l}{2}  -\frac{i \omega}{2H} - \frac{ {\nu}}{2}, \frac{3}{4} +\frac{l}{2}  -\frac{i \omega}{2H} + \frac{ {\nu}}{2}, \frac{3}{2}+l ; H^2 r^2   \right), 
\label{staticmode1}
\ee
and
\bea
\Phi_{\omega l m}^{(2)}(r) &=& r^{-l-1} (1-H^2 r^2)^{\frac{-i \omega}{2H}} \times \nonumber\\
&{}&_2F_1\left(\frac{1}{4} +\frac{l}{2}  -\frac{i \omega}{2H} - \frac{ {\nu}}{2}, \frac{1}{4} +\frac{l}{2}  -\frac{i \omega}{2H} + \frac{ {\nu}}{2}, \frac{1}{2}-l ; H^2 r^2   \right),
\label{staticmode2}
\eea
where ${}_2F_1(a,b,c;z) $  is the Gauss hypergeometric function. The  regularity of the solution at $Hr =0$ requires us to choose the first solution in \eq{staticmode1} and discard the second \cite{Pascu:2013wla}. So the regular solution is:
\bea
v_{\omega l m}(r,\theta , \phi) = N_{\omega} e^{- i \omega \tau} Y_{lm}(\theta,  \phi) \Phi_{\omega l m}^{(1)}(r).
\eea
The normalization constant $N_\omega$ has to be fixed in terms of the standard Klein-Gordon inner product. This is straightforward though somewhat algebraically involved. One can show that (see Appendix \ref{app:derd}) the correct normalization leads to:
\bea
|N_{\omega}|^2 = \frac{H^2}{\pi \omega}\left(1+\frac{\omega^2}{H^2}\right). \label{Norm2}
\eea
Finally note that the mode functions in static co-ordinates are related in a simple manner to the mode functions in Painlev\'{e} co-ordinates, through \eq{painstatic}.
So the mode function obtained above continues to be useful in the Painlev\'{e} co-ordinates as well. We stress that quantum field theory built from these modes correspond to a vacuum state which has positive frequency modes with respect to cosmic time $t$, a fact which does not seem to have been properly appreciated in the literature. 

The modes $v_\omega$, which are positive frequency with respect to the static time co-ordinate $\tau$ (with time dependence $\exp(-i\omega \tau)$), define the static vacuum while the modes $f_k$ in \eq{fkdSmassive} and \eq{masslessdSmodes} define the Bunch-Davies vacuum, for massive and massless fields respectively. There exists a non-trivial Bogoliubov transformation between these modes and the Bunch- Davies vacuum will contain static frame `particles'. (This is similar to inertial vacuum containing Rindler `particles'). These Bogoliubov coefficients can be computed using the standard Klein-Gordon scalar product. Such a calculation is drastically simplified by choosing the spacelike hypersurface (on which the scalar product is computed) to be close to the horizon $r=H^{-1}$ where only the $s$-mode makes significant contribution.  Rewriting the Bunch-Davies modes in the static co-ordinate system, one can compute the Bogoliubov coefficients in a relatively straightforward manner (see e.g \cite{Singh:2013pxf}) 
and show 
that 
\begin{equation}
  k|\beta_{\omega k}|^2 = \frac{\beta}{(e^{\beta\omega} - 1)};\hspace{5pt}\beta = \frac{2\pi}{H}.
  \label{thermal1}
\end{equation} 
This shows that the Bunch-Davies vacuum appears to be thermally populated by the static frame particles with temperature $H/2\pi$. This thermal factor will come up later on when we study the power spectrum of vacuum noise.

\section{The Wightman function}

The quantum field theory of a free field in any spacetime is contained in the Wightman function.  It would therefore be logical to work entirely in terms of this function which satisfies the wave equation. Unfortunately, not all solutions of the wave equation $(\Box - m^2)G =0$ qualify as  legitimate Wightman functions because a generic solution will not have the structure $G(x_2,x_1) = \bk{0}{\phi(x_2)\phi(x_1)}{0}$ for a field operator $\phi(x)$ (satisfying the same equation $(\Box - m^2)\phi = 0$) for any normalizable state $\ket{0}$ in a Hilbert space. To circumvent this difficulty, one should either (a) impose specific boundary conditions on the solutions to $(\Box - m^2)G =0$ (usually called the Hadamard condition) to construct a legitimate Wightman function or (b) construct the fundamental solutions to the equation $(\Box - m^2)\phi = 0$, define a suitable vacuum state in terms of, say, the positive frequency solutions to this equation (or by some other criteria) and construct the Wightman function 
from its definition $G(x_2,x_1) = \bk{0}{\phi(x_2)\phi(x_1)}{0}$. 

The approach (a) is particularly useful to construct the Euclidean Green's function which satisfies the equation, $(\Box - m^2)G =\delta$ with a delta function source, (which, on analytic continuation, gives the Feynman Green's function, rather than the Wightman function but one can construct the latter from the former). There is an elegant way of implementing this program using an analogy with electrostatics in $D=5$ Euclidean space, leading to a simple integral representation for the Euclidean Green's function. We will first describe this procedure in the next section \ref{sec:5ded}, and then obtain the same result --- by a brute force method, based on approach (b) --- in Sec.\ref{sec:intrepforw}.

\subsection{Euclidean de Sitter Green's function from $D=5$ electrostatics}\label{sec:5ded}

Consider a 5-D Euclidean flat space time, with the line element (in standard polar co-ordinates) being
$
ds^2=dr^2+r^2~d\Omega_4^2
$
where, $d\Omega_4^2$ is the metric on an unit 4-sphere. The electrostatic potential $\phi$ produced by a charge distribution $\rho(\mathbf{x})$ in this space satisfies the 5-dimensional Poisson equation:
$
-\nabla^2_5\phi = \rho
$
where, $\nabla^2_5$ is the Laplacian in 5D Euclidean space. The solution to this equation  is given by

\begin{eqnarray}
\phi(\mathbf{x})=\int_{R^5} \mathrm{d}V'~\rho(\mathbf{x'})G_5(\mathbf{x},\mathbf{x}');\qquad
G_5(\mathbf{x},\mathbf{x}')=\frac{1}{8\pi^2}\frac{1}{|\mathbf{x}-\mathbf{x'}|^{3}},
\label{gensolution}
\end{eqnarray}
where, $G_5(\mathbf{x},\mathbf{x}')$ is the Green's function corresponding to a delta function source in 5-D Euclidean space.
The expanded form of the Poisson equation, on the other hand, appears as
\begin{eqnarray}
	-\frac{1}{r^4}\partial_r(r^4\partial_r\phi)-\frac{1}{r^2}\nabla^2_{S}\phi=\rho,
\label{polar_GFeq}
\end{eqnarray}
where we have used the polar co-ordinates and $\nabla^2_S$ is the Laplacian on a unit 4-sphere. Our  aim is to  connect this electrostatic Poisson equation to the equation satisfied by the Euclidean Green's function for a massive scalar field on a 4-sphere $S_4(R)$ of an arbitrary radius $R$. This is possible because the 5-D Laplacian in \eq{polar_GFeq}  has two parts: (i) one involving radial derivatives and (ii) one involving `polar' derivatives. If we are interested in the Laplacian of a field $\phi(r,\Omega_4)$ evaluated at the surface of a 4-sphere of radius R, we would get:
\begin{eqnarray}
\nabla^2_5\phi\rvert_{S_4(R)}= \frac{1}{R^4}\partial_r(r^4\partial_r\phi)\rvert_R+\frac{1}{R^2}\nabla^2_{S}\phi(R,\Omega_4).
\label{laplacian_restricted}
\end{eqnarray}
We can  reduce the first term in \eq{laplacian_restricted} to the mass term appearing in the Green's function equation, provided
\begin{eqnarray}
\frac{1}{R^4}\partial_r(r^4\partial_r\phi)\rvert_R=-m^2\phi(R,\Omega_4).
\end{eqnarray} 
If we look for the separable solution for the field, satisfying this condition, we obtain
\begin{eqnarray}
\phi(r,\Omega_4)= r^{-3/2-\nu}f(\Omega_4),
\label{powerpotential}
\end{eqnarray}
where, $\nu\equiv \pm \sqrt{9/4-m^2R^2}$ and $f(\Omega_4)$ is an arbitrary function of the angular co-ordinates and $C$ is a dimensionful constant. We want this potential to satisfy
\begin{eqnarray}
	-R^{-2}\nabla_S^2\phi(R,\Omega_4)+m^2\phi(R,\Omega_4)=R^{-4}\delta_S(\Omega_4,\Omega_4'),
\end{eqnarray}
at the radius $R$ of the sphere, where, $\nabla_S^2$ is the Laplacian operator on a unit 4-sphere. \textit{But this is just the defining equation for the Green's function $G_S$ for a massive scalar field on $S_4(R)$!} 

At any other general point $(r, \Omega_4)$ the Green's function equation yields
\bea
 r^{-4} \left(\frac{r}{R}\right)^{1/2-\nu}\left[m^2R^2 f(\Omega_4) + \nabla^2_{S}  f(\Omega_4)  \right] = \rho({\bf x}).
\eea
The separability condition of the field forces the charge density to adopt a form
\bea
\rho({\bf x}) = k r^{-4} \left(\frac{r}{R}\right)^{1/2-\nu}\delta_S(\Omega_4,\Omega_4'),
\eea
for some constant $k$.  Moreover, we want that the
 $\rho(R)=\delta_S(\Omega_4,\Omega_4')/R^4$. This equation fixes the constant $k$ to unity yielding the form for the charge density
\begin{eqnarray}
\rho(\mathbf{x})=r^{-4} \left(\frac{r}{R}\right)^{1/2-\nu}\delta_{S}(\Omega_4,\Omega_4'),
\label{form_of_rho}
\end{eqnarray}
to be the 
 source for the potential $\phi(r,\Omega_4)$ given in \eq{powerpotential}.
Therefore the explicit expression for $G_S(\Omega_4,\Omega_4')=\phi(R,\Omega_4)$ can be obtained immediately from \eq{gensolution}, leading to:
\begin{eqnarray}
	G_S(\Omega_4,\Omega_4')=\phi(R,\Omega_4)&=& \frac{1}{8\pi^2}\int_0^{\infty}\left(\frac{r'}{R}\right)^{1/2-\nu}\frac{r'^4 \mathrm{d}r'}{(R^2-2Rr'\cos\theta+r'^2)^{3/2} r'^4},\\
	&=& \frac{1}{8\pi^2R^2}\int_{0}^{\infty} \mathrm{d}s\frac{s^{1/2-\nu}}{(s^2-2s\cos\theta+1)^{3/2}},
\label{ed5final}	
\end{eqnarray}
where, $\theta$ is the angle subtended at the origin by the arc joining $(R,\Omega_4)$ and $(R,\Omega_4')$ and $s=r'/R$. This is indeed the Euclidean Green's function (which, on analytic continuation will normally lead to the Feynman Green's function) in the de Sitter space with $R=1/H$ (In order to smoothly go over to the form of  Green's function of the de Sitter going to appear in the paper later on, we choose $\nu\equiv + \sqrt{9/4-m^2R^2} $. We could have chosen any signature for $\nu$, as the Green's function (or the potential $\phi(R,\Omega_4)$) turns out to be symmetric in $\nu$).
The integral in \eq{ed5final}, as we shall see later in Sec. \ref{sec:asideEGF}, can be expressed in terms of the Gauss hypergeometric function.
We shall now obtain the same result by {\it somewhat less elegant} methods using the mode functions.

\subsection{Massive de Sitter and massless power-law}\label{sec:intrepforw}

From the definition $G(x_2,x_1) = \bk{0}{\phi(x_2)\phi(x_1)}{0}$ of the Wightman function, it is obvious that it can be expressed as a mode sum in the vacuum state associated with the modes. So the  
 Wightman function for the Bunch-Davies vacuum is given by:
\begin{eqnarray}
G(\eta, \bm{x};\eta', \bm{x}') &=& \langle 0, \rm BD \rvert \hat{\phi}(\eta, \bm{x})\hat{\phi}(\eta', \bm{x}')\lvert 0, \rm BD\rangle, \\
&=& \frac{\pi\beta^2}{4 (2\pi)^3} (\eta\eta')^{q+\frac{1}{2}} \int\limits_{\mathbb{R}^3}\mathrm{d}^3 k\ \mathrm{H}_\nu^{(1)}(-k\eta) \mathrm{H}_\nu^{(2)}(-k\eta') e^{i\bm{k}\cdot(\bm{x}-\bm{x}')},
\end{eqnarray}
where $\beta=(H/q)^q$ and we have  rewritten $\mathrm{H}_\nu^{(1)*}$ in terms of $\mathrm{H}_\nu^{(2)}$.
Evaluating the angular part of the $\bm{k}$-integral gives, with $\rho = \lvert \bm{x}-\bm{x}'\rvert$:
\begin{equation}
\label{WightmanG_HankelSine_01}
G(\eta,\eta',\rho) = \frac{\beta^2}{8\pi \rho}(\eta\eta')^{q+\frac{1}{2}}\int\limits_0^\infty k\mathrm{d} k\ \mathrm{H}_\nu^{(1)}(-k\eta) \mathrm{H}_\nu^{(2)}(-k\eta') \sin(k\rho).
\end{equation}
Our aim is to reduce this expression to a simple integral representation so that we can ascertain its properties and --- in particular --- decide when this integral exists. This requires rather involved algebraic manipulations with special care, to handle the subtleties involved in the convergence of various expressions. We will outline the steps here delegating the details to Appendix \ref{app:dere}. 

The first step is to use a standard  integral representation for Hankel functions (see 10.9.10, 10.9.11 of \cite{DLMF}) and write the integrand in \eq{WightmanG_HankelSine_01} as a double integral. Such a procedure is usually used to obtain an integral representation for the products like $\mathrm{K}_\nu(iz)\mathrm{K}_\nu(iz')$ in literature (see e.g., 13.71 of \cite{watson}). Unfortunately we cannot use this approach directly because the integral representation for Hankel functions have restrictions on the phase of the integrand which are difficult to incorporate. Nevertheless it is possible to manipulate the expressions carefully and arrive at the following representation for the Wightman function:
\begin{eqnarray}
\label{DefIntRep_Wightman}
G(\eta, \eta'; Z) &=&
\frac{\beta^2}{16\pi^2\sqrt{2}}(\eta\eta')^{q-1}\int\limits_{-\infty}^{\infty} \mathrm{d} u \frac{e^{-\nu u}}{\left(\cosh u - Z\right)^{\frac{3}{2}}},\\ 
&=&\frac{\beta^2}{8\pi^2\sqrt{2}}(\eta\eta')^{q-1}\int\limits_{0}^{\infty} \mathrm{d} u \frac{\cosh\nu u }{\left(\cosh u - Z\right)^{\frac{3}{2}}},
\label{CoshIntRep_Wightman0}
\end{eqnarray} 
where 
\begin{equation}
Z = \frac{\eta^2 + \eta'^2-\rho^2}{2\eta\eta'} = 1+ \frac{(\eta - \eta')^2-\rho^2}{2\eta\eta'}, 
\label{Zdef111}
\end{equation}
is related to the geodesic distance $\ell(x,x')$ by
$H\ell(x,x') \equiv \cos^{-1} Z(x,x')$; see \eq{zfrw}.
It is also implicitly understood that $(\eta-\eta')$ is to be treated as the limit of  $(\eta-\eta'-i\epsilon)$ when  $\epsilon\to0^+$ to obtain the Wightman function  while the Feynman Green's function is obtained by treating $(\eta-\eta')^2$ as the limit of $(\eta-\eta')^2-i\epsilon$.

Another useful integral representation, involving polynomials in the integration variable, can be obtained by substituting $s = e^u$ in
\eq{DefIntRep_Wightman} leading to:
\begin{equation}
G(\eta, \eta'; Z) = \frac{\beta^2}{8\pi^2}(\eta\eta')^{q-1}\int\limits_{0}^{\infty} \mathrm{d} s \frac{s^{\left(\tfrac{1}{2}-\nu\right)}}{\left(s^2-2Zs+1\right)^{\frac{3}{2}}}.
\label{intreppoly}
\end{equation}
On analytic continuation, in the de Sitter limit of $q=1$, this expression  reduces to \eq{ed5final}, obtained earlier from the $D=5$ electrodynamics.
We immediately see that, when the integral exists, $G(x,x')$ has the structure $G(x,x')=(\eta\eta')^{q-1}G_{\rm dS}[\ell(x,x')]$. This is precisely what we concluded in Sec. \ref{sec:hiding} from the fact that massless fields in power-law universes can be mapped to massive field in a de Sitter universe by a field redefinition, which accounts for the $(\eta\eta')^{q-1}$ factor; see \eq{ggrelation}.

It is instructive to  verify that this expression has the correct flat spacetime limit when $H\to0$. This can be done directly from \eq{intreppoly} along the following lines:
Changing the variable  to  $u=H^{-1}(s-1)$, setting $q=1$ and taking the $H\rightarrow0$ limit we get, 
\begin{eqnarray}
G(\eta, \eta'; Z) &\approx& \frac{\beta^2}{8\pi^2}\int_{-H^{-1}}^{\infty} H  \mathrm{d}u \frac{(Hu+1)^{-i\frac{m}{H}}}{\left((Hu+1)^2-2Z(Hu+1)+1\right)^{\frac{3}{2}}},\\
&=&\frac{1}{8\pi^2}\int_{-\infty}^{\infty} \mathrm{d}u \frac{e^{-imu}}{(u^2-\sigma^2)^{3/2}}+\mathcal{O}(H),\\
&=&\frac{im}{4\pi^2\sigma}\mathrm{K}_{1}(-im\sigma)+\mathcal{O}(H),
\label{ftwrtm}
\end{eqnarray} 
where  $\sigma^2\equiv-\ell^2\equiv \Delta t^2 - \Delta x^2$ with appropriate $i\epsilon$ prescriptions to define the integral leading to Wightman and Feynman Green's functions in flat spacetime. The last result, for example,  follows from a standard integral representation of $\mathrm{K}_{1}$ leading to Feynman Green's functions. (Incidentally, the penultimate line gives a simple integral representation for Feynman/Wightman functions which does not seem to be well-known. This is also derived from the more standard expression in Appendix \ref{app:derB1}. )

The integral representation in \eq{intreppoly} is particularly useful to study the convergence properties of the integral and we shall turn to this issue in a moment. But first we will relate these results to more conventional expressions used in the literature. It is possible to use 
the integral representations of associated Legendre function $P_\lambda^\mu(z)$ (see e.g., 8.713(3) of \cite{gd}) to write the Wightman function as:
\begin{equation}
G(\eta, \eta'; Z) = \frac{\beta^2 (\eta \eta')^{q-1}}{8\pi^2}\frac{1}{(Z^2-1)^{\frac{1}{2}}}\Gamma(\tfrac{3}{2}+\nu)\Gamma(\tfrac{3}{2}-\nu)P_{\nu-\frac{1}{2}}^{-1}(-Z).
\label{GinP}
\end{equation}
This result is only applicable (in the sense of yielding finite expressions) for $\lvert\nu\rvert < 3/2$. We will also assume that this relation can be analytically continued to values of $Z$ such that $\mathrm{Re} Z \geq 1$, with the $i\epsilon$ term ensuring that the integrand has no pole on the path of integration.

Alternatively, we can also obtain an expression for the Wightman function, in terms of the Gaussian hypergeometric function ${_2F_1}(a,b;c;z)$, using the relation between associated Legendre and hypergeometric functions (see e.g., 8.702 of \cite{gd}) to get:
\begin{equation}
G(\eta, \eta'; Z) = \frac{\beta^2(\eta \eta')^{q-1}}{8\pi^2}\frac{\Gamma(\tfrac{3}{2}+\nu)\Gamma(\tfrac{3}{2}-\nu)}{1-Z} {_2F_1}\left(\frac{1}{2}-\nu,\frac{1}{2}+\nu;2;\frac{1+Z}{2}\right).
\label{GinF1}
\end{equation}
This can be further transformed  using the identity, 
${_2F_1}(a,b;c;z) = (1-z)^{c-a-b}{_2F_1}(c-a,c-b;c;z)$ (see 9.131(1) of \cite{gd})
to arrive at
\begin{equation}
G(\eta, \eta'; Z) = \frac{\beta^2(\eta \eta')^{q-1}}{16\pi^2}\Gamma(\tfrac{3}{2}+\nu)\Gamma(\tfrac{3}{2}-\nu){_2F_1}\left(\frac{3}{2}+\nu,\frac{3}{2}-\nu;2;\frac{1+Z}{2}\right),
\label{GinF2}
\end{equation}
so that  the $Z$-dependence is completely contained within a hypergeometric function.

We have seen earlier in Sec. \ref{sec:hiding} that the theory will exhibit pathologies for $|\nu|>3/2$. It is straight forward to see that such a pathology arises in the context of the Wightman function and show that the Wightman function does not exist if $|\nu|>3/2$. This is because, while the mode functions are well-defined the integral over $ \mathrm{d}^3k$ required to define the Wightman function does not exist if $|\nu|>3/2$.
From \eq{DefIntRep_Wightman}, we see that the integral diverges for both  $\nu \leq -3/2$ and for $\nu \geq 3/2$; that is it exists only for $-3/2<\nu<3/2.$
The divergence in both ranges outside this band is due to the infrared behaviour of the mode functions i.e. the behaviour of the Hankel functions near $k = 0$. We can see this from their limiting forms for $z\to 0$ as given in (see e.g 10.7.7 of \cite{DLMF}):
\begin{equation}
\mathrm{H}_\nu^{(1)}(z) \simeq -\mathrm{H}_\nu^{(2)}(z) \simeq -\frac{i}{\pi}\Gamma(\nu)\left(\frac{z}{2}\right)^{-\nu}.
\end{equation}
We also have (see 10.4.6 of \cite{DLMF}) the relation $\mathrm{H}_{-\nu}^{(1)}(z) = e^{\nu\pi i}\mathrm{H}_\nu^{(1)}(z)$, $\mathrm{H}_{-\nu}^{(2)}(z) = e^{-\nu\pi i}\mathrm{H}_\nu^{(2)}(z)$. Thus, we have, for real $\nu$,
\begin{equation}
\mathrm{H}_{\nu}^{(1)}(z), \mathrm{H}_{\nu}^{(2)}(z) \rightarrow z^{-\lvert\nu\rvert}\ \text{ for }\ z \to 0.
\end{equation}
For our case, $z = k\eta$, and an integral of the product of Hankel functions over $ \mathrm{d}^3 k \sim k^2 \mathrm{d} k$ goes as $k^{2-2\lvert\nu\rvert}\mathrm{d} k$ near $k = 0$, which is divergent for $\lvert\nu\rvert \geq \frac{3}{2}$. This leads to a rather interesting situation: viz, the Wightman function exists in Fourier ($\bm k$) space but not in the real ($\bm x$) space because the integral over $ \mathrm{d}^3k$ diverges. As we shall see later, this allows the power spectra to exist even though the Wightman function does not.

The fact that Wightman function does not exist whenever $\nu$ lies outside the band $-3/2 < \nu < 3/2$, translates, in  the context of power law expansion with $a(t)\propto t^p$, to the condition that $p < 2/3$ for the Wightman function to exist. From \eq{tp3} we see that this requires $w>0$ (with $w=0$, corresponding to pressure-less dust, being the limiting case) for the Wightman function and hence the QFT to exist. Clearly, the de Sitter spacetime, corresponding to the limiting case of $w=-1$ with logarithmic divergence, is also pathological in this context --- which is a well known result in the literature. The reason for this result, however, is usually thought to be the breakdown of de Sitter invariance for the vacuum state of a massless field. While such states exist for massive fields in de Sitter, it is well known that no de Sitter invariant vacuum state exists in the massless limit. 
But as we see, (also see  \cite{FordParker, Janssen:2008dp, Janssen:2008dw, Janssen:2008px, Higuchi:2017sgj}) the massless scalar field has a diverging Wightman function --- and hence, strictly speaking, the quantum field theory does not exist --- not only for $w=-1$ but also for all negative values of $w$; that is, for all $w<0$ with $w=-1$ being just a special case. For values of negative $w$ other than $-1$, the spacetime does not posses any special symmetries or any analogue of de Sitter invariance. Hence it does not make sense to attribute the divergence of Wightman function for $-1 < w < 0$ to any specific lack of symmetry or invariance. Since $w=-1$ is just a limiting value of this band, it seems more logical to think of the {\it divergence of Wightman function in the case of a de Sitter as just a special case of the general feature which arises whenever the source for the Friedmann universe has negative pressure. }

\subsection{Massless de Sitter as a limiting case}\label{sec:series}

The massless scalar field in de Sitter --- obtained either as $m\to0$ limit of a massive field in de Sitter or as the $q\to1$ limit of a massless field in a power law universe with $a\propto \eta^{-q}$ will correspond to the $\nu=(1/2)+q=3/2$ limit which is at the edge of the pathology band. From \eq{intreppoly}, say, it is obvious that the Wightman function diverges in this case as well. We will consider this limit as arising from $m\to0$ limit of a massive field in de Sitter and determine the nature of the divergence in the Wightman function.

To do this, we start with the 
 Wightman function for a massive scalar field in de Sitter background, written in terms of the hypergeometric function, as:
\begin{eqnarray}\label{explicitGF}
	G(Z)=\frac{H^2}{16\pi^2}\Gamma(c)\Gamma(3-c)~_2F_1\left(c,3-c,2;\frac{1+Z}{2}\right),
\end{eqnarray}
where, $c(3-c)=m^2/H^2\equiv3\epsilon$ and consider its limiting for for small $\epsilon$, by obtaining an expansion of $G(Z)$ in powers of $m^2$, or equivalently, $\epsilon$. It is possible to do this in two separate ways both which, of course leads to the same conclusion. The first approach is to use the series representation for the hypergeometric function  given by:
\begin{eqnarray}\label{defineF0}
	~_2F_1(A,B,C;x)=\sum_{n=0}^{\infty}\frac{(A)_{n}(B)_{n}}{(C)_n}\frac{x^n}{n!},
\end{eqnarray}
where, $(y)_n\equiv\Gamma(y+n)/\Gamma(y)$. This is defined inside  the unit disc $|x|<1$ in a straightforward manner and outside this domain, the function is defined by an analytic continuation. It is then possible to show (see Appendix \ref{app:derf}) that the Wightman function has the series expansion:
\begin{eqnarray}\label{seriesforG}
	G(Z)=\frac{3H^4}{8m^2\pi^2}-\frac{H^2}{8\pi^2}\left[\frac{-1}{1-Z}+\log\left\{(1-Z)\lambda\right\}\right]+\mathcal{O}(\epsilon^2).
\end{eqnarray}
where, $\lambda=e^2/2$.
The second, alternative,  approach is to rewrite the integral representation in \eq{intreppoly} in the form of a series expansion:
\begin{eqnarray}
	G(Z)&=& \frac{H^2}{8\pi^2}\int_{0}^{\infty} \mathrm{d}s\frac{(s^2-2Zs+1)^{-3/2}}{s^{1-\epsilon}}
	\propto\int_{0}^{\infty} \mathrm{d}s\frac{(s^2+1)^{-3/2}}{s^{1-\frac{\epsilon}{2}}}\left[1-\frac{2Zs}{s^2+1}\right]^{-3/2},\nonumber\\
	&=& \int_{0}^{\infty}\frac{(s^2+1)^{-3/2}}{s^{1-\frac{\epsilon}{2}}}\sum_{n=0}^{\infty}\left[\frac{(-1)^nx(s)^n\Gamma(-1/2)}{\Gamma(n+1)\Gamma(-1/2-n)}\right] \mathrm{d}s,
	\end{eqnarray}
where, $x(s) = 2 Zs/s^2+1$.
It is now easy to show that the divergence arises from the $n=0$, which behaves as $1/\epsilon$ near $\epsilon\approx0$, while the rest of the terms can be summed up. This leads to the final result which is the same as in \eq{seriesforG}. 

In the case of a power law universe we get a similar result with $m$ replaced by the effective mass $m_{\rm \rm eff}$ and we perform the series expansion in $\epsilon_{\rm \rm eff}=m_{\rm \rm eff}^2/(3H^2)$. Here,  
 in the expression for the Wightman function, in addition to the combination of the Gamma functions and the hypergeometric function, there is a term of the form 
 \begin{equation}
(H^2\eta\eta')^{-\epsilon_{\rm \rm eff}}\approx 1-\epsilon_{\rm \rm eff}\log(H^2 \eta\eta')+\mathcal{O}(\epsilon_{\rm eff}^2),                                                                                                      \end{equation} 
as is clear, for example from \eq{GinF1}. Hence, the Wightman function ends up with  the following expansion in powers of $m_{\rm eff}^2$.
\begin{eqnarray}\label{seriesforGpowerlaw}
	G(Z,\eta,\eta')=\frac{3H^4}{8m_{\rm eff}^2\pi^2}-\frac{H^2}{8\pi^2}\left\{\frac{-1}{1-Z}+\log\left[(1-Z)(H^2 \eta\eta')\lambda\right]\right\}+\mathcal{O}(\epsilon^2).
\end{eqnarray} 
There are several features which are noteworthy about the results in \eq{seriesforG} and \eq{seriesforGpowerlaw} which we shall now comment about. 

One would have thought that  the massless scalar field in de Sitter can be approached through two possible  limits: It can be thought of as the $m\to 0$ limit of the massive field in de Sitter [which we have called Case A].  We can also think of it as the limit of a massless field in a power law universe with $a(\eta) \propto \eta^{-q}$ in the limit of $q\to 1$ [which we have called  Case B]. One would have naively expected these two limits to lead to the same Wightman function, which would indeed have been the case \textit{if only} the Wightman function, for a massless scalar field in de Sitter, was well-defined. Unfortunately the Wightman function diverges in this limit irrespective of whether the limit is taken in Case A or in Case B. To give any meaning to such a divergent quantity we need to introduce some kind of regularization procedure and it is \textit{not} guaranteed that the result will be independent of the regularization scheme which we choose.\footnote{Of course, the divergence of the Wightman 
function in this limit is an infrared divergence; it is illegal to ``subtract out'' any \textit{infrared} divergence in a field theory --- a point which should be kept in mind in these discussions.}

When we think of massless scalar field in de Sitter as the $m\to 0$ limit of a massive field, we arrive at the expression in \eq{seriesforG}.  The entire divergence is contained in the first term \textit{and it is independent of co-ordinates}. In other words, the derivative  
\begin{eqnarray} \label{dgdz1}
 \frac{\diff G_{\rm dS}(Z)}{\diff Z} &=& \frac{3\beta^2}{8\pi^2}(\eta\eta')^{q-1}\int\limits_{0}^{\infty} \mathrm{d} s~~\frac{s^{3/2-\nu}}{\left(s^2-2 sZ+1\right)^{5/2}},\\
 &=& \frac{m^2}{64\pi^2}\Gamma(\tfrac{3}{2}+\nu)\Gamma(\tfrac{3}{2}-\nu){_2F_1}\left(\frac{5}{2}+\nu,\frac{5}{2}-\nu;3;\frac{1+Z}{2}\right),
 \label{dgdz}
\end{eqnarray} 
is well-defined, finite and is de Sitter invariant in the sense that it depends only on $Z$.

All other derivatives with respect to co-ordinates can be obtained by multiplying this expression by the derivatives of $Z$ with respect to the co-ordinates. The fact that the derivative expression remains finite when $m\to0, \nu\to 3/2$ can be directly verified. If we put  
$\nu=3/2-\epsilon$, where $\epsilon$ is a small positive quantity, then we can easily show that (see Appendix \ref{app:derf})
\begin{eqnarray}
\frac{\diff G(\eta, \eta'; Z)}{\diff Z} 
&=& \frac{3\beta^2}{8\pi^2}(\eta\eta')^{q-1}\int\limits_{0}^{\infty} \mathrm{d} s~~\frac{s^{\epsilon}}{\left(s^2-2 sZ+1\right)^{5/2}},\\
&=& \frac{3\beta^2}{8\pi^2}(\eta\eta')^{q-1}\left[-\frac{Z-2}{3 (Z-1)^2}\right]+\mathcal{O}(\epsilon),
\label{tp111}
\end{eqnarray}
(Note that the integrals evaluated above are convergent
only for $Z<1$ and the $\epsilon$ series of $G$ for other ranges of $Z$ can be obtained by analytical continuation). The term in the square bracket in \eq{tp111} is just the derivative of $-(1-Z)^{-1}+\log(1-Z)$, which appears in \eq{seriesforGpowerlaw}. So integrating \eq{tp111} will reproduce the de Sitter invariant $G(Z)$ in the case of $q=1$, with an undetermined integration constant.

These features suggest that it may be natural to define   the Wightman function of a massless scalar by this procedure, of working with  $\diff G/\diff Z$ and integrating the expression.
But given the fact that the result in \eq{seriesforG} contains a logarithmic term $(H^2/8\pi^2) \ln (1-Z)$ and a divergent constant $3H^4/8m^2\pi^2$, it follows that neither term is well-defined \footnote{The divergent term plays an important role in the calculation of the vacuum expectation value of the stress energy tensor using the point-splitting method. In the presence of a non-minimal curvature coupling, there is also an ambiguity in the massless minimally coupled limit (see e.g. \cite{Kirsten:1993ug}) of 
this expectation value. Appendix \ref{app:derf0} briefly discusses this issue.} .  We can (i) add and subtract a function $(H^2/8\pi^2) \ln F(x,x')$ to this expression, where $F(x.x')$ is an arbitrary function of the co-ordinates,
(ii) modify the logarithmic dependence to the form $\ln [F(x,x')(1-Z)]$, and (iii) change the form of the divergent term by including a factor $(1+\ln F)$. This means that the co-ordinate dependence of the Wightman function can be modified at will because it is formally infinite. 

In general, there is no physical motivation at all to introduce such a function $F(x,x')$ into the divergent expression. But this is precisely what happens when we treat the massless scalar field in de Sitter as arising from the limit $q\to 1$ of a power law. We see from \eq{seriesforGpowerlaw} that it contains an extra, co-ordinate-dependent, factor 
\begin{equation}
 \frac{H^2}{8\pi^2} \ln (\eta\eta') \rightarrow \frac{H^3}{8\pi^2} (t+t'),
\end{equation} 
which indicates a secular growth proportional to the cosmic time $t$. If we think of the expectation value of $\phi^2(x)$ as arising from the coincidence limit of the Wightman function, then this term will contribute a secular growth:
\begin{equation}
 \bk{0,\rm BD}{\phi^2(x)}{0, \rm BD}_{\rm secular} = \frac{H^3}{4\pi^2} t,
\end{equation} 
which is a well known, ancient, result in this subject \cite{Ford:1989mf, Dolgov:1994cq, Ford:1997hb, Krotov:2010ma}. Our analysis shows that the secular growth does \textit{not} arise when we treat the Wightman function as the $m\to 0$ limit of the massive de Sitter unless one adds by hand a function $F(x,x')$ to reproduce the result in \eq{seriesforGpowerlaw}. This difference between the two approaches needs to be emphasized even though the discussion has all the well known arbitrariness which creeps in when one tries to make sense of expressions which are infrared divergent. 

From a physical point of view one can argue equally well for both the approaches. The approach based on $m\to 0$ limit in de Sitter has the virtue of preserving manifest de Sitter invariance all throughout with all the co-ordinate dependence of the theory contained in $Z$. In fact, if we had worked with the expression for $\diff G/\diff Z$  obtained from \eq{intreppoly}, then we would have obtained the result in \eq{dgdz} and everything would have been de Sitter invariant. This makes sense, for example,  \textit{if one believes that a scalar field $\phi$ is not directly observable and only its derivatives have physical meaning}. In that case, one simply discards the divergent term as well as any secular growth term.

On the other hand, one could equally well argue that our realistic universe was/is never in an \textit{exact} de Sitter phase. From this point of  view, a power law expansion $a\propto t^p$ with a finite but large $p$ represents a more realistic situation. The de Sitter limit is then best understood as the limit $q \to 1$ of a power law universe. If we take this route, the secular growth term is a physical feature. 

In this approach, however, there is no question of de Sitter invariance of the vacuum state or its breaking because we are never considering the exact de Sitter invariance. 
In the literature, the secular growth term is often attributed to the breaking of the de Sitter invariance, which does not seem to be a tenable interpretation from \textit{either} point of view.  We believe there is scope for more investigation in this regard especially if scalar fields are unobservable and only spatial derivatives of scalar fields make physical sense.

\section{The  Wightman function and its geodesic Fourier transform}\label{sec:deqforG}

\subsection{The differential equation for $G(Z)$}\label{sec:deqforG1}

The two-point function satisfies the Klein-Gordon equation $(\square-m^2)G=0$ on both its arguments. We will now consider the case in which $G$ depends on the co-ordinates only through the geodesic distance $\ell(x,x')$ or, equivalently, through $Z(x,x')$ e.g., $Z=\cos H\ell$ for spacelike separations). This can happen, for example, for a massive field in de Sitter if we choose a vacuum state which respects the de Sitter invariance. (In fact the expression in \eq{GinF2}, for example, has this property). So we should be able to obtain the Wightman function by (a) looking for the solutions to  the differential equation $(\square-m^2)G=0$ which depend only on $Z$ and (b) imposing proper boundary conditions. 

The task in (a) is simplified drastically if we use the geodesic co-ordinates introduced in Sec. \ref{sec:co-geo}. In fact, when we look for solutions to $(\square-m^2)G=0$ in this co-ordinate system we are actually looking for static (no $\tau$-dependence), spherically symmetric (no $\theta, \phi)$ dependence) solutions. 
It is straightforward to show that the equation then reduces to the form in \eq{D156} reproduced here for convenience:
\begin{equation}
\label{InvariantdSGreen}
(Z^2-1)\frac{\mathrm{d}^2 G}{\mathrm{d} Z^2} + 4Z\frac{\mathrm{d} G}{\mathrm{d} Z} + \mu^2G = 0,
\end{equation}
where $\mu = {m/H}$.
For some of our future applications, and for taking the $H\to0$ limit, it is convenient to rewrite this equation in terms of the variable $L(x,x')$ related to $Z(x,x')$ by
$Z = 1+(1/2)L^2H^2$ 
(Recall that $L^2=(1/H^2\eta\eta')[(\eta-\eta')^2-\Delta x^2]$ which goes over to
 the Minkowski line interval $A(x,x') \equiv L_M^2(x,x') = \Delta t^2 - \Delta x^2$ between the two events when $H\to 0$). In terms of $L^2$ the differential equation becomes: 
\begin{equation}
(4L^2 + H^2L^4)\frac{\mathrm{d}^2 G}{\mathrm{d} (L^2)^2} + (8+4H^2L^2)\frac{\mathrm{d} G}{\mathrm{d} (L^2)} + m^2 G = 0.
\label{gwrtl}
\end{equation}

As a warm-up for the analysis in the de Sitter spacetime let us first consider this equation and its solution in the limit of $H\to0$, i.e, in standard flat spacetime quantum field theory. In the limit of $H\to0$ the \eq{gwrtl} reduces to:
\begin{equation}
\label{G_L2FT_H0A}
4L^2\frac{\mathrm{d}^2 G}{\mathrm{d} (L^2)^2} + 8\frac{\mathrm{d} G}{\mathrm{d} (L^2)} + m^2 G = 0,
\end{equation}
which is indeed the correct equation for the Minkowski Green's functions $G_M(x,x')$ when it is a function of 
only $A$ i.e. $G_M(x,x') = G_M(A)$ --- a fact that be directly verified  using the geodesic co-ordinates (see \eq{sphrin}) in the flat spacetime.
The general solution (we do not assume $m^2 > 0$ in this analysis) is given by:
\begin{equation}
G(L^2) = g_+ \frac{\mathrm{K}_1(i\sqrt{m^2 L^2})}{\sqrt{m^2 L^2}} + g_- \frac{\mathrm{K}_1(-i\sqrt{m^2 L^2})}{\sqrt{m^2L^2}}.
\end{equation}
The correct solution can be chosen by considering the asymptotic behaviour and demanding that the correlations must vanish large spacelike intervals. From the asymptotic behaviour of the modified Bessel function $\mathrm{K}_\nu(z)$ from (see 10.25.3 of \cite{DLMF})
\begin{equation}
\mathrm{K}_\nu(z) \sim \sqrt{\frac{\pi}{2z}} e^{-z},\ \lvert z\rvert \rightarrow \infty, \lvert\mathrm{arg}(z)\rvert < \tfrac{3\pi}{2}
\end{equation}
it follows that $\mathrm{K}_1(iz)$ diverges as $\mathrm{Im}\ z \to \infty$. (We take $\sqrt{z}$ to denote the square root of $z$ in the upper half plane ($\mathrm{arg}(\sqrt{z})\in [0,\pi)$)). This means that while the solution $\mathrm{K}_1(-i\sqrt{m^2L^2})$ approaches zero for large spacelike (for $m^2 > 0$) or timelike (for $m^2 < 0$) separations (i.e., as $m^2L^2 \to -\infty$) the  $\mathrm{K}_1(i\sqrt{m^2L^2})$ diverges and must be discarded. This allows us to pick up the correct solution except for an arbitrary constant. 

Similar difficulties related to spurious solutions --- but much less obvious --- arise in the case of de Sitter spacetime in which $G(Z)$ satisfies \eq{InvariantdSGreen}.
Let us now get back to this equation and discuss its solutions.
The two linearly independent solutions of \eq{InvariantdSGreen} can be expressed in terms of  associated Legendre functions or hypergeometric functions. Explicitly,
the most general solution to \eq{InvariantdSGreen} can be written as
\begin{eqnarray}\label{gensolmassive}
	G_{m}(Z)=A_{+}~~_2F_1\left(c,3-c,2;\frac{1+Z}{2}\right)+A_{-}~~_2F_1\left(c,3-c,2;\frac{1-Z}{2}\right),
\end{eqnarray}
where, $A_{+}$ and $A_{-}$ are constants and  $c(3-c)=m^2/H^2$. If we set $A_{-}=0$ and choose $A_+$ appropriately we reproduce  the standard Wightman function, given by \eq{explicitGF}, which has the Hadamard form. Equivalently, by demanding that $G$ should have the Hadamard form we could have determined the constants $A_\pm$ and thereby determined the Wightman function.

It is of  interest to ask what happens to the $m\to0$ limit in this approach because we know that the correct Wightman function diverges in that case. But the two terms in \eq{gensolmassive} can go to finite values in this limit. We find that, when $m\to0$ the expression in \eq{gensolmassive} reduces to the form:
\begin{eqnarray}
G_{0}(Z)=A+B {\left[\frac{2Z}{1-Z^2}+\log\left(\frac{1+Z}{1-Z}\right)\right]}\equiv A+BW(Z).
\label{falseG}
\end{eqnarray}
The function $W(Z)$ is obtained by a particular choice of constants in \eq{gensolmassive} followed by using the $m\rightarrow0$ limit of the expression:
\begin{eqnarray}
	W(Z)=\lim_{m\rightarrow 0}\left\{ \frac{1}{2}\Gamma(c)\Gamma(3-c)\left[_2F_1\left(c,3-c,2;\frac{1+Z}{2}\right)-~_2F_1\left(c,3-c,2;\frac{1-Z}{2}\right)\right]\right\}.
\end{eqnarray}
It appears that we have succeeded in obtaining a de Sitter invariant (i.e., the result depends only on $Z$), finite, Wightman function for a massless scalar field in de Sitter contrary to what we found earlier. This is, of course, not true. The expression in \eq{falseG} is finite, de Sitter invariant and satisfies the Klein-Gordon equation but it is \textit{not} a Wightman function. This is because it cannot be expressed as a mode sum which is a necessary requirement for a Wightman function. The best one can do for $G_{0}(Z)$ is to express it as an \textit{in-out} matrix element of a time-ordered product of the form (following \cite{Wetterich:2015gya, Fukuma:2013mx})
\begin{equation}
iG_F^{\text{out/in}}(x,x') = \langle 0, +\infty\rvert T\hat{\phi}(x)\hat{\phi}(x')\lvert 0, -\infty\rangle,
\end{equation}
where $\lvert 0, -\infty\rangle = \lvert 0, \rm BD\rangle$ is the Bunch-Davies vacuum at early times, and $\lvert 0, +\infty\rangle$ is the vacuum at late times, defined via instantaneous Hamiltonian diagonalization in the Friedmann co-ordinates. This expression is valid for  massive as well as massless fields and is a hybrid (Feynman Green's function-like) object. In terms of mode functions, this reduces to the integral \cite{Fukuma:2013mx}:
\begin{equation}
iG_F^{\text{out/in}}(x,x') = \frac{1}{4\pi}\frac{(\eta\eta')^{\frac{3}{2}}}{\rho} \int_0^\infty \mathrm{d} k\ k \sin(k\rho)\mathrm{J}_\nu(-k\eta_+)\mathrm{H}_\nu^{(2)}(-k\eta_-),
\label{intforfalseG}
\end{equation}
where $\eta_+ = \text{max}(\eta, \eta'), \eta_- = \text{min}(\eta, \eta')$ due to the time ordering. For definiteness, we take $\eta > \eta'$, so that the above expression is then a positive in-out ``Wightman'' function. The integral can be evaluated to give  \cite{Fukuma:2013mx} % (see Appendix \ref{app:derg})
\begin{equation}
G^{\text{out/in}}(x,x') = \frac{1}{4\pi^2}(Z^2-1)^{-\frac{1}{2}} \mathcal{Q}^{1}_{\nu - \frac{1}{2}}(Z),
\end{equation}
where
$\mathcal{Q}_\lambda^\kappa(z)$ is the associated Legendre function of the second kind.
In the massless limit, $\nu = \frac{3}{2}$ and the associated Legendre function becomes:
\begin{equation}
\mathcal{Q}^1_1(z) = i\left(\frac{z}{\sqrt{z^2-1}} + \frac{\sqrt{z^2-1}}{2}\log\left(\frac{1-z}{1+z}\right)\right),
\end{equation}
leading to the two-point function \cite{Wetterich:2015gya} :
\begin{equation}
G^{\text{out/in}}(x,x') \propto \left(\frac{Z}{1-Z^2}+\frac{1}{2}\log\left(\frac{1+Z}{1-Z}\right)\right).
\end{equation}
We see that this is proportional to $W(Z)$ in \eq{falseG}. So the result we found in \eq{falseG} is a transition element (an in-out function) rather than an expectation value (The constant $A$ in  \eq{falseG} is immaterial because, in the massless limit, the equation $\square G=0$ has $G=$ constant as one of the solutions \cite{Wetterich:2015gya}).

The algebraic reason for this expression $G^{\text{out/in}}(x,x')$ to be finite while the correct Wightman function diverges for $m=0$ is the following. The function  $G^{\text{out/in}}(x,x')$ is given by the integral in \eq{intforfalseG} involving a product of Bessel and Hankel functions  while the one for the Wightman function is given by \eq{WightmanG_HankelSine} involving a product of two Hankel functions. (This is because one in-vacuum is changed to an out-vacuum).
The Bessel function can be written in terms of the Hankel functions as follows:
\begin{equation}
\mathrm{J}_\nu(-k\eta_\pm) = \frac{1}{2}\left(\mathrm{H}_\nu^{(1)}(-k\eta_\pm) + \mathrm{H}_\nu^{(2)}(-k\eta_\pm)\right).
\end{equation}
For the massless case, we have $\nu = \frac{3}{2}$ and 
\begin{equation}
\mathrm{H}_{\frac{3}{2}}^{(1/2)}(-k\eta_+) = \sqrt{\frac{2}{\pi}}\frac{e^{\mp ik\eta_+}(\mp i+k\eta_+)}{(-k\eta_+)^{\frac{3}{2}}},
\end{equation}
we get a term:
\begin{equation}
-\frac{H^2}{2\pi^2\rho}\int\limits_0^\infty\mathrm{d} k \frac{e^{ik\rho}-e^{-ik\rho}}{2i} e^{ik(\eta_+ + \eta_-)}\left(\eta_+\eta_- + \frac{i(\eta_+ + \eta_-)}{k} - \frac{1}{k^2}\right),
\end{equation}
This is the same as the expression due to $\mathrm{H}_\nu^{(1)}$, with the formal replacement $\eta_+ \rightarrow -\eta_+$, which results in $Z \rightarrow -Z$,  with an overall negative sign.
So  the divergent contributions in  these two individual terms  (which are $\eta$-independent) cancels out leading to a finite expression. The in/out Wightman function is then
\begin{eqnarray}
G^{\text{out/in}} &=& \frac{H^2}{8\pi^2}\left(\frac{1}{Z-1}-\frac{1}{-Z-1} + \log(1-Z) - \log(1+Z)\right), \\
&=& -\frac{H^2}{8\pi^2}\left(\frac{2Z}{Z^2-1} + \log\left(\frac{1-Z}{1+Z}\right)\right),
\end{eqnarray}
which is the expression found earlier.

\subsection{Aside: Analytic continuation Euclidean Green's function}\label{sec:asideEGF}

In Sec.\ref{sec:5ded} we obtained an expression for the Euclidean Green's function by mapping the problem to that of $D=5$ electrostatics. In this section we revisit the derivation of Euclidean Green's function, now as an explicit solution to the differential equation (in the spirit of Sec. \ref{sec:deqforG1}) and highlight some aspects of its analytic continuation. 

We begin with the metric in the geodesic co-ordinates (see \eq{C11}), analytically continued into the Euclidean sector by introducing Euclidean time $\tau_E$, with $\tau = -i\tau_E$.
\begin{equation}
\mathrm{d} s^2 = \frac{\sin^2(H\ell)}{H^2}\mathrm{d}\tau_E^2 + \mathrm{d} \ell^2 + \frac{\sin^2(H\ell)}{H^2} \cos^2 \tau_E \mathrm{d} \Omega_2^2.
\label{egc}
\end{equation}
This also corresponds to using the  Euclidean time co-ordinate for the 5D embedding Minkowski spacetime, wherein the de Sitter manifold becomes a sphere of radius $1/H$. The geodesic distance $\ell$ is then restricted to lie between $0$ and $\pi/H$.
The flat spacetime limit $H \to 0$ of \eq{egc} correctly leads to the  Euclidean version of the spherical Rindler metric, given by:
\begin{equation}
\lim_{H\rightarrow0} \mathrm{d} s^2=\ell^2\mathrm{d} \tau_E^2+\mathrm{d}\ell^2+\ell^2\cos^2\tau_E \mathrm{d}\Omega_2^2.
\end{equation} 
Introducing $Z = \cos(H\ell) \in [-1,1]$ (corresponding to the range $\ell = \pi / H$ to $\ell = 0$), the line element in \eq{egc} becomes:
\begin{equation}
\mathrm{d} s^2 = (1-Z^2)\mathrm{d} \tau_E^2 + \frac{1}{1-Z^2}\mathrm{d} Z^2 + (1-Z^2)\cos^2\tau_E \mathrm{d} \Omega_2^2.
\end{equation}

The Green's functions for a massive scalar field obeying the (Euclidean) Klein-Gordon equation satisfies the differential equation:
\begin{equation}
-\left(\frac{1}{\sqrt{g}}\partial_\mu\left(\sqrt{g} g^{\mu\nu}\partial_\nu\right) - m^2\right)_x G(x,x') = \frac{\delta^{(4)}(x-x')}{\sqrt{g(x')}}.
\label{Euclidean_KG_Greens}
\end{equation}
Out of all possible solutions to this equation, we would like to pick one which has the correct ultraviolet behaviour. To decide this behaviour, we first obtain a constraint on the ultraviolet behaviour of the Green's function using this equation.

Consider a region $R: x' \in R$. Integrating the \eq{Euclidean_KG_Greens} with respect to $x$ over the region $R$, and using the divergence theorem gives
\begin{equation}
-\int\limits_{\partial R} \mathrm{d}^3 x\ \sqrt{g}\ n^\mu (\partial_x)_\mu G(x,x') + m^2 \int\limits_{R} \mathrm{d}^4 x\ \sqrt{g} G(x,x') = 1,
\end{equation}
where $n^\mu$ are the unit normals to the boundary $\partial R$ of $R$. In the limit of $R$ shrinking to $x'$, if $G(x,x') = G(\ell)$, where $\ell$ is the geodesic distance between $x$ and $x'$ (which is unique for points that approach each other) we can choose $\partial R$ as a surface of all points at a constant value of $\ell$ around $x'$. The 4D solid angle is given by $\Omega_4 = 2\pi^2$, and $\mathrm{d}^3 x \sqrt{g} \sim \Omega_4 \ell^3$; $\mathrm{d}^4 x \sqrt{g} \sim (1/4)\Omega_4 \ell^4$, which is negligible in comparison to the boundary term. We then require that, for $l \to 0$,
\begin{equation}
\frac{\mathrm{d}}{\mathrm{d} \ell}G(\ell) \sim -\frac{1}{\Omega_4 \ell^3}.
\end{equation}
This can be integrated to give (keeping only the leading divergence):
\begin{equation}
G(\ell) \sim \text{const.}+ \frac{1}{2\Omega_4 \ell^2} = \text{const.}+\frac{1}{4\pi^2\ell^2},
\end{equation}
which is the requirement that the leading order divergence of a rotation-invariant Green's function behaves like a massless flat spacetime Green's function.
We can, therefore, look at rotation-invariant solutions of the homogeneous version of \eq{Euclidean_KG_Greens} in the region excluding a neighborhood of $x = x'$, and impose the above ultraviolet behaviour on the solution near this point. 

Taking advantage of the rotation invariance in the Euclidean de Sitter spacetime, we choose $x'$ such that $Z' = 1$ corresponding to $\ell' = 0$. Then, $\ell$ also serves as the geodesic distance between the two points. We  seek a Green's function that is rotation invariant (i.e. depends only on $\ell$ or equivalently, $Z$) so that $G(x,x') = G(Z)$. Then the equation satisfied by $G(Z)$ will be that satisfied by the hypergeometric function in the variables $(1+Z)/2$ except for an extra Dirac delta on the right hand side. The most general solution to this equation is given by
\begin{equation}
G(Z)=A~_2F_1\left(\frac{3}{2}-\nu,\frac{3}{2}+\nu;2;\frac{(1+Z)}{2}\right)+B~_2F_1\left(\frac{3}{2}-\nu,\frac{3}{2}+\nu;2;\frac{(1-Z)}{2}\right),                                                                          \end{equation} 
where $A,B$ are two constants.
We will now demand that  the solution (i) should be smooth except at $Z=1$ and (ii) should reduce to the massless, Euclidean, Green's function in flat space-time in the $l\rightarrow0$ limit. Condition (i) implies that $B=0$. Condition (ii) will give us the correct normalization $A$. Using the known limits (see 15.3.12 of \cite{as}), we see that as $l^2\rightarrow 0$
\begin{eqnarray}
A~_2F_1\left(\frac{3}{2}-\nu,\frac{3}{2}+\nu;2;\frac{(1+Z)}{2}\right) \approx \frac{A}{\Gamma\left(\frac{3}{2}-\nu\right)
\Gamma\left(\frac{3}{2}+\nu\right)}\left(\frac{4}{H^2l^2}\right).
\end{eqnarray}
Comparing this with the UV behavior of the Euclidean flat Green's function, we get
\begin{eqnarray}
A=\frac{\Gamma\left(\frac{3}{2}-\nu\right)\Gamma\left(\frac{3}{2}+\nu\right)H^2}{16\pi^2}.
\end{eqnarray}
Therefore, the Euclidean Green's function for $dS_4$ is given by
\begin{eqnarray}
G(Z)=\frac{\Gamma\left(\frac{3}{2}-\nu\right)\Gamma\left(\frac{3}{2}+\nu\right)H^2}{16\pi^2}~_2F_1\left(\frac{3}{2}-\nu,\frac{3}{2}+\nu;2;\frac{(1+Z)}{2}\right);\quad (\mathrm{for} |Z|<1).
\end{eqnarray}
One could have also found this by directly integrating \eq{ed5final}.

To find the Lorentzian version we have to analytically continue $G(Z)$ to the domain $-\infty <Z<\infty$. Recall that (the principle branch of) $~_2F_1$ has a branch-cut from $Z=1$ to $Z=\infty$. Therefore, the analytical continuation of $G(Z)$ with different choices of the orientation of this branch-cut will give us different Lorentzian Green's functions. The fact that the order of operators appearing in the correlator matters in the Lorentzian theory manifests itself as the multivaluedness of $G(Z)$ that results from the presence of the branch-cut. The branch-cut difference, hence is  a measure of the commutator Green's function. Let us briefly recall how this leads to the different Green's functions. 

From the definition of $Z$, it is easy to see that, for a fixed $|\Delta \mathbf{x}|$, the  $G(Z(\eta,\eta',\Delta\mathbf{x}))$ has branch cuts starting from $\Delta\eta=\pm|\Delta\mathbf{x}|$ to $\Delta\eta=\pm\infty$ in the complex $\Delta\eta$-plane. The transformation $\Delta\eta\rightarrow-\Delta\eta$ will not be accompanied by any change in the operator ordering inside vacuum expectation value $\braket{~~}$ for the positive(negative) Wightman  Function. On the other hand, under the transformation $\Delta\eta\rightarrow-\Delta\eta$, the operator ordering flips for (Anti-)Feynman Green's function. This implies that the branch, for the analytic continuation of $G(Z)$ that corresponds to the positive(negative) Wightman function, should be such that we can rotate from $\Delta\eta\rightarrow e^{\pm i\pi}\Delta\eta$ without touching a branch-cut. While, that for the (Anti-)Feynman Green's function should be such that as we rotate from $\Delta\eta\rightarrow e^{\pm i\pi}\Delta\eta$ we pass through a branch-cut 
at most once. Therefore, the Feynman Green's function corresponds to choosing the branch-cut of the hypergeometric function to lie just below the real axis. On the other hand, the Wightman  Green's function is obtained by orienting the branch-cut along the ray of slope $\textrm{tan}[\varepsilon~\textrm{sign}(\Delta\eta)]$ that starts from $Z=1$ to $Z=e^{i\varepsilon\textrm{sign}(\Delta \eta)}\infty$. This clearly is equivalent to the standard $i\varepsilon$ prescription.

\subsection{The geodesic Fourier transform of the Wightman function}\label{sec:geoftofW}

Two other functions  which we will make use of are (i) the Fourier transform $\mathcal{G}(Q)$ of $G(Z)$ with respect to $Z$ in terms of a conjugate variable $Q$ (which may be called \textit{geodesic momentum}, since it essentially arises as a Fourier transform conjugate of the geodesic distance) as well as 
(ii) the Fourier transform $\overline{\mathcal{G}}(K)$  of $G(Z)$ with respect to $L^2$ in terms of a conjugate variable $K$.
Since $Z=1+H^2L^2/2$, these two are
related in a simple manner:
\begin{equation}
\mathcal{G}(Q) = \frac{H^2e^{-iQ}}{2}\ \overline{\mathcal{G}}\left(\frac{H^2Q}{2}\right).
\end{equation}
While one can write down the differential equations satisfied by these functions, by Fourier transforming \eq{InvariantdSGreen}, their solutions will again suffer from the kind of ambiguities we saw earlier. (In fact the situation is worse because the Fourier transforms have to be defined carefully in the complex plane.) A simpler route to obtaining these results is by Fourier transforming an  integral representation for $G(Z)$, say the one in \eq{DefIntRep_Wightman}, reproduced below for convenience: 
\begin{equation}
\label{Gplusdef}
G(\eta, \eta'; Z) = \frac{\beta^2}{16\pi^2\sqrt{2}}(\eta\eta')^{q-1}\int\limits_{-\infty}^{\infty} \mathrm{d} u \frac{e^{-\nu u}}{\left(\cosh u- Z\right)^{\frac{3}{2}}},
\end{equation}
with respect to $Z$.
In fact this approach has the advantage that,
formally, we can now treat $Z$ and $\eta$,$\eta'$ as independent variables so that we need not assume the entire co-ordinate dependence of $G$ is through $Z$. We are, therefore,  interested in the Fourier transform of $G$ with respect to $Z$, and to be precise with resepect to $Z_{\epsilon}$ defined as
\begin{equation}
Z_{\epsilon} = 1+\frac{(\eta-\eta'-i\epsilon)^2 - \Delta x^2}{2\eta\eta'}   = Z -\frac{\epsilon^2}{2 \eta \eta'} -i\epsilon\frac{\eta-\eta'}{\eta\eta'},                                                           
\end{equation} 
in the limit of $\epsilon \rightarrow 0^+$.
The Fourier transform with respect to this complex variable can be evaluated from an integral transform defined as
\begin{equation}
\widetilde{G}^+(\eta, \eta'; Q) = \int\limits_{-\infty - i\epsilon(\eta,\eta')}^{\infty -i \epsilon(\eta,\eta')}\mathrm{d} Z_{\epsilon}\ e^{-iQZ_{\epsilon}}G^+(\eta, \eta'; Z_{\epsilon}).
\end{equation}
The integral transform yields (see Appendix \ref{app:deri} for details)
\begin{eqnarray}
G^+(\eta, \eta'; Q) = -\frac{H^2}{4\sqrt{2\pi}}e^{\frac{i\pi}{2}(\frac{1}{2}-\nu)}Q^{\frac{1}{2}}\mathrm{H}_{\frac{3}{2}}^{(2)}(Q)s_{\eta}\theta(-s_{\eta}Q),
\end{eqnarray}
with $s_{\eta}=\text{sgn}(\eta-\eta')$. A similar transform can be defined for the power law case as well which yields the above result with an additional factor of $(\eta \eta')^{q-1}$. For the de Sitter case, this is of interest as it is  $\eta$-independent, and depends  on $Q$ as well as $\text{sgn}(\eta-\eta')$

It is relatively straightforward to obtain the flat spacetime limit of the Fourier transform of the Wightman function  by taking the 
$H\to 0$ limit and obtain the geodesic Fourier transform of the Minkowski Wightman function (see Appendix \ref{app:deri} for details).
We find that:
\begin{equation}
\overline{\mathcal{G}}(K) = \frac{1}{2\pi i}e^{\frac{im^2}{4K}}\theta(-K).
\end{equation}
 As a check of the calculation we can obtain the Minkowski Wightman function in real space by Fourier transforming this expression. One can indeed show that the Fourier transform gives (see Appendix \ref{app:deri}):
\begin{eqnarray}
G(L^2) = \frac{i}{4\pi^2} \sqrt{\frac{m^2}{L^2}} \mathrm{K}_{1}\left(\sqrt{-m^2L^2}\right) 
= \frac{i}{4\pi^2} \sqrt{\frac{m^2}{L^2}} \mathrm{K}_{1}\left(-i\sqrt{m^2L^2}\right),
\end{eqnarray}
which is the Wightman function for a massive scalar field in Minkowski spacetime. In the massless limit, using 
the property
 $\mathrm{K}_\nu(z) \sim (1/2)\Gamma(\nu)\left(z/2\right)^{-\nu}$ as $z\to 0$, we recover the familiar result for the massless Wightman function, i.e.,
\begin{equation}
G(L^2)\rvert_{m = 0} = -\frac{1}{4\pi^2L^2}.
\end{equation}
We will make use of these results later on, in evaluating the power spectra in flat spacetime.

\section{Power spectra of the vacuum noise: from mode functions}

We shall next address the task of characterizing the quantum fluctuations in de Sitter spacetimes in terms of suitably defined power spectra. In the literature this issue is often discussed in the context of inflationary cosmology and the usual definition for power spectra is based on spatial Fourier transform in conformal Friedmann co-ordinates. In the language of Killing vector fields introduced in Sec. \ref{sec:psgen} this corresponds to using the Killing vectors which represents spatial translational symmetry. 
The vacuum state is usually taken to be  the Bunch-Davies vacuum which can be thought of as de Sitter invariant in a limiting sense.
This leads to the result: 
\begin{equation}
 P(k,\eta) = \frac{H^2}{2 (2 \pi)^3 k^3} \, (1+k^2\eta^2).
 \label{psfone}
\end{equation} 
In the terminology of the standard literature, we will like to point out that what we define as the power spectrum will be the power spectrum amplitude in the Killing space (along which the Fourier transforms will be naturally defined). The standard power spectrum $\tilde{P}(k)$ will be obtained from our results through 
\bea
\tilde{P}(k) = \Omega_qk^q P(k),
\eea
where $\Omega_q$ is the solid angle of the space of Killing vectors ($q -$ in number) with respect to which the Fourier transforms have been carried out to obtain $P(k)$, which for the Bunch Davies vacuum 
\bea
\tilde{P}(k) =\Omega_3k^3 \frac{H^2}{2 (2 \pi)^3 k^3} \, (1+k^2\eta^2) =\frac{H^2}{4 \pi^2}(1+k^2\eta^2),
\eea
which expectedly gives the scale invariant power spectrum in the super-Hubble limit $k \eta \rightarrow 0$ \cite{Parker:2009uva, gravitation, Baumann:2009ds}.

It is also possible to define the power spectrum of fluctuations, for the same Bunch-Davies vacuum, by using the Killing vector corresponding to translational invariance in cosmic time $t$ or equivalently, the static time $\tau$. This will lead to the following result (which we will soon derive), when we use $e^{-i\omega \tau}$ for the definition of the Fourier transform, as in \eq{pkilling2}:
\begin{equation}
 P_-(\omega,\bm{0}) = \frac{H^2}{4\pi^2 \omega} \, \left(1+ \frac{\omega^2}{H^2}\right) \frac{e^{-\pi \omega/H}}{2\sinh (\pi\omega/H)}
 = \frac{H^2}{4\pi^2 \omega} \, \left(1+ \frac{\omega^2}{H^2}\right)n(\omega),
 \label{psfone0}
\end{equation}
where $n(\omega)\equiv[e^{\beta\omega}-1]^{-1}$ with $\beta^{-1}=H/2\pi$ is the Planckian number density at de Sitter temperature $H/2\pi$.
On the other hand, if we use $e^{+i\omega \tau}$  for the definition of the Fourier transform as in \eq{pkilling}, we will get:
\begin{equation}
 P_+(\omega,\bm{0}) = \frac{H^2}{4\pi^2 \omega} \, \left(1+ \frac{\omega^2}{H^2}\right) \frac{e^{\pi \omega/H}}{2\sinh (\pi\omega/H)}
 = \frac{H^2}{4\pi^2 \omega} \, \left(1+ \frac{\omega^2}{H^2}\right)[1+n(\omega)].
 \label{psftwo}
\end{equation}
In other words, we have
\bea
 \tilde{P}_\pm(\omega)= \Omega_1\omega P(\omega,\bm{0}) = \frac{H^2}{4\pi^2} \, \left(1+ \frac{\omega^2}{H^2}\right) \frac{e^{\pm \pi \omega/H}}{2\sinh (\pi\omega/H)},
\eea
according to the two conventions, leading to either $n(\omega)$ or $1+n(\omega)$. This is the subtlety we mentioned while discussing the definitions in \eq{pkilling} and \eq{pkilling2}.  Either of the definitions can be used but with the following understanding: The switch in the sign of $\omega$ changes absorption to emission during physical processes and brings in the spontaneous emission term which leads to the $1$ in the $(1+n)$.   (We  remind the reader again that, the two conventions do not lead to different results if the Killing vector is spacelike). 

There is also  a third possibility for the choice of an observer in de Sitter spacetime. One can define a vacuum state adapted to the static co-ordinates, introduce the corresponding two point function, and work out the resulting power spectrum  as seen by, say, an observer at the origin using the Killing vector field corresponding to translational invariance in $\tau$. This leads to the result which is very close to the previous one and is given by
\bea
 P_\pm(\omega,\bm{0}) = \pm \frac{H^2}{4\pi^2 \omega} \, \left(1+ \frac{\omega^2}{H^2}\right) \theta(\pm\omega), \\
 \tilde{P}_\pm(\omega)= \Omega_1|\omega| P_\pm(\omega,\bm{0}) =   \frac{H^2}{4\pi^2} \left(1+ \frac{\omega^2}{H^2}\right)\theta(\pm\omega).
 \label{psfthree}
\eea
The difference between the two Fourier transforms are clear in this context . For one choice, with ($e^{-i\omega \tau}$), the power spectrum vanishes for positive frequencies, while for the other choice, with $e^{+i\omega \tau}$, it survives. So in a sense, the second convention picks up the quantum correlations of the vacuum for positive frequencies which the other convention does not. This might favor the second convention. But it is the first choice of definition,
which uses \eq{pkilling2}, that agrees with the response of an Unruh-deWitt detector  \cite{Unruh:1976db, Garbrecht:2004du}. Since we do not want an inertial detector to spontaneously get excited in inertial vacuum, one can argue that the inertial vacuum should not have power in positive frequencies.  So with this convention,  the Unruh-deWitt detector  (UDD) response rate ${\cal R}(\omega) \equiv P_-(\omega) =P_+(-\omega)$, which in turn ensures that an inertial detector in the inertial vacuum does not spontaneously get excited. From this point of view, the first choice  captures the \textit{operational} part of the quantum correlation.
In summary, both conventions provide us with some information about the nature of field fluctuations
and there is a simple physical interpretation (involving spontaneous and induced emissions in a thermal state) which connects \eq{psftwo} with \eq{psfthree}.

\subsection{Bunch-Davies vacuum}

The simplest case, worked out several times in the literature corresponds to the power spectrum in Bunch Davies vacuum evaluated in the conformal Friedmann co-ordinates by taking a spatial Fourier transform of the equal time Wightman function:
\be
{P}(k)  \equiv  \int \frac{ \mathrm{d}^3{\bf x}}{(2\pi)^3}\ e^{i {\bf k}\cdot{\bf x}}\langle0, \rm BD|\phi(x)\phi(x')|0 , \rm BD\rangle 
                          = \frac{|{f}_k(\eta)|^2}{(2\pi)^3},
\ee
where $f_k(\eta)$ are the mode functions in \eq{fkdSmassive}. This works for even a massive field in de Sitter or a massless field in power-law universe. We will be concerned with the case of massless field in de Sitter for which, using the simple mode functions in \eq{masslessdSmodes}, the power spectrum is obtained as :
\be
{P}(k)=
 \frac{H^2}{2(2\pi)^3 k^3}\left(1 + k^2 \eta^2\right). 
 \label{PS01}
\ee
For the scales $k\eta \ll 1$, the power spectrum becomes scale invariant and its amplitude
\be
\tilde{P}(k)=4 \pi k^3 {P}(k) =
 \frac{H^2}{4 \pi^2}\left(1 + k^2 \eta^2\right),
 \ee
approaches ${H^2/4 \pi^2}$ as argued above, a result which can be obtained in many other ways as well
(As mentioned earlier, one can also compute the power spectrum of the Bunch-Davies vacuum by using a Fourier transform with respect to static time co-ordinate as well. This will lead to the expression in \eq{psftwo}, as we will see soon).

In spite of the fact that the result in \eq{PS01} is beaten to death in the literature one particular aspect of it seems not to have been emphasized. To bring this out, let us rewrite \eq{PS01} introducing the cosmic time with $\eta=-(1/H)\exp{(-Ht)}$ obtaining:
\be
{P}(k)=\frac{1}{(2\pi)^3 }\left[\frac{1}{2 k}e^{-2Ht}+\frac{H^2}{2 k^3}\right]
=P_{\rm flat}(k)e^{-2Ht}+\frac{H^2}{2(2\pi)^3 k^3},
 \label{PS01again}
\ee
where $P_{\rm flat}(k)\equiv 1/(2(2\pi)^3 k)$ is the power spectrum of flat spacetime vacuum noise --- as can be seen by taking the $H\to0$ limit in \eq{PS01again} or directly by Fourier transforming the flat spacetime, equal time Wightman function, $G_{\rm flat}\propto 1/|\bm x|^2$ with respect to $\bm x$ --- and the second term $(H^2/2(2\pi)^3 k^3)$ an \textit{irreducible vacuum noise in de Sitter spacetime}. We see from \eq{PS01again} that the exponential expansion reduces the flat spacetime vacuum noise and, eventually, the de Sitter noise wins out. In other words there is a minimum vacuum noise $4\pi k^3P(k)_{\text{min}}=(H^2/4\pi^2)$ in the de Sitter spacetime. We will come back to this discussion in the next section.

The result in \eq{PS01} tells us that, for a massless field in de Sitter spacetime,  the Fourier transform of the Wightman function exists even though the Wightman function itself does not. This is of course, obvious from the fact that power spectrum goes as $k^{-3}$ near $k=0$ so that its Fourier transform, which gives the correlator in real space, does not exist. But we know that the Wightman function in real space has the structure in \eq{seriesforG} for small enough, nonzero mass. We should, therefore, be able to compute the Fourier transform of 
\eq{seriesforG} and obtain the same power spectrum as in \eq{PS01}. We will now describe how this calculation proceeds since it illustrates some features related to the divergence in the massless case.

We will write the $m\to 0$ limit of the Wightman function in de Sitter spacetime (corresponding to $\nu \rightarrow (3/2)^-$ and $q\to1$),  as follows:
\begin{equation}
G(\eta,\eta';\Delta x) = \frac{3H^4}{8\pi^2 m^2} - \frac{1}{4\pi^2 L^2} - \frac{H^2}{8\pi^2}\log(L^2 H^2 c),
\end{equation}
where $L^2 =(H^2\eta\eta')^{-1}[(\eta-\eta')^2-\rho^2]$ (which reduces to the Minkowski interval as $H\rightarrow 0$) and $c$ is an undetermined constant. We have added this constant because of the ambiguity in separating the divergent and finite terms in $G$, mentioned earlier. Setting $\eta = \eta'$ gives, with $L^2 = -(\rho^2/H^2\eta^2) $
\begin{equation}
G(\eta,\eta; \rho) = \frac{3H^4}{8\pi^2 m^2} + \frac{H^2\eta^2}{4\pi^2\rho^2} - \frac{H^2}{8\pi^2}\log\left(c\frac{\rho^2}{\eta^2}\right).
\end{equation}
To obtain the power spectrum we have to Fourier transform it with respect to spatial co-ordinates. Performing the angular integrals
and noting that
 $G$ is invariant under $\rho\rightarrow-\rho$, we can reduce the expression to the form:
\begin{eqnarray}
P(k,\eta) = \frac{2\pi}{ik(2\pi)^3}\int\limits_{-\infty}^\infty x \mathrm{d} x\ e^{ikx} G(\eta,\eta;x) 
= \frac{H^2}{(2\pi)^4 ik} \int\limits_{-\infty}^\infty \mathrm{d} x\ e^{ikx}\left(\frac{\eta^2}{x} - x\log \lvert x\rvert + \text{const.}\right),
\label{pk1}
\end{eqnarray}
where ``$\text{const.}$" includes all terms that are independent of $x$. Their contribution is only to the $k=0$ part of the power spectrum and we will ignore them with the understanding that we are only interested in $P(k, \eta)$ for $k>0$. To proceed further we shall use two results:
\begin{eqnarray}
\int\limits_{-\infty}^{\infty}\mathrm{d} x\ \frac{1}{x} e^{ikx} = i\pi\frac{\lvert k\rvert}{k};\qquad 
\int\limits_{-\infty}^{\infty}\mathrm{d} x\ e^{ikx} \log\lvert x\rvert = -\frac{\pi}{\lvert k\rvert} - 2\pi\gamma\delta(k),
\end{eqnarray}
where $\gamma$ is the Euler-Mascheroni constant. These hold in a distributional sense and we 
  differentiate the latter equation, with respect to $k$, to obtain the Fourier transform of our interest
\begin{eqnarray}
\int\limits_{-\infty}^{\infty} e^{ikx} x\log\lvert x\rvert = -i\frac{\mathrm{d}}{\mathrm{d} k}\int\limits_{-\infty}^{\infty}\mathrm{d} x\ e^{ikx} \log\lvert x\rvert 
= -i\frac{\pi}{k^2} + 2\pi i\gamma\delta'(k).
\end{eqnarray}
Using these in \eq{pk1} and noting that for $k>0$,  the delta functions and derivatives of delta functions (including the contribution from the Fourier transforms of constants) are irrelevant, we get:
\begin{eqnarray}
P(k, \eta) = \frac{H^2}{2(2\pi)^3\pi ik} \left(i\pi \eta^2 + \frac{i\pi}{k^2}\right)
= \frac{H^2}{2(2\pi)^3k^3}(1+k^2\eta^2),
\end{eqnarray}
which agrees with the previous result. 

Incidentally the same approach can also be used to determine the power spectrum in the case of a massless field in a power law universe with $q\approx1$. In this case the series expansion of Wightman function to the lowest orders  in $q-1$ looks like
\begin{eqnarray}
G(\eta,\eta';\rho) &= \frac{H^2}{8\pi^2 (q-1)} - \frac{1}{4\pi^2 L^2} - \frac{H^2}{8\pi^2}\log(\eta\eta'L^2 H^2 c), \\
 &= \frac{H^2}{8\pi^2 (q-1)} - \frac{1}{4\pi^2 L^2} - \frac{H^2}{8\pi^2}\log(\eta\eta') - \frac{H^2}{8\pi^2}\log(L^2 H^2 c).
 \label{chk3}
\end{eqnarray}
As far as the Fourier transform of the Wightman function with respect to $\rho$ is concerned, the additional term involving $\log(\eta\eta')$ merely contributes like a constant i.e; only to the $k=0$ case via a delta function and get back the same result for $k>0$ to the lowest order.

\subsection{Cosmic (static) Vacuum}

As we mentioned earlier, an observer at $r=0$ in the static co-ordinates is a geodesic observer and has the same conceptual status as an observer at $\bm x=0$ in the conformal Friedmann co-ordinates. In the latter, the conventional choice for the vacuum state is the Bunch Davies vacuum. Since the spatial homogeneity is manifest in the Friedmann co-ordinates the power spectrum is defined by Fourier transform with respect to the spatial co-ordinates in this system.  In the static co-ordinate system, on the other hand, there is no spatial homogeneity but we now have manifest invariance with respect to time translations. Therefore the Wightman function in the static vacuum $\langle  0, \rm ss|\phi(r,\tau_1)\phi(r,\tau_2)|0, \rm ss\rangle$ --- while not de Sitter invariant --- can only depend on the time difference $\tau=\tau_1-\tau_2$. 
So for a geodesic observer at the origin, we can  define the power spectrum of quantum fluctuations using the Fourier transform with respect the time difference
\begin{equation}
 {P}_\pm(\omega)  \equiv \int_{-\infty}^{\infty} \frac{ \mathrm{d}\tau}{2\pi} e^{\pm i\omega\tau}\langle  0, \rm ss|\phi(0,\tau_1)\phi(0,\tau_2)|0, \rm ss\rangle.
\end{equation} 
Using the mode functions obtained in Sec. \ref{sec:staticmodes}, we find that the power spectrum for Painlev\'{e} observers is given by:
\begin{equation}
{P}_\pm(\omega)=  \pm|\phi_{\omega}(0)|^2  |Y_{00}|^2 \theta(\pm \omega)
= \pm|N_{\omega}|^2 |Y_{00}|^2  \left |{}_2F_1\left(-\frac{i \omega_0}{2H}, \frac{3}{2}  -\frac{i \omega_0}{2H}, \frac{3}{2} ; 0   \right)\right|^2\theta(\pm \omega). \label{PS02A}
\end{equation}
Using \eq{Norm2}, we obtain
\bea
{P}_\pm(\omega)    = \pm\theta(\pm \omega)\frac{H^2}{4\pi^2\omega}\left(1+\frac{\omega^2}{H^2}\right),
\eea
since ${}_2F_1\left(a,b,c,0\right) =1 $.  The result shows that for small $\omega/H=k/H \ll 1,$ the power spectrum
becomes independent of $\omega$, i.e., again becomes scale invariant and  resembles the form in \eq{PS01}. Since the Painlev\'{e} and static observers are related through trivial Bogoliubov coefficients, the same power spectrum expression
remains true for the static observers \cite{Agullo:2008ka}.

To summarize, we  see that computation has been done for two different  vacuum states. Yet, the form of power spectrum remains the same to the leading order and even the correction is similar in structure.
The $r=0$ observer is a geodesic observer and can be identified with a co-moving Friedmann observer at $\bm x=0$. Such a geodesic observer can now perform two different physical operations, viz., either analyzing spatial correlations in the Bunch Davies vacuum state, or analyzing the temporal correlations in a totally different static vacuum state. Yet the low frequency limits of these two operations are exactly the same. 
Given the fact that for massless fields there is no natural de Sitter invariant vacuum, it is rather striking that these observers, different in all regards, share a common infrared feature.

\section{Power spectra of the vacuum noise: an alternative approach}\label{sec:altpow}

We will next consider an alternative route to the power spectra mention in Sec. \ref{sec:psgen}.
We have seen that, in general,  the power spectrum is defined as a Fourier transform of the Wightman function with respect to the Killing parameter through \eq{pkilling}. When the Wightman function depends on the co-ordinates only through the geodesic distance (which will be the case for spacelike separations), we can introduce its Fourier transform with respect to $Z$ (or with respect to $L^2$) and express $G(Z)$ in terms of $\mathcal{G}(Q)$ (or $\overline{\mathcal{G}}(K)$), The power spectrum in \eq{pkilling} then becomes:
\begin{equation}
  P(\omega;x^a_\perp)=  \int_{-\infty}^\infty \frac{ \mathrm{d}\lambda}{2\pi}\, \exp(i\omega \lambda)\, G[Z(\lambda,x^a_\perp)]
 =\int_{-\infty}^\infty \frac{ \mathrm{d}Q}{2\pi}\, \mathcal{G}(Q) f(\omega, Q,x^a_\perp),
  \label{pkilling1}
 \end{equation} 
where we have defined the function:
\begin{equation}
f(\omega, Q,x^a_\perp)\equiv\int_{-\infty}^\infty  \mathrm{d}\lambda\, \exp[i\omega \lambda+iQZ(\lambda,x^a_\perp)]. 
\label{F1}
\end{equation} 
A similar result can be obtained if we work with the Fourier transform $\overline{\mathcal{G}}(K)$ of $G(L^2)$ with respect to $L^2$.
We see that \eq{pkilling1} nicely separates the quantum dynamics in a specific vacuum (completely contained in $\mathcal{G}(Q)$) from the geometrical symmetries of the spacetime (contained in $f(\omega, Q,x^a_\perp)$). For a given $\mathcal{G}(Q)$, changing the relevant Killing vector will change $f(\omega, Q,x^a_\perp)$ and thus give different power spectra. 

However, one should notice that such a clean break up of geometry and quantum dynamics depends strongly on the Wightman function being a function of the geodesic distance. This remains true for spacelike separated events (which we can put on a constant time surface). However, in general the Wightman function depends apart from $Z$ on the temporal co-ordinates of the events as well (through  an -$i\epsilon$ prescription). Therefore, whenever, we have a timelike killing vector in the spacetime and we choose a co-ordinate system where the parameter $\lambda$ is the parameter along the timelike direction the \eq{pkilling1} gets modified to

\begin{eqnarray}
  P(\omega;x^a_\perp) &=&  \int_{-\infty}^\infty \frac{ \mathrm{d}\lambda}{2\pi}\, \exp(i\omega \lambda)\, G[\lambda, Z(\lambda,x^a_\perp)],\nonumber\\
 &=& \int_{-\infty}^\infty \frac{ \mathrm{d}Q}{2\pi}\int_0^{\infty}\frac{ \mathrm{d}\lambda}{2\pi}\left[\mathcal{G}(Q,\lambda)e^{iQ Z(\lambda) + i \omega \lambda} + \mathcal{G}(Q,-\lambda)e^{iQ Z(-\lambda) - i \omega \lambda}\right].
   \label{pkilling3}
 \end{eqnarray} 
Thus, the clean break up as was the case for the spacelike vector goes away and the power spectrum along a timelike Killing vectors has to be obtained through a careful evaluation of the Fourier transform of the Wightman function. We shall now see how this works out in flat spacetime as well as in de Sitter spacetime.

\subsection{Warm-up: inertial vacuum in Flat spacetime}\label{sec:warmup}

As a warm up, we will use the above technique to construct the power spectrum of inertial vacuum evaluated using two different Killing vector fields. The first Killing vector corresponds to (i) translations in the Minkowski time co-ordinate $t$ and the second Killing vector corresponds to (ii) invariance under Lorentz boosts, which maps into invariance under translations of the Rindler time co-ordinate $\tau$. In both cases we will use the Fourier transform of the Green's function in the inertial vacuum given by:
\bea
\tilde{G}(K)= \int \mathrm{d}L_{\epsilon}^2 G(L_{\epsilon}^2) e^{-i kL_{\epsilon}^2};
\eea
leading to
\begin{equation}
\tilde{G}(K)=\left(\frac{s_t}{2\pi i}\right) \theta(- s_tK),
\end{equation}
with $s_t = \text{sgn}(t-t')$. The second result is obtained for the massless case we are interested in, where the Wightman function gets the form $G(L_{\epsilon}^2) =-1/4 \pi^2 L_{\epsilon}^2$.

Let us start with the first case which is almost trivial. 
The invariant geodesic distance is just $L_{\epsilon}^2=t_{\epsilon}^2\equiv (t_2-t_1 -i \epsilon)^2$ so that we  now need to compute the integral:
\begin{eqnarray}\label{int_rep_for_pomega}
P_\pm(\omega)&=&\int_{-\infty}^{\infty}\frac{ \mathrm{d}t}{2\pi}\int_{-\infty}^{\infty} \mathrm{d} K \tilde{G}(K)e^{i(Kt_{\epsilon}^2 \pm \omega t)},\\
&=&\int_{-\infty}^{\infty}\frac{ \mathrm{d}t}{2\pi}\int_{-\infty}^{\infty} \mathrm{d} K \left(\frac{s_t}{2\pi i}\right) \theta(- s_tK) e^{i(Kt_{\epsilon}^2+\omega t)},
\end{eqnarray}
Integrating the above expression yields the expression  (for details, see Appendix \ref{app:derh}) :
\begin{eqnarray}
P^{\text{inertial}}_\pm(\omega)=\pm\frac{\omega}{4 \pi^2}\theta(\pm\omega)= \frac{|\omega|}{4 \pi^2}\theta(\pm\omega),
\label{inpowspec}
\end{eqnarray}
which is the standard result.

Let us now turn to the case of determining the power spectrum in the inertial vacuum with respect to the boost Killing vector. 
To find $L^2$ in this  case, we introduce the (spherical) Rindler co-ordinates $(\tau, \xi, \theta, \varphi)$, in terms of Minkowski spherical polar co-ordinates $(t,r, \theta, \varphi)$ (with the same angular co-ordinates $\theta, \varphi$) by:
\begin{eqnarray}
t = a^{-1}e^{a\xi}\sinh(a\tau);\qquad
r = a^{-1}e^{a\xi}\cosh(a\tau).
\end{eqnarray}
The geodesic distance $L^2$ between two points along the Killing trajectory, $(\xi,\theta,\varphi) = \text{const.}$ (which actually gives the integral curve for the boost Killing vector) is then given by:
\begin{equation}
L^2 = \frac{2e^{2a\xi}}{a^2}(\cosh(a\Delta\tau)-1)\equiv \bar{A}(\cosh(a\Delta\tau)-1).
\end{equation}
For convenience, we choose $\xi = 0$ ($\bar{A} = \frac{2}{a^2}$) so that the proper time along the trajectory of such an observer is given by $\tau$. 
It is simpler to work with a Fourier transform of $G$ wit respect to $L^2$ in this case.
%The function corresponding $f$ (see \eq{F1}) is now given by:
%\begin{equation}
 %f(K)=\int\limits_{-\infty}^{\infty}\mathrm{d}\Delta\tau\ e^{iKL^2+i\omega\Delta\tau}
% =\frac{i\pi}{a}e^{\frac{\pi\omega}{2a}}   e^{-iKA} \mathrm{H}_{-\frac{i\omega}{a}}^{(1)}(kA)
%\end{equation} 
The power spectrum  (for details, see Appendix \ref{app:derh}) expression is evaluated to obtain:
\begin{eqnarray}
P^{\text{Rindler}}_\pm(\omega) 
= \frac{\omega}{4\pi^2}\left(\frac{e^{\pm\frac{\pi\omega}{a}}}{e^{\frac{\pi\omega}{a}}-e^{\frac{-\pi\omega}{a}}} \right).
\label{rinps1}
\end{eqnarray}
At this stage we came back again to the contrast of the conventions of Fourier transform. With the convention $e^{-i \omega t} $ one obtains the Rindler power spectrum to be
\begin{eqnarray}
P^{\text{Rindler}}_-(\omega) 
= \frac{\omega}{4\pi^2}\left(\frac{1}{e^{\frac{2\pi\omega}{a}}-1} \right) = \frac{\omega}{4\pi^2}(n_{\omega}),
\label{rinps}
\end{eqnarray}
which is the Rindler power spectrum and $n_{\omega}$ is the well known thermal spectrum for the Rindler observer. So as promised, this convention agrees with the power spectrum defined through the response of the UDD.
On the other hand, with the Fourier transform convention using $e^{i \omega t},$ we end up with 
\begin{eqnarray}
P^{\text{Rindler}}_+(\omega) 
= \frac{\omega}{4\pi^2}\left(\frac{e^{\frac{2\pi\omega}{a}}}{e^{\frac{2\pi\omega}{a}}-1} \right) = \frac{\omega}{4\pi^2}(1+ n_{\omega}).
\label{rinps0}
\end{eqnarray}
This result has a straight forward interpretation in terms of detector response. The upward transition rate of the detector corresponds to absorption of quanta and is proportional to $n(\omega)$. This is correctly captured by \eq{rinps} which uses the Fourier transform convention corresponding to the response of UDD. But when we use the opposite convention, we are flipping the sign of $\omega$, which physically corresponds to a downward transition. Such an emission process, as is well-known, has a spontaneous rate, given by the  $\omega/4\pi^2$ term in \eq{rinps0}, as well as an induced emission term $(\omega/4\pi^2)n_{\omega}$ (proportional to the number of ambient quanta). This is precisely what we get in 
\eq{rinps0}.

If we do not want to think in terms of UDD response then, we could interpret \eq{rinps0} as follows.
This definition  takes account of the inertial vacuum correlation (and the corresponding power spectrum)  and then supplements it with the thermal power spectrum due to ambient quanta.
(It is easy to verify that in the limit of vanishing acceleration $a \rightarrow 0$, we recover the inertial power spectrum \eq{inpowspec}, since $n_{\omega} \rightarrow -  \theta(-\omega)$ in this limit.)
That is, the power spectrum of the vacuum fluctuations, evaluated by Fourier transform along the boost Killing vector trajectory has a supplementary thermal character over and above the inertial power spectrum. 

In the above analysis we took $\xi=0$ for simplicity. At non-zero $\xi$ it makes better sense to evaluate the Fourier transform with respect to the proper time at the location, viz $\tau_p\equiv \tau\exp a\xi$. This merely redshifts the frequency and thus the temperature  to $T=(a/2\pi)\exp(-a\xi)$. 

\subsection{Power Spectra in Friedmann universes}\label{sec:bdpsdS}

We know that the  Wightman function \eq{Gplusdef} for any Friedmann cosmology is a function of $Z_{\epsilon}, \eta, \eta'$ alone. So we can compute the  power spectrum along the same manner as in the case of flat spacetime power spectrum using the spatial Killing vectors. However, since the Wightman function is potentially divergent in many scenarios, as we discussed previously, it will not be advisable to compute the Wightman explicitly first and then do the Fourier transform as we could do in flat spacetime. We will first demonstrate the evaluation of power spectrum, through the spatial Killing vectors in a generic Friedmann cosmology, then we will carry out the exercise for a special power law i.e., the de Sitter spacetime. Using the three spatial Killing vectors, we can define the power spectrum as
\bea
P(\bm{k},\eta)  = \frac{\beta^2}{16\pi^2\sqrt{2}}(\eta\eta')^{q-1}\int \frac{ \mathrm{d}^3 {\bf \rho}}{(2\pi)^3} 
 \int\limits_{-\infty}^{\infty} \mathrm{d} u \frac{e^{-\nu u}}{\left(\cosh u- Z\right)^{\frac{3}{2}}}e^{i {\bf k}\cdot{\bm\rho}},
\eea
where $\bm\rho=\bm x- \bm x'$.
Since, we have spatial Killing vectors we will evaluate the Fourier transform at an equal $\eta$ surface in conformal co-ordinates
\bea
Z=1+\frac{(\Delta \eta)^2 -\rho^2}{2 \eta \eta'}\big|_{\eta=\eta'} = 1-\frac{\rho^2}{2 \eta^2},
\eea
which leads to the expression
\bea
P(\bm{k},\eta)  = \frac{\beta^2}{16\pi^2\sqrt{2}}(\eta)^{2q-2}\int  \frac{ \mathrm{d}^3{\bm\rho}}{(2\pi)^3} 
 \int\limits_{-\infty}^{\infty} \mathrm{d} u \frac{e^{-\nu u}}{\left(\cosh u- 1+\frac{\rho^2}{2 \eta^2}\right)^{\frac{3}{2}}}e^{i {\bf k}\cdot{\bm\rho}}.
\eea
Using the identity
\bea
\frac{1}{b^{\frac{3}{2}}}=\frac{1}{\Gamma\left[\frac{3}{2}\right]}\int_0^{\infty} \mathrm{d}s \ s^{\frac{1}{2}}e^{-s b},
\eea
for positive $b$, we can write
\bea
P(\bm{k},\eta)  = \frac{\beta^2}{16\pi^2\sqrt{2}}(\eta)^{2q-2}\int  \frac{ \mathrm{d}^3{\bm\rho}}{(2\pi)^3} \frac{1}{\Gamma\left[\frac{3}{2}\right]}
\int_0^{\infty}  \mathrm{d}s \ s^{\frac{1}{2}} \int\limits_{-\infty}^{\infty} \mathrm{d} u e^{-\nu u} e^{\left[-s\left(\cosh u- 1+\frac{\rho^2}{2 \eta^2}\right)\right]}e^{i {\bf k}\cdot{\bm\rho}}.
\eea
Performing the spatial integrals, we can write the expression as
\bea
P(\bm{k},\eta)  = \frac{\beta^2}{16\pi^2\sqrt{2}}(\eta)^{2q-2}\frac{(2 \pi \eta)^{\frac{3}{2}}}{\Gamma\left[\frac{3}{2}\right]}
\int\limits_{-\infty}^{\infty} \mathrm{d} u e^{-\nu u}\int_0^{\infty}  \mathrm{d}s \ s^{\frac{1}{2}}  e^{-2s\sinh^2{\left( \frac{u}{2}\right)}}e^{-\frac{k^2 \eta^2}{2 s}}.
\eea
Again, using $e^{u/2}=z$, one can carry out the $u-$ integration casting the expression into
\bea
P(\bm{k},\eta)  = 2\frac{\beta^2}{16\pi^2\sqrt{2}}(\eta)^{2q-2}\frac{(2 \pi \eta)^{\frac{3}{2}}}{\Gamma\left[\frac{3}{2}\right]}
\int_0^{\infty} \frac{ \mathrm{d}s}{s} e^s e^{-\frac{k^2 \eta^2}{2 s}}  K_{-\nu}(s),
\eea
with $K_{\nu}(s)$ being the Bessel function of order $\nu$. Finally the left-over $s-$ integration can be done to obtain
\bea
P(\bm{k},\eta)  &=& 2\frac{\beta^2}{16\pi^2\sqrt{2}}(\eta)^{2q+1}\frac{(2 \pi)^{\frac{3}{2}}}{\Gamma\left[\frac{3}{2}\right]}\frac{\pi^2}{2}\frac{1}{\sin^2{\pi \nu}}\left(J_{\nu}(k\eta)^2 + J_{-\nu}(k\eta)^2 -2\cos{\pi\nu}J_{\nu}(k\eta) J_{-\nu}(k\eta)\right),\nonumber\\
 &=& 2\frac{\beta^2}{16\pi^2\sqrt{2}}(\eta)^{2q+1}\frac{(2 \pi)^{\frac{3}{2}}}{\Gamma\left[\frac{3}{2}\right]}\frac{\pi^2}{2}|H_{\nu}^{(2)}(k\eta)|^2,
\eea
recovering the power spectrum \cite{gravitation} which clearly relates the power spectrum of a massless field in Friedmann universe to that of a massive scalar field in the de Sitter spacetime. 

Incidentally, we can also straight away calculate the power spectrum of scalar field through the $q\rightarrow 1$ limit, yielding
\bea
P^{\text{dS}}(\bm{k},\eta) = 2\frac{H^2}{16\pi^2\sqrt{2}}(\eta)^{3}\frac{(2 \pi)^{\frac{3}{2}}}{\Gamma\left[\frac{3}{2}\right]}\frac{\pi^2}{2}|H_{\nu}^{(2)}(k\eta)|^2,
\eea
which in the massless limit becomes
\bea
P^{\text{dS}}(\bm{k},\eta) = \frac{H^2}{2(2 \pi)^3k^3}(1+k^2\eta^2).
\eea
For completeness we will just show an explicit calculation for the de Sitter massless case, where we use different Killing directions (spacelike and timelike), as done in the flat spacetime, which builds trust in the mechanism developed in this work.

\subsubsection{Bunch Davies vacuum in de Sitter spacetime}

We first start with the spatial Killing vectors, as before. Though this time (being a spatial Fourier transform), we decompose the power spectrum into totally geometric and state dependent pieces as we argued previously. For this purpose, we just need to use \eq{pkilling1} with appropriate $\mathcal{G}(Q)$ and $f$.
However, we have to be careful about one subtlety that the Bunch Davies vacuum is a natural vacuum for co-moving observers, for whom the time direction is not a Killing direction, but the spatial directions are. Thus, we need to convert the integral transforms on a spacelike surface rather than a timelike one.
On such a surface the  Fourier transform $\mathcal{G}(Q)$ of the Wightman function is given by (see \ref{app:deri}):
\begin{equation}
\mathcal{G}(Q) = -\frac{\beta^2}{4\sqrt{2\pi}}e^{-\frac{i\pi}{2}} \left(\theta(-Q)Q^{\frac{1}{2}} \mathrm{H}_{\frac{3}{2}}^{(2)}(Q)\right),
\label{gplusq}
\end{equation}
where we have  set $s_\eta = 1$  \footnote{Or equivalently, $s_\eta =-1$ as well for that matter. In fact one should choose $|\eta-\eta'|= \epsilon$ and gradually take $\epsilon \rightarrow 0$ rather than directly putting $s_\eta =0$. Of course, for spacelike separated events, the Wightman function does not vanish, so neither should its Fourier transform.}.
Similar to the Minkowski spacetime, in the de Sitter spacetime also, we have two natural choices (although corresponding to two different sets of observers) which are: (a) the integral curves of the Killing vectors related to spatial translation for comoving observers and (b) the integral curve of the Killing vector related to translation along the static time co-ordinate $\tau$. We will consider both these possibilities. 

The situation in (a) is best handled in the  conformal Friedmann co-ordinates, in which the power spectrum  is given by a spatial Fourier transform of the Wightman function with respect to the comoving separation between two points at the same time $\eta = \eta'$. 
Since $Z$ in these co-ordinates is given by $Z = 1-(\rho^2/2\eta^2)$,
the function $f$ reduces to the triple Gaussian integral:
\begin{equation}
f= e^{iQ}\left( \int\limits_{\mathbb{R}^3}\frac{\mathrm{d}^3 \bm{\rho}}{(2 \pi)^3} e^{-\frac{iQ}{2} \frac{\rho^2}{\eta^2}}e^{i\bm{k}\cdot \bm{\rho}}\right)=(-\eta)^3 \left(\frac{2\pi}{iQ}\right)^{\frac{3}{2}} e^{iQ}e^{\frac{i}{2Q}k^2\eta^2}.
\label{fgauss}
\end{equation} 
On using \eq{gplusq} and \eq{fgauss} in \eq{pkilling1} the power spectrum reduces to (with the notation $\xi = -Q$)
\begin{equation}
P(\bm{k},\eta) = \frac{\pi H^2}{4\pi(2 \pi)^3 i}(-\eta)^3 e^{-\frac{3\pi i}{4}} \int\limits_0^\infty \frac{\mathrm{d} \xi}{\xi^{\frac{3}{2}}} \xi^{\frac{1}{2}} \mathrm{H}_{\frac{3}{2}}^{(2)}(-\xi) e^{-i\xi}e^{-\frac{i}{2\xi}k^2\eta^2}.
\end{equation}
Using the explicit from of $\mathrm{H}_{\frac{3}{2}}$ and 
writing $\xi  = \frac{1}{u^2}$ the integral becomes:
\begin{equation}
P(\bm{k},\eta) = -\frac{H^2}{\sqrt{2\pi}(2 \pi)^3} (-\eta)^3 e^{-\frac{3\pi i}{4}} \int\limits_0^\infty \mathrm{d} u\ (1+iu^2)e^{-\frac{i}{2}k^2\eta^2u^2}.
\end{equation}
This is straight forward to evaluate using
\begin{equation}
\int\limits_0^\infty \mathrm{d} u\ (1+iu^2)e^{-\frac{i}{2(2 \pi)^3}k^2\eta^2u^2} = i\sqrt{\frac{\pi}{2}} \frac{1+k^2\eta^2}{(ik^2\eta^2)^{\frac{3}{2}}}.
\end{equation}
Using the fact that $\eta < 0$ to simplify the denominator and substituting this in the expression for the power spectrum, we recover the familiar result:
\begin{equation}
P(\bm{k},\eta) = \frac{H^2}{2(2 \pi)^3k^3}(1+k^2\eta^2).
\label{pkfinal1}
\end{equation}

\noindent We will now repeat this analysis for case (b), using the integral curves of the Killing vector corresponding to time translation symmetry in the static co-ordinates. Using the spatial homogeneity of de Sitter manifold we can always choose the two points on the Killing trajectory to have the co-ordinates  $(\tau_1,R = 0)$ and $(\tau_2,R=0)$. Breaking the $\mathcal{G}(Q)$ into two regions of positive temporal separation and negative temporal separations respectively , the expression for the power spectrum (see Appendix \ref{app:derj}) is obtained as
\begin{equation}
P_\pm(\omega) = \frac{H^2}{4\pi^2 \omega}\left(1+\frac{\omega^2}{H^2}\right)\frac{e^{\pi\frac{\pm \omega}{H}}}{2\sinh\left(\pi\frac{\omega}{H}\right)}
\label{chk2},
\end{equation}
again leading to
\begin{eqnarray}
P_+(\omega) &=& \frac{H^2}{4\pi^2 \omega}\left(1+\frac{\omega^2}{H^2}\right)\frac{e^{\pi\frac{ \omega}{H}}}{2\sinh\left(\pi\frac{\omega}{H}\right)}=\frac{H^2}{4\pi^2 \omega}\left(1+\frac{\omega^2}{H^2}\right)(1+ n_\omega)
\label{chk21},\\
P_-(\omega) &=& \frac{H^2}{4\pi^2 \omega}\left(1+\frac{\omega^2}{H^2}\right)\frac{e^{-\pi\frac{ \omega}{H}}}{2\sinh\left(\pi\frac{\omega}{H}\right)} =\frac{H^2}{4\pi^2 \omega}\left(1+\frac{\omega^2}{H^2}\right)n_\omega, \label{chk22}
\end{eqnarray}
where 
\begin{equation}
\qquad n_\omega = \frac{1}{e^{\beta \omega} - 1}; \qquad \beta^{-1} = \frac{H}{2\pi}.
\end{equation} 
In addition to the power spectrum $P_{\rm static}(\omega)$  we found earlier in the static vacuum state we have a supplement through $n_\omega$, the number density of thermal quanta with temperature $H/2\pi$. We will now discuss the implications of this result.

\subsection{Comments on the vacuum noise}

The quantum fluctuations in the vacuum state is described, for a free field, in terms of the Wightman function which can be thought of as a correlation function in real space. Whenever a suitable Fourier transform can be defined we can associate a power spectrum with this correlation function which quantifies the amount of quantum vacuum noise. We have described in Sec. \ref{sec:psgen}  (and elaborated in Sec. \ref{sec:altpow})  how this can be achieved in terms of Killing vector fields. The resulting power spectrum (see \eq{pkilling}) depends on the conjugate variable, introduced in the Fourier transform   with respect to the Killing parameter of the integral curves of the Killing vector fields. The definition in \eq{pkilling} is generally covariant but, of course, depends on (i) the choice of the vacuum state and (ii) the choice of the Killing vector field. 

In the context of flat spacetime, the natural choice for the vacuum state is the inertial vacuum. Two natural choices for the Killing vector fields correspond to (a) translation along the inertial time direction and (b) the Lorentz boost.  We found in Sec. \ref{sec:warmup} that the cases (a) and (b) lead to the power spectrum of the form in \eq{inpowspec} and \eq{rinps}. For positive frequencies the power spectrum with respect to time translation vanishes while the power spectrum defined using boost Killing vector leads to well known thermal fluctuations in the Rindler frame, in one of the convention. Further, in yet another convention of power spectrum for a timelike Killing vector, there is a non-zero power spectrum of the vacuum for the inertial observer, which adds  to the thermal fluctuation in the case of Rindler observer. Clearly, whatever may be the convention of evaluation of power spectrum, the \textit{minimum value} of the vacuum noise for positive frequencies, as measured by these power 
spectra,
 is zero, corresponding to the case (a). 

In Sec. \ref{sec:bdpsdS},  we performed exactly the same analysis with different choices for the vacua and Killing vector fields in the de Sitter spacetime. The results obtained earlier are summarized below  for three different situations:
\begin{equation}
 P =
 \begin{cases}
  \frac{H^2}{2 (2\pi)^3 k^3} ( 1 + k^2 \eta^2) &\qquad \text{(BD; homogeneity)}\\
  { }\\
  \frac{H^2}{4\pi^2 \omega}\left(1 + \frac{\omega^2}{H^2}\right) &\qquad \text{(static; time translation)}\\
  { }\\
   \frac{H^2}{4\pi^2 \omega}\left(1 + \frac{\omega^2}{H^2}\right) ( 1 + n_\omega) &\qquad \text{(BD; time translation)}
 \end{cases}
 \label{tpnote1feb14}
\end{equation} 
As indicated, the first result used Bunch-Davies vacuum and Killing vectors corresponding to spatial homogeneity of the de Sitter manifold. The second uses static vacuum with the Killing vector corresponding to translations in static time co-ordinate; the last one is obtained when we use Bunch-Davies vacuum and the Killing vector corresponding to translations in static time co-ordinate.

The first result is well known in literature in the context of inflationary perturbations. It leads to the scale invariant power spectrum $4\pi k^3P(k) \approx (H^2/4\pi^2)$ in the infrared limit of $k\to 0$. 
We will comment briefly on the second and third results. Note that in obtaining the second and third results we are using the same Killing vector field to define the power spectrum but two different vacua. The factor $(1+\omega^2/H^2)$ in both arises due to purely kinematic reasons in both results and has been noticed --- and discussed --- in couple of earlier works \cite{Singh:2013pxf,Garbrecht:2004du} in connection with detector response in de Sitter, as well as static observer power spectrum \cite{Agullo:2008ka}.  The third result then shows that the power spectrum of the Bunch-Davies vacuum is enhanced with respect to the power spectrum $P_{\rm static}(\omega)$ of the static vacuum by the factor $1+n_\omega$. This has a natural interpretation of the enhancement of the spontaneous process by an induced process proportional to the presence of thermal quanta $n_\omega$. This interpretation makes sense because it is well known that the static and Bunch-Davies vacua are indeed related by a thermal Bogoliubov 
coefficient.

What is really interesting about the three results is that in the de Sitter spacetime --- unlike in flat spacetime --- there is a \textit{residual vacuum noise} which persists in the infrared limit. All the expressions in \eq{tpnote1feb14} diverge in the infrared limit because of the phase space volume element $k^{-3}, \omega^{-1}$ etc. But if we  multiply $P(k)$ by these relevant phase space factors to evaluate the amplitude of the power spectrum and obtain e.g., $k^3 P(k), \omega P(\omega)$ in the three cases (upto solid angle factors),  we get a vacuum noise having a value of $H^2/4 \pi$ for the first two cases while it has a diverging character for the third case (Bunch Davies vacuum and the static observer) due to the $n_\omega$ factor.
In all the  cases, there exists a minimum vacuum noise. Its value is $H^2/4 \pi$ in the for first two cases, while in the third case the minimum value is $\sim 1.26 H^2/4 \pi $ and occurs for $\omega_{min} \sim 0.37 H$ 
(If we adopt the second convention for the power spectrum,  the infrared limit of the vacuum noise  diverges in the third case, but for negative frequencies where the power spectrum survives has support. However, unlike the first convention, this has a vanishing ultraviolet character. Thus, In any case there is a minimum of vacuum noise  at the infrared end of the spectrum, which is determined by the curvature $H^2$ of the de Sitter spacetime.)

This fact suggests an interesting conjecture for de Sitter vacuum fluctuations which we will now describe: Let $G(x_1,x_2)$ be a Wightman function in any suitably defined vacuum state in de Sitter. Let $C(\lambda)$ be an integral curve of a Killing vector field $\xi^a(x)$ in the spacetime with $\lambda$ being the Killing parameter which varies along the full real line. Then, the power spectrum expressed in terms of a variable $\Omega$ conjugate to the Killing parameter $\lambda$ can be defined as the Fourier transform in \eq{pkilling}. The infrared limit of this expression, multiplied by a phase space measure $\mathcal{M}(\Omega)$, will then seem to have a minimum value. It appears that 
\begin{equation}
\mathcal{P}_{\rm IR} = P(\Omega) \mathcal{M}(\Omega) = \mathcal{M}(\Omega) \int_{-\infty}^\infty \frac{ \mathrm{d}\lambda}{2\pi}\ G[x(\lambda_1=\lambda), x(\lambda_2=0)] e^{ i \Omega \lambda} \ge \frac{H^2}{4\pi^2}.
\end{equation}
We have explicitly verified this result in the three cases above, but for a general proof one needs to sharpen the idea of the measure  $\mathcal{M}(\Omega)$. We hope to address this question in a future publication. 

\section{Conclusions}

The study of quantum fields in Friedmann universe is a mature subject with decades of literature. We have revisited several issues in this subject and have been able to obtain some fresh insights. We provide a summary of the results in this paper below:
\begin{itemize}
  \item {\bf Quantum dynamics of a field in one Friedmann universe is same as that of another field in another Friedmann universe}: We have shown that there is a dynamical equivalence between massive scalar field (say $\phi,m$) on a cosmological Friedmann background (with scale factor, say, $a$) with another scalar field $(\psi,m')$ in another Friedmann universe (with scale factor $b$). In particular, a massless scalar field in any power law cosmology can be mapped to a massive field in a de Sitter spacetime where the mass of the  field in the de Sitter spacetime is determined by the power-law coefficient $q$ of the original scale factor.
  This result is of importance because  it links all the  features of the massless fields in any power law cosmology to a massive field in de Sitter, which  is well-studied and fairly well-understood. This opens up a new line of attack on  QFT  in different phases of the Universe (e.g. radiation era, matter era etc) when the expansion factor can be approximated as a power law. We plan to take up these aspects  in an upcoming work.
  Another important lesson learnt in this result is the ``omnipresent''  de Sitter character of the geometry even in any  power law cosmology, which a massless field always ``feels'' and responds to. 
 
 \item {\bf Massless field in de Sitter:} It is an age-old result in this subject that there is no de Sitter invariant vacuum for a massless scalar field. This reflects in the divergence of the two-point function for massless fields. One can show that such a vacuum state and a well defined correlation function do exist for \textit{massive} fields which become pathological in the massless limit. There are multiple reasons suggested in the literature for the  occurrence of this divergence but all of which seems to linked to breakdown of a symmetry. 
 
 The present study shows that this pathology may have an origin which is quite different from the lack of de Sitter invariance for massless vacuum. We have shown that such (and more severe) divergences for massless scalar fields occur  in other power law cosmologies as well whenever the equation of state parameter $w$ is negative. Power-law Friedmann universes  corresponding to $-1<w<0$ have \textit{no special symmetries} but still exhibit results similar to those we find when $w=-1$ corresponding to the de Sitter. So if we think of $w=-1$ as a limiting case in this band then the pathology has no special relationship to de Sitter invariance or its breakdown. It exhibits a special case of a \textit{more general pathology} which arises whenever the \textit{pressure of the source becomes negative}.
 
 That is, we show that the non-existence of the QFT (indicated by divergent two-point functions) is intimately coupled with the character of the source supporting the geometry. Whenever, the Friedmann geometry is sourced by a negative pressure fluid ($w <0$) the massless field suffers a divergence. This result, therefore becomes a quantum response of the matter field to the negative pressure source and --- as a special case --- to the acceleration of the universe. {\it Further the existence of a  well-defined Hilbert space for a massless scalar field is incompatible with the acceleration of universe} which requires $w<-1/3$. This incompatibility must leave observational imprints in cosmology over the era of accelerated expansions (even today). In fact we show that the pressure-less dust limit ($\omega \rightarrow 0^-$) has similar divergent character as the  exponential expansion as a limit. This result also opens up an independent line of research, which we hope to undertake in a subsequent paper.

 \item {\bf  Power Spectrum through Killing Directions :} Since the Power spectrum is  one of the well-accepted and observationally  useful characterizations of  a fluctuating quantum field, it is important to have a  geometric understanding of its definition. We explain how Killing vectors provide a natural way of characterizing the quantum fluctuations of fields and their correlations. We develop the machinery to obtain the power spectrum in the de Sitter universe through its Killing vectors and extend the results to  any other (including power law) cosmology. We show the equivalence of using any Killing direction in the flat as well as the de Sitter spacetime. This provides us a geometric definition of the power spectrum and allows us to  explore interesting cosmological contexts using this tool.

 \item {\bf Persistent vacuum noise:} We have shown that the cosmological spacetimes --- in particular, the de Sitter spacetime --- host a minimum vacuum noise (vacuum power spectrum) which is related to its curvature. This persistent noise is revealed when we probe  the quantum correlators at points separated by infinite Killing affine parameter (that is to say, large wavelength limit in the power spectrum). The de Sitter spacetime  always has a $\sim H^2$ vacuum noise
 which reflects itself also in the standard scale invariant  power spectrum. We see that this is a lower bound and it is possible to enhance this noise by adopting trajectories for which the chosen state does not remain the natural vacuum. Through inertial/Rindler observer correspondence in Minkowski spacetime and comoving/static observer correspondence in the de Sitter spacetime, we show how the persistent vacuum noise gets enhanced by {\it stimulated emission} in two different contexts. 
 
\end{itemize}

 The main results of this research, outlined above, open up new approaches for investigations  which are  rich in terms of possibilities, some of which we hope to explore in future studies.

\section*{Acknowledgements}
Research of KL is supported by INSPIRE Faculty Fellowship grant by Department of Science and Technology, Government of India. The research of TP is partially supported by the J.C. Bose research grant of the Department of Science and Technology, Government of India. KR is supported by Senior Research Fellowship of the Council of Scientific \& Industrial Research (CSIR), India. AV would like to thank IUCAA for hospitality during part of this work. 

%\newpage

\appendix

\section*{Appendix}

\section{ Supplementary material - mathematical details}\label{app:supp}

\subsection{Derivation of \eq{C27} and related results}\label{app:dera}

We know that for spacelike separations, the parameter $Z =\cos{(H \ell)}$ is related to the conformal co-ordinate separation (from a base point $(\eta_0, {\bf x_0})$) like
\bea
\cos{(H \ell)} = \frac{\eta^2 + \eta_0^2 -({  \bf \Delta  x })^2}{2 \eta \eta_0}.
\eea
Since  the separation is spacelike, then we can trade $({\bf \Delta x })^2 \equiv r^2$ for $\ell$ as
\bea
r &=&\sqrt{\eta^2 + \eta_0^2 -2 \eta \eta_0 \cos{(H \ell)}},\nonumber\\
dr &=& \frac{(\eta -\eta_0\cos{(H \ell)})}{\sqrt{\eta^2 + \eta_0^2 -2 \eta \eta_0 \cos{(H \ell)}}} \mathrm{d}\eta  -\frac{H \eta \eta_0\sin{(H \ell)} }{\sqrt{\eta^2 + \eta_0^2 -2 \eta \eta_0 \cos{(H \ell)}}} \mathrm{d}\ell. \label{dSELL}
\eea
Thus, substituting the expressions \eq{dSELL} in the de Sitter line element in conformal co-ordinates we get
\bea
ds^2 &=& -\frac{1}{H^2 \eta^2}\left(d \eta^2 - dr^2 -r^2 d\Omega^2 \right), \nonumber\\
&=& -\frac{d \eta^2}{H^2 \eta^2}\frac{ \eta_0^2\sin^2{(H \ell)} }{(\eta^2 + \eta_0^2 -2 \eta \eta_0 \cos{(H \ell)})} +\frac{\eta_0 ^2 \sin^2{(H \ell)}d \ell^2 }{(\eta^2 + \eta_0^2 -2 \eta \eta_0 \cos{(H \ell)})} \nonumber\\
&-& \frac{2\eta_0 \sin{(H \ell)}(\eta -\eta_0\cos{(H \ell)}) \mathrm{d}\eta \mathrm{d}\ell }{(\eta^2 + \eta_0^2 -2 \eta \eta_0 \cos{(H \ell)})}
+ \frac{(\eta^2 + \eta_0^2 -2 \eta \eta_0 \cos{(H \ell)})}{H^2 \eta^2} \mathrm{d}\Omega^2. \label{EtaElLE}
\eea
The $\eta, \ell$ sector of the line element can further be diagonalized by going to a new time co-ordinate $\tau \equiv \tau(\eta, \ell),$
such that the function $\tau$ satisfies the following condition
\bea
\partial_{\eta} \tau g^{01} + \partial_{\ell} \tau g^{11} =0,
\eea
with $g^{\mu \nu}$ marking the inverse metric for \eq{EtaElLE}.
Further, the integrability condition will force us to choose
\bea
\partial_{\eta} \tau &=& -k(\tau,\ell)g^{11}, \nonumber\\
\partial_{\ell} \tau &=& k(\tau,\ell)g^{01},
\eea
with a function $k(\tau,\ell)$ satisfying the condition (for integrability) 
\bea
\partial_{\ell}\left( k(\tau,\ell)g^{11}\right) + \partial_{\eta}\left( k(\tau,\ell)g^{01} \right) = 0.
\eea
Thus, we obtain,
\bea
k(\eta,\ell) = \frac{\eta_0}{\eta \sqrt{\eta^2 + \eta_0^2 -2 \eta \eta_0 \cos{(H \ell)}}},
\eea
giving
\bea
\tau = \cosh^{-1}\left(- \frac{\sqrt{\eta^2 + \eta_0^2 -2 \eta \eta_0 \cos{(H \ell)}}}{\eta\sin{H \ell}} \right),
\eea
which leads us to the line element 
\begin{equation}
ds_H^2=-\frac{\sin^2(H\ell)}{H^2}d\tau^2+d\ell^2+\frac{\sin^2(H\ell)}{H^2}\cosh^2\tau d\Omega_2^2,
\end{equation} 
relating the conformal Friedmann co-ordinates ($\eta, r, \theta, \phi$) to the geodesic co-ordinates $(\tau, \ell, \theta, \phi)$ as
 \begin{equation}
 \cosh\tau=-\frac{2\eta_0 r}{\sqrt{4\eta^2\eta_0^2-(\eta^2+\eta_0^2-r^2)^2}};\qquad
 \cos(H\ell)=\frac{\eta^2+\eta_0^2-r^2}{2\eta\eta_0}; \qquad
 \end{equation} 
as given in \eq{cot1}.

%\newpage
\subsection{Derivation of results in Sec. \ref{sec:hiding}}\label{app:derb}
We start with a massless scalar field $\phi$ in a Friedmann universe characterized by a scale factor $a$, which is minimally coupled to gravity
\bea
{\cal A} = \frac{1}{2} \int  \mathrm{d}^4 x \sqrt{-g} [- \partial^{i}\phi\partial_{i}\phi-m^2 \phi^2]. \label{ScalarActionFriedmann1}
\eea
Since in the conformal co-ordinates $\sqrt{-g}= a(\eta)^4$, we can write the action \eq{ScalarActionFriedmann1} as 
\bea
{\cal A} &=& \frac{1}{2} \int  \mathrm{d}^4 x \sqrt{-g} \left[\frac{1}{a^2}\left(\dot{\phi}^2-(\partial_\mu \phi)^2 \right)-m^2 \phi^2\right],\\
&=& \frac{1}{2} \int  \mathrm{d}^4 x \left[a^2\left(\dot{\phi}^2-(\partial_\mu \phi)^2 \right)-m^2 a^4\phi^2\right],\label{ScalarActionFriedmann1_1}
\eea
where overdot ($\dot{\hspace{0.25 cm} }$) denotes the derivative w.r.t. the conformal time co-ordinate $\eta$.
Let us define another scalar field $\psi$ through $ \phi = {\cal F}(\eta) \psi$, with some arbitrary function ${\cal F}(\eta)$.
Therefore,
\bea
\dot{\phi} &=& \dot{\psi}{\cal F} + \psi \dot{{\cal F}},\\
a^2\dot{\phi}^2 &=& a^2 \dot{\psi}^2{\cal F}^2 + a^2 \psi^2 \dot{{\cal F}}^2 + 2 a^2 \psi\dot{\psi} {\cal F} \dot{{\cal F}},\\
                &=& a^2 \dot{\psi}^2{\cal F}^2 + a^2 \psi^2 \dot{{\cal F}}^2 + \frac{d}{d \eta} [a^2 \psi^2 {\cal F} \dot{{\cal F}}]-\psi^2\frac{d}{d \eta}[a^2  {\cal F} \dot{{\cal F}}].
\eea
Therefore, we can write \eq{ScalarActionFriedmann1_1} as
\bea 
{\cal A} = \frac{1}{2} \int  \mathrm{d}^4 x \left[  a^2 \dot{\psi}^2{\cal F}^2 + a^2 \psi^2 \dot{{\cal F}}^2 + \frac{d}{d \eta} [a^2 \psi^2 {\cal F} \dot{{\cal F}}]\right. \nonumber\\
\left.-\psi^2\frac{d}{d \eta}[a^2  {\cal F} \dot{{\cal F}}] - (a^2{\cal F}^2 \partial_\mu \psi)^2-m^2 a^4{\cal F}^2 \psi^2 \right].
\eea
Now, the divergence term can be thrown at the boundary to vanish under the variation of the action, thus the effective part of the action which will contribute to the equation of motion will be
\bea
{\cal A} = \frac{1}{2} \int  \mathrm{d}^4 x \left[  a^2 \dot{\psi}^2{\cal F}^2 + a^2 \psi^2 \dot{{\cal F}}^2-\psi^2\frac{d}{d \eta}[a^2  {\cal F} \dot{{\cal F}}] - (a^2{\cal F}^2 \partial_\mu \psi)^2-a^4{\cal F}^2 m^2 \psi^2 \right], \\
  = \frac{1}{2} \int  \mathrm{d}^4 x  \left[ a^2 {\cal F}^2\left(\dot{\psi}^2 - (\partial_\mu \psi)^2\right)- a^4 {\cal F}^4\left( \frac{m^2}{{\cal F}^2}+ \frac{1}{a^4 {\cal F}^4}\frac{d}{d \eta}[a^2  {\cal F} \dot{{\cal F}}] -\frac{\dot{{\cal F}}^2}{a^2 {\cal F}^4}\right) \psi^2 \right].\label{ScalarActionFriedmann1_2}
\eea
Therefore, we see that if we define a new parameter $b(\eta) =a(\eta) {\cal F}(\eta)$ the action becomes that of a scalar field in a Friedmann universe with scale factor $b$
\bea
{\cal A} =  \frac{1}{2} \int  \mathrm{d}^4 x  \left[ b^2\left(\dot{\psi}^2 - (\partial_\mu \psi)^2\right)- b^4 m_{\text{\rm eff}}^2 \psi^2 \right],\label{ScalarActionFriedmann1_3}
\eea
with an effective mass
\bea
 m_{\text{\rm eff}}^2 =\left( \frac{m^2}{{\cal F}^2}+ \frac{1}{a^4 {\cal F}^4}\frac{d}{d \eta}[a^2  {\cal F} \dot{{\cal F}}] -\frac{\dot{{\cal F}}^2}{a^2 {\cal F}^4}\right). \label{MassRelation}
\eea
The above expression relates mass of a scalar field in one Friedmann universe to the mass of another scalar field in yet another Friedmann universe. However, for a given mass of the system in one universe, we need to find which ${\cal F}$ to use in order to land into a new cosmology and then calculate the new mass in the new universe from this function ${\cal F}$ and the old $`a$'. In other words, for a given $`a$' we need to find ${\cal F}$ which is a consistent solution of \eq{MassRelation}. We should note that the LHS of \eq{MassRelation} is $\eta$ independent parameter in a theory, thus it restricts the choice of ${\cal F}$.

For example, if we start with a massless scalar field $m=0$ in a power law cosmology $a(\eta) = \eta^{-q}$ and let ${\cal F}=\eta^{-k}$  (upto constant rescalings of $H$ respectively) for a consistent solution of \eq{MassRelation}, we get
\bea
a^2{\cal F}^2 m_{\text{\rm eff}}^2 = 2qk\eta^{-2} + k(k+1) \eta^{-2} \\
m_{\text{\rm eff}}^2 \equiv M^2 = \eta^{-2(q+k) +2}=2qk +k(k+1).
\eea
Clearly, for $\eta-$independent ${\cal M}$ we should have $k=1-q$. Thus, the mass of the field turns out to be
\bea
M^2 = 2-q-q^2 =(2+q)(1-q),
\eea
in units of $H^2$.
\subsubsection{Non-minimal coupling}

The above results generalize to case in which the 
 the action also has a term involving curvature coupling, i.e., for the action of the form:
\bea
{\cal A} = \frac{1}{2} \int d^4 x \sqrt{-g} \left[  -\partial^{i} \phi \partial_{i} \phi - m^2 \phi^2 -\xi R \phi^2\right],
\eea
When we carry out the algebra, we see that the additional, curvature coupling, term transforms to 
\bea
\sqrt{-g} \xi R \phi^2 = 6\xi a \ddot{a} \phi^2 \rightarrow 6 \xi b \ddot{b} \psi^2  - 6 \xi b^4 \left( 
\frac{2b \dot{b} {\cal F} \dot{\cal F}}{b^4 {\cal F}} -\frac{2b^2 (\dot{\cal F})^2 }{b^4 {\cal F}^2} + \frac{b^2 
 \ddot{\cal F}}{b^4 {\cal F}}\right)\psi^2,\nonumber\\
 = \sqrt{-\tilde{g}} \xi\tilde{R} \psi^2 - 6\xi b^4 \left( 
\frac{2b \dot{b} {\cal F} \dot{\cal F}}{b^4 {\cal F}} -\frac{2b^2 (\dot{\cal F})^2 }{b^4 {\cal F}^2} + \frac{b^2 
 \ddot{\cal F}}{b^4 {\cal F}}\right)\psi^2,
\eea
where $\tilde{g}, \tilde{R}$ are calculated in the Friedmann universe with scale factor $b$.
Therefore, the full action transforms to 
\bea
{\cal A} = \frac{1}{2} \int d^4 x \sqrt{-\tilde{g}} \left[  -\partial^{i} \psi \partial_{i} \psi - m_{\text{eff}}^2 \psi^2 -\xi \tilde{R} \psi^2\right],
\eea
with 
\bea
m_{\text{eff}}^2= \frac{m^2}{{\cal F}^2} +\frac{1-6\xi}{a^4 {\cal F}^4}\left[2 a \dot{a} {\cal F}\dot{\cal F} + a^2 {\cal F}\ddot{\cal F} \right].
\eea
We can proceed exactly as before and map (i) the massless  theory with curvature coupling in a power-law Friedmann universe to (ii) a massive theory with the same curvature coupling parameter in the de Sitter spacetime. in this case, (with $m=0$) we will find that:
\bea
m_{\text{eff}}^2 \equiv M^2 = (1-6\xi)(2-q-q^2) = (1-6\xi) (2+q)(1-q),
\eea
again, in units of $H^2$. We hope to explore the non-minimal coupling in detail in a future publication.
\subsubsection{General conformal transformation in 4D}
This structure of effective mass can easily be generalized to any two spacetimes related with conformal transformation
\bea
g_{ab} \rightarrow \tilde{g}_{ab}(x)  &=& \Omega^2(x) g_{ab}(x)\\
\phi(x) \rightarrow \psi(x) &=& \Omega^{-1}(x) \phi(x),
\eea
where in the equation of motion
\bea
-\frac{1}{\sqrt{-g}}\partial_{a}\left[\sqrt{-g} g^{ab}\partial_{b}\phi \right] + m^2 \phi =0, \hspace{0.2 in} \text{becomes}\\
-\frac{1}{\sqrt{-\tilde{g}}}\partial_{a}\left[\sqrt{-\tilde{g}} \tilde{g}^{ab}\partial_{b}\psi \right] + \left[m^2 \Omega^{-2} + 2\frac{(\partial \Omega)^2}{\Omega^2} - \Omega^{-1}\square \Omega \right]\psi = 0.
\eea
For a choice $\Omega = e^{-\Lambda}$ we identify the new mass parameter as
\bea
M^2 =\Box \Lambda + (\partial \Lambda)^2 + m^2 e^{2\Lambda}.
\eea

%\newpage
\subsection{Derivation of the $H\to0$ limit of mode functions}\label{app:derc}
\subsubsection{de Sitter mode functions}
We consider the de Sitter metric $ \mathrm{d} s^2 = \mathrm{d} t^2 - e^{2Ht}\mathrm{d}\mathbf{x}^2 $, and define the conformal time coordinate $\eta = -e^{-Ht}/H$.
The (early time) positive frequency plane-wave modes for a scalar field of mass $m$ in de Sitter spacetime are given by Hankel functions $\mathrm{H}_\nu^{(1)}(z)$ in the conformal time $\eta$:
\begin{equation}
u_{\mathbf{k}}(\eta, \mathbf{x}) =  f_k(\eta)\frac{e^{i\mathbf{k}\cdot\mathbf{x}}}{(2\pi)^{\frac{3}{2}}},
\end{equation}
where $k = \sqrt{\mathbf{k}\cdot\mathbf{k}}$ and, with $\nu = \sqrt{3/4-m^2/H^2}$ 
\begin{equation}
f_k(\eta) = \frac{\sqrt{\pi}H}{2} e^{\frac{i\nu\pi}{2}} (-\eta)^{\frac{3}{2}} \mathrm{H}_\nu^{(1)}(-k\eta).
\end{equation}

Now, we will analyze the behaviour of the above positive frequency mode function as $H\rightarrow 0$, which corresponds to the limit of Minkowski spacetime in the metric.

The parameter $\nu$ approaches infinity along the positive imaginary axis as $H\rightarrow 0$. In this limit, we have
$ \nu \approx im/H$. For notational convenience, we will introduce $\mu = m/H > 0$ for a massive field, so that $\nu \approx i\mu$.
The conformal time is also affected by this limit; we must consider both an  $\mathcal{O}(H^{-1})$ (the dominant term) and $\mathcal{O}(H^0)$ part to arrive at any conclusions about the dependence of the modes on the cosmological time $t$:
\begin{equation}
\eta \approx -\frac{1}{H} + t = -\frac{1}{H}(1-Ht). 
\end{equation}
The limiting form of $f_k(\eta)$, now expressed as a function of $t$, is therefore:
\begin{equation}
\label{fasymp}
f_k(t) \approx \frac{\sqrt{\pi} H}{2} e^{-\frac{\mu\pi}{2}}\left(\frac{1-Ht}{H}\right)^{\frac{3}{2}} \mathrm{H}_{i\mu}^{(1)}\left(\frac{k}{H}(1-Ht)\right).
\end{equation}

Next, we define $z = \frac{k}{m}(1-Ht)$, a positive real number that remains finite as $H\rightarrow 0$. The Hankel function can now be expressed as:
\begin{equation}
\mathrm{H}_{i\mu}^{(1)}\left(\frac{k}{H}(1-Ht)\right) = \mathrm{H}_{i\mu}^{(1)}(\mu z).
\end{equation}

We are therefore essentially interested in the asymptotic form of $\mathrm{H}_{i\mu}^{(1)}(\mu z)$ for fixed $z$ (the ${\cal O}(H)$ correction to $z$ is small, allowing such a treatment) as $\mu \rightarrow \infty$.

From \cite{dunster}, we use the leading term in inverse powers of $\mu$ of the asymptotic series for a Hankel function of this form:
\begin{equation}
\mathrm{H}_{i\mu}^{(1)}(\mu z) \sim \left(\frac{2}{\pi\mu}\right)^{\frac{1}{2}} e^{\frac{\pi\mu}{2}} e^{-\frac{i\pi}{4}} (1+z^2)^{-\frac{1}{4}} e^{i\mu\xi(z)} \ ;\ \ \mu \rightarrow \infty,
\end{equation}
where \begin{equation}
\xi(z) = (1+z^2)^{\frac{1}{2}} + \ln\left(\frac{z}{1+(1+z^2)^{\frac{1}{2}}}\right).
\end{equation}
Plugging this into Eq.(\ref{fasymp}),
\begin{align}
f_k(t) &\sim \frac{e^{-\frac{\pi\mu}{2}}}{2} \sqrt{\frac{\pi}{H}} (1-Ht)^{\frac{3}{2}} \mathrm{H}_{i\mu}^{(1)}(\mu z),\\
&\sim \frac{e^{-i\frac{\pi}{4}}}{\sqrt{2m}} (1+z^2)^{-\frac{1}{4}} e^{i\mu\xi(z)}.
\end{align}
In arriving at the second line, we've canceled out factors of $H$ occurring outside $\mu$, which results in a leading order term in the factor multiplying the exponential in $\xi(z)$ that is $O(H^0)$, and therefore finite and non-vanishing in the $H\rightarrow 0$ limit.

Now,  in Minkowski spacetime, the frequency of a plane wave mode with wave-vector $\mathbf{k}$ for a massive scalar field is given by:
$\omega_k = \sqrt{k^2 + m^2}$.
We will find it useful to express $(1+z^2)^{\frac{1}{2}}$ in terms of $\omega_k$, up to ${\cal O}(H)$ which gives us
\begin{equation}
(1+z^2)^{\frac{1}{2}} \approx \frac{\omega_k}{m} - \frac{k^2}{m\omega_k}Ht.
\end{equation}

This substitution gives the familiar normalization factor in $f_k(t)$ for Minkowski space:
\begin{equation}
f_k(t) \sim \frac{e^{-i\frac{\pi}{4}}}{\sqrt{2\omega_k}} e^{i\mu\xi(z)}.
\end{equation}

Now, we must consider the behaviour of $\xi(z)$. Writing $\mu = m/H$, we see that the non-vanishing contributions to the phase are from the  $\mathcal{O}(H^0)$ and $\mathcal{O}(H^1)$ terms in $\xi(z)$. By using $\ln(1+x) \approx x$ for small $\lvert x\rvert$, we find the ${\cal O}(H)$ term:
\begin{align}
{\cal O}(H) \text{ term in } \xi(z) : \left(-\frac{k^2}{m\omega_k} - 1 + \frac{k^2}{\omega_k(m+\omega_k)}\right)Ht = -\frac{\omega_k}{m} Ht.
\end{align}
The $\mathcal{O}(H^0)$ term may be found by setting $z = k/m $ in $\xi(z)$, giving
\begin{equation}
\xi\left(\frac{k}{m}\right) = \frac{\omega_k}{m} + \ln\left(\frac{k}{\omega_k + m}\right).
\end{equation}
The $H\rightarrow 0$ limit of $f_k(t)$ therefore turns out to be the time-dependent part of the normalized  plane wave positive frequency mode functions in Minkowski spacetime:
\begin{equation}
f_k(t) \sim \left[e^{-\frac{i\pi}{4}+\frac{i}{H}\left(\omega_k + m\ln\left(\frac{k}{\omega_k + m}\right)\right)}\right] \frac{e^{-i\omega_k t}}{\sqrt{2\omega_k}}  \ ;\ \ H \rightarrow 0,
\end{equation}
the expression in square brackets being an irrelevant (though divergent) phase factor, which can be accounted for in the normalizing constant for the original de Sitter mode functions $u_{\mathbf{k}}(\eta, \mathbf{x})$.

\subsubsection{Power law mode functions}
The plane wave modes in the power law case $a(t) = (1+(Ht/p))^{p}$ with conformal time 
\begin{align}
\eta &= -\frac{1}{H}\frac{p}{p-1}a^{-\frac{p-1}{p}}(t) = -\frac{1}{H}\frac{p}{p-1}\left(1+\frac{Ht}{p}\right)^{1-p},
\end{align}
are given by:
\begin{equation}
u_{\mathbf{k}}(\eta, \mathbf{x}) = \frac{\sqrt{\pi}}{2}\left(\frac{(p-1)H}{p}\right)^{\frac{p}{p-1}}e^{\frac{i\nu\pi}{2}}(-\eta)^{\nu}\mathrm{H}_\nu^{(1)}(-k\eta) \frac{e^{i\mathbf{k}\cdot\mathbf{x}}}{(2\pi)^{\frac{3}{2}}},
\end{equation}
with $\nu = (p/(p-1))+(1/2)$.

There are two 'independent' ways of taking the limit of Minkowski spacetime i.e. $a_{p}(t) \rightarrow 1$, by varying $p$ and $H$: $p \rightarrow 0$ and $H \rightarrow 0$.

First, we will consider the $p \to 0$ limit. In this limit, $p/(p-1) \to 0$, and $\nu \to (1/2)$.

The factor $\left(p/(p-1)\right)^{p/(p-1)}$ has a nontrivial limiting form. Writing $p = 0 + \epsilon$ for small, positive $\epsilon$, this factor is approximately \begin{align}
\left(\frac{p-1}{p}\right)^{\frac{p}{p-1}} &= \left(1-\frac{1}{\epsilon}\right)^{\frac{\epsilon}{\epsilon-1}} \\
&\approx \epsilon^\epsilon \\
&= e^{\epsilon\log\epsilon} \to 1.
\end{align}

We also have $H^{\frac{p}{p-1}} \to 1$. The conformal time $\eta$ reduces to $t$ in this limit as expected (without any additive constants, unlike the $H\to 0$ case).

For the Hankel function, as the limiting value of $\nu$ is  $1/2$, we use the well known expression in terms of elementary functions for Hankel functions of order half (10.6.2 of \cite{DLMF}):
\begin{equation}
\mathrm{H}_{\frac{1}{2}}^{(1)}(z) = -i\sqrt{\frac{2}{\pi z}}e^{iz}.
\end{equation}

We therefore have:
\begin{equation}
\mathrm{H}_{\frac{1}{2}}^{(1)}(-k\eta) = -i\sqrt{\frac{2}{\pi}}\frac{e^{-ik\eta}}{\sqrt{-k\eta}}.
\end{equation}

With $\eta \to t$, and canceling out factors of $\sqrt{-\eta}$ and multiplying other constant factors, we get the positive frequency Minkowski plane wave modes:
\begin{equation}
u_{\mathbf{k}}(t,\mathbf{x}) \to \frac{1}{(2\pi)^{\frac{3}{2}} \sqrt{2k}} e^{i\mathbf{k}\cdot\mathbf{x}-ikt}.
\end{equation}

Now, we will also look at the $H\to 0$ limit. In this limit, the conformal time goes as
\begin{equation}
\eta \rightarrow t - \frac{1}{H}\frac{p}{p-1},
\end{equation}
with the second (constant) term being dominant (and divergent) in this limit.

The order $\nu$ of the Hankel function remains unaffected, but its argument now approaches infinity. We must therefore use the asymptotic form (10.2.5 of \cite{DLMF})
\begin{equation}
\mathrm{H}_{\nu}^{(1)}(z) \sim \sqrt{\frac{2}{\pi z}} e^{\frac{i\pi\nu}{2}}e^{-\frac{i\pi}{4}}e^{iz}.
\end{equation}
This gives (retaining only relevant terms in the asymptotic form - which will turn out to be the leading constant term in the prefactor, and when substituted in the mode function gives a finite $H$-independent factor; expanding $\eta$ up to $\mathcal{O}(H^0)$ in the exponents then gives a nonvanishing $t$-dependence in the limit)
\begin{equation}
\mathrm{H}_{\nu}^{(1)}(-k\eta) \to \sqrt{\frac{2}{\pi}}\left(\frac{H(p-1)}{kp}\right)^{\frac{1}{2}} e^{\frac{i\pi\nu}{2}}e^{-\frac{i\pi}{4}} e^{i\frac{k}{H}\frac{p-1}{p}} e^{-ikt}.
\end{equation}

The only non-vanishing contribution from the $(-\eta)^\nu$ factor is due to the leading constant term. Putting all this together, we get:
\begin{equation}
u_{\mathbf{k}}(t,\mathbf{x}) \rightarrow \frac{e^{\frac{i\pi\nu}{2}}e^{-\frac{i\pi}{4}} e^{i\frac{k}{H}\frac{p-1}{p}}}{(2\pi)^{\frac{3}{2}} \sqrt{2k}} e^{i\mathbf{k}\cdot\mathbf{x}-ikt},
\end{equation}
which, up to a constant phase (including a divergent part) that can be accounted for by redefining the normalizing factor for the modes, are the positive frequency Minkowski plane wave modes.

\subsection{Derivation of \eq{Norm2}}\label{app:derd}
We have already obtained the regular mode function to be of the form
\bea
v_{\omega l m}(r,\theta , \phi) &=& N^{(1)}_{\omega l m} e^{- i \omega t} Y_{lm}(\theta,  \phi)r^{l} (1-H^2 r^2)^{\frac{-i \omega}{2H}}  \nonumber\\
&\times&\left[  {}_2F_1\left(\frac{3}{4} +\frac{l}{2}  -\frac{i \omega}{2H} - \frac{\nu}{2}, \frac{3}{4} +\frac{l}{2}  -\frac{i \omega}{2H} + \frac{\nu}{2}, \frac{3}{2}+l ; H^2 r^2   \right) \right].
\eea
In the zero-angular momentum case $l=0$ the mode function (s-wave) for a massless field becomes
\bea
v_{\omega}(r,\theta , \phi) &=& N_{\omega} e^{- i \omega \tau}\Phi_{\omega}(r) Y_{00}(\theta,  \phi) \nonumber\\
&=& N_{\omega} e^{- i \omega t} Y_{00}(\theta,  \phi) \left[ (1-H^2 r^2)^{\frac{-i \omega}{2H}} {}_2F_1\left(-\frac{i \omega}{2H}, \frac{3}{2} -\frac{i \omega}{2H}, \frac{3}{2} ; H^2 r^2   \right) \right].
\eea

In order to fix the normalization constant $N_{\omega}$, we normalize the mode function at the (achronal)  hypersurface : future null horizon $Hr = 1$ where the mode function assumes a form
\bea
v_{\omega}(r,\theta , \phi)|_{Hr =1} = N_{\omega}\left[A_{\omega} e^{-i \omega u} + B_{\omega} e^{-i \omega v}  \right], \label{NearHorizon}
\eea
where $u = \tau -r_*$ and $v=\tau+r_*$ and 
\bea
r_* = \frac{1}{2H}\log{\left[\frac{1+Hr}{1-Hr}\right]}.
\eea
The constants $A_{\omega}$ and $B_{\omega}$ are related as $B_{\omega} = A^{*}_{\omega}$ and
\bea
A_{\omega} = \frac{\Gamma\left[\frac{3}{2}\right] \Gamma\left[\frac{i \omega}{H}\right]}{\Gamma\left[\frac{i \omega}{2H}\right]\Gamma\left[\frac{3 + \frac{i \omega}{H}}{2}\right]}2^{-\frac{i \omega}{H}}. \label{Coeff1inF}
\eea
The inner product of the mode function with itself gives
\bea
(v_{\omega},v_{\omega'}) = -4 \pi i \int   \mathrm{d}u r^2 v_{\omega} \overleftrightarrow{\partial}_u v_{\omega'}^{*}\big|_{Hr=1}
\eea
For such mode functions as in \eq{NearHorizon}, the self normalization yields
\bea
-i\int_{-\infty}^{\infty}  \mathrm{d}u (v_{\omega} \overleftrightarrow{\partial}_u v_{\omega'}^{*} ) = 4 \pi\omega H^{-2} |N_{ \omega} A_{ \omega} |^2 \delta(\omega -\omega') +   4 i \pi\omega H^{-2} N_{ \omega} A_{ \omega} N_{ \omega'}^* B_{ \omega'}\delta(\omega)\nonumber\\
 -4 i \pi\omega'H^{-2} N_{ \omega} B_{ \omega} N_{ \omega'}^* A_{ \omega'}\delta(\omega'),
\eea
which in the case $$\lim_{\omega \rightarrow 0} \omega N_{ \omega} A_{ \omega} \rightarrow 0,$$ fixes the normalization to be
\bea
|N_{\omega}|^2 = [4 \pi \omega H^{-2} |A_{\omega}|^2]^{-1}, \label{NormConst1}
\eea
which using \eq{Coeff1inF} gives,
\bea
|N_{\omega}|^2 = \frac{H^2}{\pi \omega}\left(1+\frac{\omega^2}{H^2}\right).
\eea
%\newpage

\subsection{Derivation of results in Sec. \ref{sec:intrepforw}}\label{app:dere}
We consider a massive scalar field $\phi$ obeying the Klein-Gordon equation in de Sitter spacetime  ($a_{\rm dS}(t) = e^{Ht}$; $t\in \mathbb{R}$) and a massless scalar field in a power-law expanding spacetime ($a_{p}(t) = (1+Ht/p)^{p}$; $t,p\in [1,\infty)$) (both in (3+1)-D), where $t$ is the cosmic time coordinate and $a(t)$ is the corresponding function in the general Friedmann metric $\mathrm{d} s^2 = \mathrm{d} t^2 - a^2(t)\mathrm{d} \mathbf{x}^2$, with $\mathbf{x}\in\mathbb{R}^3$. We note that with this choice of a power law metric, $\lim_{p\rightarrow\infty} a_p(t) = a_{\rm dS}(t)$, and we obtain the de Sitter spacetime as a limiting case.

We define the conformal time by $\eta = \int \mathrm{d} t/a(t)$, with the integration constant chosen so that $\eta_{\rm dS} = -e^{-Ht}/H$  for the de Sitter case and $$\eta_{p} = -\frac{1}{H}\frac{p}{p-1}a_p^{-\frac{p-1}{p}}(t)$$ for the power law case. Both have the same range: $\eta \in (-\infty, 0)$ with $\eta\rightarrow -\infty$ corresponding to $t\rightarrow -\infty$ and $\eta\rightarrow 0$ to $t \rightarrow \infty$, for $p > 1$.% We will therefore not distinguish between the two conformal time co-ordinates, as far as notation is concerned.
The positive (unit-)norm modes (in the Klein-Gordon norm sense) corresponding to the Bunch-Davies vacuum $\lvert 0,\rm BD\rangle$ are  given by:
\begin{equation}
u_{\mathbf{k}}^{\rm dS}(\eta, \mathbf{x}) = \frac{\sqrt{\pi} H}{2}e^{\frac{i\nu\pi}{2}}(-\eta)^{\frac{3}{2}}\mathrm{H}_\nu^{(1)}(-k\eta) \frac{e^{i\mathbf{k}\cdot\mathbf{x}}}{(2\pi)^{\frac{3}{2}}},
\end{equation}
with $\nu = \sqrt{9/4-m^2/H^2}$ for the de Sitter case, and
\begin{equation}
u_{\mathbf{k}}^{p}(\eta, \mathbf{x}) = \frac{\sqrt{\pi}}{2}\left(\frac{(p-1)H}{p}\right)^{\frac{p}{p-1}}e^{\frac{i\nu\pi}{2}}(-\eta)^{\nu}\mathrm{H}_\nu^{(1)}(-k\eta) \frac{e^{i\mathbf{k}\cdot\mathbf{x}}}{(2\pi)^{\frac{3}{2}}},
\end{equation}
with $\nu = 1/2+p/p-1$ for the massless power law case, where $\mathrm{H}_\nu^{(1)}(z)$ are Hankel functions of the first kind (with the Hankel functions of the second kind $\mathrm{H}_\nu^{(2)}(z)$ appearing analogously in the negative norm modes, which are given by the complex conjugates of these in these cases (for both real and imaginary $\nu$)).

Introducing the notation $q = \text{min}(1, \text{Re}\ \nu-1/2)$ i.e.
\begin{equation}
q = 
\begin{cases}
1, &\text{ for de Sitter} \\
\frac{p}{p-1}, &\text{ for power law},
\end{cases}
\end{equation}
and $\beta = \left(H/q\right)^{q}$, we see that both types of mode functions can be written using the same expression:

\begin{equation}
u_{\mathbf{k}}(\eta, \mathbf{x}) = \frac{\sqrt{\pi} \beta}{2} e^{\frac{i\pi\nu}{2}}(-\eta)^{q+\frac{1}{2}} \mathrm{H}_\nu^{(1)}(-k\eta).
\end{equation}
The positive Wightman function\footnote{We define the `positive' Wightman function as $G^+(x,y) = G(x,y)$ and the `negative' Wightman function as $G^-(x,y) = G(y,x)$.} in the Bunch-Davies vacuum is given by:
\begin{align}
G^+(\eta, \mathbf{x};\eta', \mathbf{x}') &= \langle 0, \rm BD \rvert \hat{\phi}(\eta, \mathbf{x})\hat{\phi}(\eta', \mathbf{x}')\lvert 0, \rm BD\rangle, \\
&= \int\limits_{\mathbb{R}^3} \mathrm{d}^3 k\ u_{\mathbf{k}}(\eta, \mathbf{x}) u^\ast_{\mathbf{k}}(\eta', \mathbf{x}'), \\
&= \frac{\pi\beta^2}{4 (2\pi)^3} (\eta\eta')^{q+\frac{1}{2}} \int\limits_{\mathbb{R}^3}\mathrm{d}^3 k\ \mathrm{H}_\nu^{(1)}(-k\eta) \mathrm{H}_\nu^{(2)}(-k\eta') e^{i\mathbf{k}\cdot(\mathbf{x}-\mathbf{x}')}.
\end{align}
Evaluating the angular part of the $\mathbf{k}$-integral gives, with $\Delta x = \lvert \mathbf{x}-\mathbf{x}'\rvert$:
\begin{equation}
\label{WightmanG_HankelSine}
G^+(\eta,\eta',\Delta x) = \frac{\beta^2}{8\pi \Delta x}(\eta\eta')^{q+\frac{1}{2}}\int\limits_0^\infty k\mathrm{d} k\ \mathrm{H}_\nu^{(1)}(-k\eta) \mathrm{H}_\nu^{(2)}(-k\eta') \sin(k\Delta x).
\end{equation}

Further, we detail the evaluation of the integral using the following integral representations of the Hankel functions (10.9.10, 10.9.11 of \cite{DLMF}):
\begin{eqnarray}
\mathrm{H}_\nu^{(1)}(z) &=& \frac{e^{-\frac{i\pi\nu}{2}}}{\pi i} \int\limits_{-\infty}^{\infty} e^{i z \cosh t - \nu t} \mathrm{d} t,\ \ \mathrm{Im}\ z > 0,\nonumber\\
\mathrm{H}_\nu^{(2)}(z') &=& -\frac{e^{\frac{i\pi\nu}{2}}}{\pi i} \int\limits_{-\infty}^{\infty} e^{-i z' \cosh t - \nu t} \mathrm{d} t,\ \ \mathrm{Im}\ z' < 0.
\end{eqnarray}
The product is then
\begin{equation}
\mathrm{H}_\nu^{(1)}(z)\mathrm{H}_\nu^{(2)}(z') = \frac{1}{\pi^2} \int\limits_{-\infty}^{\infty}\mathrm{d} t\int\limits_{-\infty}^{\infty}\mathrm{d} u\ e^{iz\cosh t - iz'\cosh u} e^{-\nu(t+u)}.
\end{equation}
Now, we will essentially follow the procedure (used in 13.71 of \cite{watson}) to derive the integral representation for the product of Modified Hankel functions $\mathrm{K}_\nu(iz)\mathrm{K}_\nu(iz')$. 
Introducing real variables $T$, $U$ such that $t = T+U$, $u = T-U$ and transforming the integrals, we get,
\begin{equation}
\mathrm{H}_\nu^{(1)}(z)\mathrm{H}_\nu^{(2)}(z') = \frac{2}{\pi^2} \int\limits_{-\infty}^{\infty}\mathrm{d} T\int\limits_{-\infty}^{\infty}\mathrm{d} U\ e^{iz\cosh(T+U) - iz'\cosh(T-U)} e^{-2\nu T}.
\end{equation}
We can manipulate the z-dependent part in the exponential to get:
\begin{equation}
z\cosh(T+U) - z'\cosh(T-U) = \frac{1}{2}\left(ze^T -z'e^{-T}\right)e^U + \frac{1}{2}\left(\frac{z^2 + z'^2-2zz'\cosh(2T)}{\left(ze^T -z'e^{-T}\right)e^U}\right).
\end{equation}
Introducing  $v = (ze^T-z'e^{-T})e^U$, $\xi = z^2 + z'^2 - 2zz'\cosh(2T)$ and defining $\tau = 2T$, we get the integral
\begin{equation}
\mathrm{H}_\nu^{(1)}(z)\mathrm{H}_\nu^{(2)}(z') = \frac{1}{\pi^2}\int\limits_{-\infty}^{\infty}\mathrm{d}\tau\int\limits_{-\infty}^{\infty}\mathrm{d} U\ e^{-\nu\tau} e^{\frac{i}{2}\left(v + \frac{\xi}{v}\right)}.
\end{equation}
Due to the dependence of the sign and range of $v$ on $\tau$, we cannot quite reduce this to a single integral over $v$ (modified Hankel functions \cite{watson}). We will therefore not (yet) integrate over $\tau$.

We may still attempt to evaluate the Wightman function by substituting this in Eqn.(\ref{WightmanG_HankelSine})
(introducing an arbitrarily small negative imaginary part to $(\eta-\eta'),$ i.e., $(\eta-\eta') \rightarrow (\eta-\eta'-i\epsilon), \epsilon>0$ that we will consider implicit for now) leading to
\begin{equation}
G^+(\eta,\eta',\Delta x) = \frac{\beta^2}{8\pi^3 \Delta x}(\eta\eta')^{q+\frac{1}{2}} \\ \int\limits_{-\infty}^{\infty}\mathrm{d}\tau\int\limits_{-\infty}^{\infty}\mathrm{d} U\int\limits_0^{\infty} k\mathrm{d} k\ e^{-ik^2\left(\cosh\tau - \frac{\eta^2+\eta'^2}{2\eta\eta'}\right)\frac{\eta\eta'}{v}}\sin(k\Delta x)  e^{\frac{iv}{2}} e^{-\nu\tau}.
\end{equation}
With $z = -k\eta$, $z' = -k\eta'$, we may also write $v = (k/H)(-H\eta + H\eta')e^U = H(-\eta + \eta')e^{\overline{U}}$, where $\overline{U} = U + \log(k/H)$. Transforming the $\mathrm{d} U \mathrm{d} k$ double integral to $\overline{U}$ is straightforward (with a unit Jacobian) and gives
\begin{equation}
G^+(\eta,\eta',\Delta x) = \frac{\beta^2}{8\pi^3 \Delta x}(\eta\eta')^{q+\frac{1}{2}} \\ \int\limits_{-\infty}^{\infty}\mathrm{d}\tau\int\limits_{-\infty}^{\infty}\mathrm{d} \overline{U}\int\limits_0^{\infty} k\mathrm{d} k\ e^{-ik^2\left(\cosh\tau - \frac{\eta^2+\eta'^2}{2\eta\eta'}\right)\frac{\eta\eta'}{v}}\sin(k\Delta x)  e^{\frac{iv}{2}} e^{-\nu\tau},
\end{equation}
with the difference being that of the three integration variables, $v$ is now only a function of $\overline{U}$.
Integrating over $k$, identifying  $\left(\eta^2+\eta'^2/2-\eta\eta'\cosh\tau\right)\equiv f$,
\begin{equation}
G^+(\eta,\eta',\Delta x) = \frac{\beta^2}{8\pi^3 \Delta x}(\eta\eta')^{q+\frac{1}{2}} \int\limits_{-\infty}^{\infty}\mathrm{d}\tau\int\limits_{-\infty}^{\infty}\mathrm{d} \overline{U}\ e^{-\nu\tau} e^{\frac{iv}{2}}\frac{\sqrt{\pi} v^{\frac{3}{2}} \Delta x }{4(-if)^{\frac{3}{2}}} e^{-iv\frac{\Delta x^2}{4f}}. 
\end{equation}
Now, let us consider the inner integral (with only the $v$-dependent factors), at a particular value of $\tau = 2T$. We make a transformation of the integration variable to $v$ from $U$ ($\mathrm{d} \overline{U} = \mathrm{d} v / v$), the former being a monotonic function of the latter. As $\overline{U}$ ranges from ($-\infty$ to $\infty$), $v$ ranges from ($0$ to $\infty$) if $(-\eta e^T +\eta'e^{-T}) > 0$, and ($-\infty$ to $0$) if $(-\eta e^T + \eta'e^{-T}) > 0$ (we neglect the $(ze^T - z'e^{-T}) =0$ case as it by itself has a vanishing contribution to the integral over $t$, being just the value of the integrand at one point).
For the former case, we have:
\begin{equation}
\int\limits_0^\infty \mathrm{d} v\ v^\frac{1}{2} e^{\frac{iv}{2}}e^{-i\frac{\Delta x^2}{4f}v} = \frac{4\sqrt{\pi}}{\left(i\frac{\Delta x^2}{f}-2i\right)^{\frac{3}{2}}}.
\end{equation}
For the latter case, with $v$ going from $0$ to $-\infty$, we make the transformation $v\rightarrow w = -v$, and convert it to the above form to obtain the final integral:
\begin{equation}
\int\limits_0^{-\infty} \mathrm{d} v\ v^\frac{1}{2} e^{\frac{iv}{2}}e^{-i\frac{\Delta x^2}{4f}v} = -i\int\limits_0^{\infty} \mathrm{d} w\ w^\frac{1}{2} e^{-\frac{iw}{2}}e^{i\frac{\Delta x^2}{4f}w} = \frac{4\sqrt{\pi}}{\left(i\frac{\Delta x^2}{f}-2i\right)^{\frac{3}{2}}}.
\end{equation}
Irrespective of the value of $\tau$, the inner integral evaluates to the same expression. Substituting this expression leaves us with an integral over $\tau$. Further expanding $f$ and defining $Z = \frac{\eta^2 + \eta'^2-\Delta x^2}{2\eta\eta'}$, we have a simple integral representation of the Wightman function:
\begin{equation}
\label{DefIntRep_Wightman1}
G^+(\eta, \eta'; Z) = \frac{\beta^2}{16\pi^2\sqrt{2}}(\eta\eta')^{q-1}\int\limits_{-\infty}^{\infty} \mathrm{d}\tau \frac{e^{-\nu\tau}}{\left(\cosh\tau - Z\right)^{\frac{3}{2}}}.
\end{equation} 
An integral representation involving polynomials in the integration variable can be obtained by substituting $s = e^\tau$ in the integrand:
\begin{equation}
G^+(\eta, \eta'; Z) = \frac{\beta^2}{16\pi^2}(\eta\eta')^{q-1}\int\limits_{0}^{\infty} \mathrm{d} s \frac{s^{\left(\tfrac{1}{2}-\nu\right)}}{\left(s^2-2Zs+1\right)^{\frac{3}{2}}}.
\end{equation} 
We can separate out the $\tau > 0$ and $\tau < 0$ parts of the integral in \eq{DefIntRep_Wightman1}, and write them together as a single integral over $\tau > 0$:
\begin{equation}
\label{CoshIntRep_Wightman}
G^+(\eta, \eta'; Z) = \frac{\beta^2}{8\pi^2\sqrt{2}}(\eta\eta')^{q-1}\int\limits_{0}^{\infty} \mathrm{d}\tau \frac{\cosh\nu\tau}{\left(\cosh\tau - Z\right)^{\frac{3}{2}}}.
\end{equation}

%\newpage
\subsection{Derivation of results in Sec. \ref{sec:series}}\label{app:derf}
\subsubsection{Derivation of \eq{seriesforG}}
The Wightman function for a massive scalar field in de Sitter background can be written in terms of the hypergeometric function.
\begin{align}\label{explicitGF1}
	G^+(Z)=\frac{H^2}{16\pi^2}\Gamma(c)\Gamma(3-c)~_2F_1\left(c,3-c,2;\frac{1+Z}{2}\right),
\end{align}
where, $c(3-c)=m^2/H^2\equiv 3\epsilon$. In this appendix, we try to find out expansion of $G^+(Z)$ in powers of $m^2$ or equivalently $\epsilon$. 

We will first derive \eq{seriesforG} using the series expansion of the hypergeometric function, for case (i) [$m\rightarrow 0$ limit of the massive field analysis]. Inside the unit disc $|x|<1$, the hypergeometric function has the following series representation. 
\begin{align}\label{defineF}
	\Gamma(A)\Gamma(B)~_2F_1(A,B,C;x)=\frac{1}{\Gamma(C)}\sum_{n=0}^{\infty}\frac{\Gamma(A+n)\Gamma(B+n)}{\Gamma(C+n)}\frac{x^n}{n!},
\end{align}
Outside this domain, the function is defined by an analytic continuation. The relevant values of parameters $A,B$ and $C$ for \eq{defineF} is given by
\begin{align}
	A=c,&&B=3-c,&&C=2,&&\text{ and }&&x=\frac{(1+Z)}{2}.
\end{align}
The solution of the defining quadratic equation for $c$ may be written as $c=3-\epsilon+\mathcal{O}(\epsilon^2)$. Therefore, to linear order in $\epsilon$, we have the following results.
\begin{align}\label{parametrerslinear}
	\frac{1}{\Gamma(C)}\frac{\Gamma(A+n)\Gamma(B+n)}{\Gamma(C+n)}&=\frac{(n+1)}{2}+\mathcal{O}(\epsilon)\,\,\,\,\,;\,\,n>0
    \\\label{parameterlinear2}
	\frac{1}{\Gamma(C)}\frac{\Gamma(A)\Gamma(B)}{\Gamma(C)}&=\frac{2}{\epsilon}+3+\mathcal{O}(\epsilon),
\end{align}
Using \eq{parametrerslinear} and \eq{parameterlinear2} in \eq{defineF}, we obtain the following expression for the hypergeometric series around $Z=-1$, to the linear order in $\epsilon$.
\begin{align}
	\Gamma(c)\Gamma(3-c)~_2F_1\left(c,3-c,2,\frac{1+Z}{2}\right)&=\frac{2}{\epsilon}+3+\sum_{n=1}^{\infty}\left[\frac{(n+2)}{2}x^n\right]+\mathcal{O}(\epsilon)
\end{align}
Inside the disk $|x|<1$, the above series converges to
\begin{align}
\Gamma(c)\Gamma(3-c)~_2F_1\left(c,3-c,2,\frac{1+Z}{2}\right)&=2 \left(\frac{1}{\epsilon }+\frac{1}{1-Z}-\log (1-Z)-2+\log (2)\right)+\mathcal{O}(\epsilon)
\end{align} 
The above expression can be analytically continued to define the $\epsilon$-expansion outside the disk $|x|<1$.
A similar analysis shows that 
\begin{align}
\Gamma(c)\Gamma(3-c)~_2F_1\left(c,3-c,2,\frac{1-Z}{2}\right)&=2 \left(\frac{1}{\epsilon }+\frac{1}{1+Z}-\log (1+Z)-2+\log (2)\right)+\mathcal{O}(\epsilon)
\end{align} 

Using these in \eq{explicitGF}, we find that
\begin{align}\label{seriesforG0}
	G^+(Z)=\frac{3H^4}{8m^2\pi^2}-\frac{H^2}{8\pi^2}\left[\frac{-1}{1-Z}+\log\left\{(1-Z)\lambda\right\}\right]+\mathcal{O}(\epsilon^2).
\end{align}
where, $\lambda=e^2/2$.
For case (ii)[$q\rightarrow 1$ limit of the $a(\eta)\sim \eta^{-q}$ power law cosmology], one can consider a series expansion in $\epsilon_{\rm eff}=m_{\rm eff}^2/(3H^2)$. Here, in the expression for Wightman function, in addition to the combination of Gamma functions and the hypergeometric function, there is a term of the form $(\eta\eta')^{-\epsilon_{\rm eff}}\approx 1-\epsilon_{\rm eff}\log(\eta\eta')+\mathcal{O}(\epsilon_{\rm eff}^2)$. Hence, the Wightman function has the following expansion in powers of $m_{\rm eff}^2$.
\begin{align}\label{seriesforGpowerlaw0}
	G^+(Z,\eta,\eta')=\frac{3H^4}{8m_{\rm eff}^2\pi^2}-\frac{H^2}{8\pi^2}\left\{\frac{-1}{1-Z}+\log\left[(1-Z)H^2(\eta\eta')\lambda\right]\right\}+\mathcal{O}(\epsilon^2).
\end{align}  

Now, we will use the integral representation \eq{intreppoly} to arrive at the same result. We will illustrate the analysis for case (i) [$m\rightarrow 0$ limit of the massive field analysis] first. The Wightman function for case (i) has the following integral representation
\begin{align}
	G^+(Z)&=\frac{H^2}{8\pi^2}\int_{0}^{\infty}\frac{(s^2-2Zs+1)^{-3/2}}{s^{1-\epsilon}} \mathrm{d}s,\\
	&\propto\int_{0}^{\infty}\frac{(s^2+1)^{-3/2}}{s^{1-\epsilon}}\left[1-\frac{2Zs}{s^2+1}\right]^{-3/2} \mathrm{d}s,\\
	&=\int_{0}^{\infty}\frac{(s^2+1)^{-3/2}}{s^{1-\epsilon}}\sum_{n=0}^{\infty}\left[\frac{(-1)^nx^n\Gamma(-1/2)}{\Gamma(n+1)\Gamma(-1/2-n)}\right] \mathrm{d}s,
	\end{align}
	where, $x=(2Zs)/(s^2+1)$. Let us consider each term in the integral.
\begin{align}
	\mathcal{I}_{n}&\equiv\frac{(-1)^n\Gamma(-1/2)}{\Gamma(n+1)\Gamma(-1/2-n)}\int_{0}^{\infty}\frac{(s^2+1)^{-3/2}}{s^{1-\epsilon}}x(s)^n \mathrm{d}s,\\
	&=\frac{(-1)^{n+1} 2^{3 n+1} \Gamma \left(\frac{1}{2} (n-\epsilon +3)\right) \Gamma \left(\frac{n+\epsilon }{2}\right)}{\Gamma \left(-n-\frac{1}{2}\right) \Gamma (2 n+2)}Z^n
\end{align}
Let us consider the $n=0$ term.
\begin{align}
	\mathcal{I}_{0}&=\frac{\Gamma \left(\frac{3-\epsilon }{2}\right) \Gamma \left(\frac{\epsilon }{2}\right)}{\sqrt{\pi }},\\
	&=\frac{1}{\epsilon}+\frac{1}{2}(-\gamma-\psi_{0}(3/2))+\mathcal{O}(\epsilon).
\end{align}

The $n>0$ terms can be written, after simplifying the Gamma function expressions, as
\begin{align}
	\mathcal{I}_{n}=\left(\frac{1}{n}+1\right)Z^n+\mathcal{O}(\epsilon).
\end{align}
Therefore, the whole expression can be written as
\begin{align}
	\sum_{n=0}^{\infty}\mathcal{I}_n=\frac{1}{\epsilon}+\frac{1}{1-Z}-\log\left[(1-Z)\lambda\right]+\mathcal{O}(\epsilon).
\end{align}
where, $\lambda=e^2/2$. After putting the right proportionality constant we see that $G^{+}(Z)$ reduces to the expression given in \eq{seriesforG}. Again, for case (ii) [$q\rightarrow 1$ limit of the $a(\eta)\sim \eta^{-q}$ power law cosmology], we can proceed exactly as in case (i), but the additional factor $(\eta\eta')^{-\epsilon_{\rm eff}}$ gives an extra $\log(\eta\eta')$ term for the Wightman functions, hence reproducing \eq{seriesforGpowerlaw}.

%\newpage
\subsubsection{Derivation of \eq{tp111}}
We begin by putting $\nu=3/2-\epsilon$ in \eq{dgdz1}, where $\epsilon$ is a small positive quantity. The integrand in \eq{dgdz1} can then be expanded as a Taylor series in $\epsilon$ to give

\begin{align}\label{expand_dgdz}
\frac{\diff G(\eta, \eta'; Z)}{\diff Z}
&=\frac{3\beta^2}{8\pi^2}(\eta\eta')^{q-1}\left[\int\limits_{0}^{\infty} \diff s~~\frac{1}{\left(s^2-2 sZ+1\right)^{5/2}}+\epsilon \int\limits_{0}^{\infty} \diff s~~\frac{\log(s)}{\left(s^2-2 sZ+1\right)^{5/2}}\right]+\mathcal{O}(\epsilon^2)
\end{align}

For $|Z|<1$ the integrals in \eq{expand_dgdz} are convergent and can be evaluated explicitly to get the following $\mathcal{O}(\epsilon)$ expression for $\diff G/\diff Z$.
\begin{align}\label{simplified_dgdz}
\frac{\diff G(\eta, \eta'; Z)}{\diff Z}
&=\frac{3\beta^2}{8\pi^2}(\eta\eta')^{q-1}\left[-\frac{Z-2}{3 (Z-1)^2}\right]+\mathcal{O}(\epsilon)
\end{align}   
The $\epsilon$-series of $\diff G/\diff Z$ for $|Z|\geq 1$ can be obtained by analytical continuation of this equation to get \eq{tp111}.

\subsection{A brief comment on  the stress-energy tensor of the massless, minimally-coupled field}\label{app:derf0}
In the de Sitter spacetime, the Ricci scalar is a constant, namely $R = 12H^2$. Thus, in the non-minimally coupled case, the additional term $\xi R\phi^2$ in the Lagrangian density can be treated as a modification of the mass term, corresponding to a scalar field with an effective mass $\overline{m}$ with $\overline{m}^2 = m^2 + \xi R$.
Therefore, the Wightman function for a massive scalar field in de Sitter spacetime, with non-minimal coupling to the curvature, can be written in essentially the same form as \eq{GinF2}
\begin{align}
G(x,x')=\frac{H^2}{16\pi^2}\Gamma \left(\tfrac{3}{2}-\overline{\nu}\right)\Gamma \left(\tfrac{3}{2}+\overline{\nu}\right)~_2F_1\left(\frac{3}{2}-\overline{\nu},\frac{3}{2}+\overline{\nu},2;\frac{(1+Z)}{2}\right),
\label{Wightman_non_minimal}
\end{align}	
where we now have
\begin{align}
\overline{\nu}=\sqrt{\frac{9}{4} - \frac{\overline{m}^2}{H^2}} = \sqrt{\frac{9}{4}-\frac{12(m^2+\xi R)}{R^2}}.
\end{align}	
Let us consider the $\overline{m}\to 0$ limit of this expression. Expanding \eq{Wightman_non_minimal} in powers of $\overline{m}$ gives a straightforward modification of \eq{seriesforG},

\begin{eqnarray}\label{seriesforG_non_minimal}
	G(Z)=\frac{3H^4}{8\overline{m}^2\pi^2}-\frac{H^2}{8\pi^2}\left[\frac{-1}{1-Z}+\log\left\{(1-Z)\lambda\right\}\right]+\mathcal{O}\left(\frac{\overline{m}^2}{H^2}\right).
\end{eqnarray}
%There is no ambiguity in the limit $(m^2,\xi)\rightarrow (0,0)$  of this expression in the sense that the result does not depend on the path taken to approach the origin of the $(m62,\xi)$ plane; but, of course,  
The constant term diverges in the $\overline{m}\to 0$ limit just as  in the case of the  minimally coupled field. The derivative $dG/dZ$ is again independent of $\overline{m}^2$ and has no divergence. It has a well defined massless, minimally-coupled, limit, and any quantity constructed \textit{only out of the first or higher derivatives} of $G(Z)$ also has an unambiguous limit.

There is, however, no well-defined limit for products of the form $m^2 G(Z)$ and $\xi R G(Z)$. when we approach the origin of the $(m^2,\xi)$ plane. These expressions will take different limiting values depending on which path we choose to approach the point $(0,0)$ in  $(m^2,\xi)$ plane. However, the sum $(m^2 + \xi R)G(Z) = \overline{m}^2G(Z)$ has the unambiguous limiting value of $3H^4/(8\pi^2)$.

To see this explicitly, consider a path $P_{y}$ in the $(m^2,\xi)$ plane that has the following limiting form as we approach the origin,
\begin{align}\label{path}
\xi\approx y\frac{m^2}{R}.
\end{align} 
Then, the $(m^2,\xi)\rightarrow (0,0)$ limits of $m^2G(Z)$ and $\xi R G(Z)$ along such a path are
\begin{align}
\lim_{(m^2,\xi)\rightarrow (0,0), P_{y}} m^2 G(Z) &=\frac{3H^4}{8(1+y)\pi^2}=\frac{R^2}{384 (y+1) \pi ^2}, \\
\lim_{(m^2,\xi)\rightarrow (0,0), P_{y}} \xi R G(Z) &=\frac{3H^4y}{8R(1+y)\pi^2}=\frac{y R^2}{384 (y+1) \pi ^2},
\end{align}
which are $y$-dependent and therefore not well-defined.

As the Einstein tensor is $G_{ab} = R_{ab} - (1/2)Rg_{ab} = -(1/4)Rg_{ab}$ in de Sitter spacetime, from \eq{seriesforG}, it is easy to obtain the following limit for
the linear combination $(A \xi G_{ab}+B m^2 g_{ab})$, where $A$ and $B$ are arbitrary constants: 
\begin{eqnarray}\label{nonunique_limit_G}
\lim_{m\rightarrow 0, P_{y}}(A \xi G_{ab}+B m^2 g_{ab})\,\,G(Z)=\left[\frac{(4B-Ay)}{1536\pi^2(y+1)}\right]R^2g_{ab},
\end{eqnarray}
We note that this limiting value is again, in general, path-dependent (except for a special case of  $A = -4B$). As we discuss below, such a linear combination with $A = 1$, $B=-1/2$ arises in the expression for the vacuum expectation value of the stress-energy tensor, thereby leading to an ambiguity in the limit $(m^2,\xi)\to(0,0)$, which has been noted in the literature (see \cite{Kirsten:1993ug}).

While obtaining the vacuum expectation value of the stress-energy tensor from $G^{(1)}(x,x') = G(x,x') + G(x',x) = 2G(Z)$ using the point splitting method (see e.g. \cite{Bernard:1986}), we will have terms in $m^2$ and $\xi$ that cannot be combined to depend only on $\overline{m}^2$, and we may expect a path-dependent limit  for the stress-energy tensor when we approach $(m^2,\xi)\to(0,0)$. In particular, from Eq. (3.2) of \cite{Bernard:1986}, we find that the terms that do \textit{not} involve derivatives of $G(Z)$ are given by
\begin{align}\label{noderiv}
\lim_{x\rightarrow x'}\left(\xi G_{ab}-\frac{m^2}{2}g_{ab}\right)\left[G(Z)-G_{ref}(x,x')\right],
\end{align}
where, $G_{ref}$ is a `reference Green's function' that is used for renormalization in the point-splitting approach. Comparing this with \eq{nonunique_limit_G}, we see that $A = 1$, $B=-1/2$ and therefore the $(m^2,\xi)\to(0,0)$ limit of this term is path-dependent.

The explicit expression for the vacuum expectation value of the renormalized stress-energy tensor $\braket{T_{ab}}_{ren}$ is given by  (as in \cite{Bunch:1978yq}),
\begin{align}
\braket{T_{ab}}_{ren}&=-\frac{g_ab}{64\pi^2}\{m^2[m^2+(\xi-1/6)R]\left[\psi(3/2-\overline{\nu})+\psi(3/2+\overline{\nu})+\log(R/2m^2)\right]
\\\nonumber&-m^2(\xi-1/6)R-m^2R/18-(\xi-1/6)^2R^2/2+R^2/2160\},
\end{align}
where the expectation value is taken in the Euclidean vacuum. The path dependence of the $(m^2,\xi)\rightarrow (0,0)$ limit of this expression stems from the following term
\begin{align}
\frac{g_{ab}}{64\pi^2}m^2(m^2+(\xi-1/6)R)\psi(3/2-\overline{\nu})=g_{ab}\frac{R^2}{1536(1 + y)\pi^2}+\mathcal{O}(m^2/H^2).
\end{align}
In \cite{Kirsten:1993ug} the authors note that the ambiguity can be traced back to the contribution of the $L=0$ mode to $\braket{T_{ab}}$. This contribution is given by
\begin{align}
\frac{-R^2}{1536}\frac{(y+2)}{(y+1)}+\mathcal{O}(m^2,\xi).
\label{contri}
\end{align}
Setting $A=1$ and $B=-1/2$ in \eq{nonunique_limit_G}, we see that the leading order contribution to \eq{noderiv} matches \textit{exactly} with that in \eq{contri}. Therefore, we can conclude that the path dependence of the  $(m^2,\xi)\rightarrow (0,0)$ limit of $\braket{T_{ab}}_{ren}$ can be attributed to those terms in $T_{ab}$ which depend explicitly on  $G$, and not from those terms involving the derivatives of $G$.  

\subsection{Derivation of results in Sec. \ref{sec:geoftofW}}\label{app:deri}
We begin with an integral representation of the positive Wightman function \eq{Gplusdef}, reproduced here:
\begin{equation}
G^+(\eta, \eta'; Z_{\epsilon}) = \frac{\beta^2}{16\pi^2\sqrt{2}}(\eta\eta')^{q-1}\int\limits_{-\infty}^{\infty} \mathrm{d}\tau \frac{e^{-\nu\tau}}{\left(\cosh\tau - Z_{\epsilon}\right)^{\frac{3}{2}}}.
\end{equation}

Formally, we may treat $Z$ and $\eta$,$\eta'$ as independent variables. We are now interested in the Fourier transform of $G^+$ with respect to $Z$, which we will define by:
\begin{equation}
\widetilde{G}^+(\eta, \eta'; Q) = \int\limits_{-\infty - i\epsilon(\eta,\eta')}^{\infty - i\epsilon(\eta,\eta')}\mathrm{d} Z_{\epsilon}\ e^{-iQZ_{\epsilon}}G^+(\eta, \eta'; Z_{\epsilon}),
\end{equation}

To evaluate this for the positive Wightman function, it is useful to recall that 
$$Z_{\epsilon} = 1+\frac{(\eta-\eta'-i\epsilon)^2 - \Delta x^2}{2\eta\eta'},$$
where $\epsilon > 0$ is an arbitrarily small positive quantity. %For brevity, we have used $s_\eta = \mathrm{sgn}(\Delta\eta)$.

The variable $Z_{\epsilon},$ then has an 
%%%%%%%%%%%%%%%%%%
%%%%%%%%%%%%%%
arbitrarily small imaginary part, given by 
$\epsilon(\eta,\eta') = -2\epsilon(\eta - \eta')/\eta \eta' $. Thus, we obtain,
\begin{equation}
\widetilde{G}^+(\eta, \eta'; Q) =  e^{Q\epsilon(\eta,\eta')}\int\limits_{-\infty}^{\infty}\mathrm{d} Z\ e^{-iQZ}G^+(\eta, \eta'; Z- i\epsilon(\eta,\eta')).
\end{equation}

%%%%%%%%%%%%%%%%%%
%%%%%%%%%%%%%%
Then, we use the following Fourier transform relations in (Exponential transforms, Elementary functions, entries 3 and 4) \cite{Erdelyi}:
\begin{eqnarray}
\int\limits_{-\infty}^{\infty}(a-ix)^{-\lambda}e^{-ixy}\mathrm{d} x &= \frac{2\pi}{\Gamma(\lambda)} y^{\lambda - 1}e^{-ay}\theta(y),\ \ &\mathrm{Re}\ a > 0, \mathrm{Re}\ \lambda > 0, \\
\int\limits_{-\infty}^{\infty}(a+ix)^{-\lambda}e^{-ixy}\mathrm{d} x &= -\frac{2\pi}{\Gamma(\lambda)} (-y)^{\lambda - 1}e^{ay}\theta(-y),\ \ &\mathrm{Re}\ a > 0, \mathrm{Re}\ \lambda > 0;
\end{eqnarray}
where $\theta(x)$ is the Heaviside unit-step function. Defining $b = ia$ and multiplying by a factor of $(-i)^\lambda$ in the former, and $b = -ia$ and multiplying by $i^\lambda$ in the latter, and changing $y \rightarrow -y$ gives more immediately applicable forms, both with $\mathrm{Re}\ \lambda > 0$:
\begin{eqnarray}
\int\limits_{-\infty}^{\infty}(b+x)^{-\lambda}e^{ixy}\mathrm{d} x = 
\begin{dcases}
\frac{2\pi}{\Gamma(\lambda)}(-i)^\lambda (-y)^{\lambda - 1}e^{ay}\theta(-y),\ \ &\mathrm{Im}\ b > 0,\\
\frac{2\pi}{\Gamma(\lambda)}i^\lambda y^{\lambda - 1}e^{-ay}\theta(y),\ \ &\mathrm{Im}\ b < 0.
\end{dcases}
\end{eqnarray}
Now, with $\lambda = \frac{3}{2}$, $b = (\cosh t + i\epsilon s_\eta)$, $x = -Z$ and $y = Q$, we get the Fourier transform of the positive Wightman function:
\begin{equation}
\widetilde{G}^+(\eta,\eta';Q) = -\frac{\beta^2}{4\pi\sqrt{2\pi}}(\eta\eta')^{q-1}e^{\frac{i\pi}{4}}(-Q)^{\frac{1}{2}}s_\eta\ \theta\left(-Qs_\eta \right)\int\limits_{-\infty}^{\infty}\mathrm{d} t\ e^{-iQ\cosh t - \nu t}.
\end{equation}
This is antisymmetric under an exchange of the two points, as $s_\eta$ is antisymmetric and $Z$ (and therefore $Q$) is symmetric under this exchange.
Recognizing the integral as a representation for the Hankel function $\mathrm{H}_\nu^{(2)}(Q)$ (10.9.11 of \cite{DLMF}), we have
\begin{equation}
\widetilde{G}^+(\eta,\eta';Q) = -\frac{\beta^2}{4\sqrt{2\pi}}(\eta\eta')^{q-1}e^{\frac{i\pi}{2}(\frac{1}{2}-\nu)}Q^{\frac{1}{2}}\mathrm{H}_\nu^{(2)}(Q)s_\eta\ \theta\left(-Qs_\eta \right).
\end{equation}
For the negative Wightman function $G^-(x,x') = G^+(x',x)$, we have $s_\eta \rightarrow -s_\eta$, and therefore,
\begin{equation}
\widetilde{G}^-(\eta,\eta';Q) = \frac{\beta^2}{4\sqrt{2\pi}}(\eta\eta')^{q-1}e^{\frac{i\pi}{2}(\frac{1}{2}-\nu)}Q^{\frac{1}{2}}\mathrm{H}_\nu^{(2)}(Q)s_\eta\ \theta\left(Qs_\eta \right).
\end{equation}
We can also write the Fourier transform of the commutator Green's function, $G^{\rm c}(x,x') = G^+(x,x') - G^-(x,x')$:
\begin{equation}
\widetilde{G}^{\rm c}(\eta,\eta';Q) = -2s_\eta\frac{\beta^2}{4\sqrt{2\pi}}(\eta\eta')^{q-1}e^{\frac{i\pi}{2}(\frac{1}{2}-\nu)}Q^{\frac{1}{2}}\mathrm{H}_\nu^{(2)}(Q).
\end{equation}
Unlike in $Z$-space, these Fourier transforms have closed form expressions in terms of well-understood functions, instead of being divergent, even for the power law case ($\nu > \frac{3}{2}$) and the massless de Sitter case ($\nu = \frac{3}{2}$).
Restricting our interest to the case of de Sitter spacetime, $q = 1$, the Wightman and commutator Green's functions' Fourier transforms are:
\begin{eqnarray}
\widetilde{G}^{\pm}_{\mathrm{dS}}(Q) &=& \mp\frac{H^2}{4\sqrt{2\pi}}e^{\frac{i\pi}{2}(\frac{1}{2}-\nu)}Q^{\frac{1}{2}}\mathrm{H}_\nu^{(2)}(Q)s_\eta \theta\left(\mp Qs_\eta \right), \\
\widetilde{G}^{\rm c}_{\mathrm{dS}}(Q) &=& -2s_\eta\frac{H^2}{4\sqrt{2\pi}}e^{\frac{i\pi}{2}(\frac{1}{2}-\nu)}Q^{\frac{1}{2}}\mathrm{H}_\nu^{(2)}(Q),
\end{eqnarray}
For the Feynman Green's function, we have $\widetilde{G}_F(Q) = -i\widetilde{G}^+(Q)\theta(s_\eta) -i \widetilde{G}^-(Q)\theta(-s_\eta)$, which gives:
\begin{eqnarray}
\widetilde{G}_{F,\mathrm{dS}}(Q) = i\frac{H^2}{4\sqrt{2\pi}}e^{\frac{i\pi}{2}(\frac{1}{2}-\nu)}Q^{\frac{1}{2}}\mathrm{H}_\nu^{(2)}(Q)\theta(-Q).
\end{eqnarray}
We can see that the Feynman Green's function  is completely $\eta$-independent, and depends only on $Q$.
The de Sitter invariant two-point functions must satisfy the differential equation:
\begin{equation}
(Z^2-1)\frac{\mathrm{d}^2 G_{\mathrm{dS}}}{\mathrm{d} Z^2} + 4Z\frac{\mathrm{d} G_{\mathrm{dS}}}{\mathrm{d} Z} + \frac{m^2}{H^2}G_{\mathrm{dS}} = 0.
\end{equation}
The formal Fourier transform of this equation with respect to $Z$, with Fourier variable $Q$ as above, reads:
\begin{equation}
\label{Gtilde_eqn}
Q^2 \frac{\mathrm{d}^2 \widetilde{G}_{\mathrm{dS}}}{\mathrm{d} Q^2} + \left(Q^2 + \frac{m^2}{H^2} - 2\right)\widetilde{G}_{\mathrm{dS}} = 0.
\end{equation}
We see that the Fourier transform of the commutator Green's function satisfies this equation, but those of the Wightman functions do not.\footnote{It is anyway not reasonable  for the Wightman function to assume dependence on $Q$ alone when the Fourier transforms also depend on $\eta$, $\eta'$ (via $s_\eta$). The $G$s are still de Sitter invariant, but there is a dependence on 'causality' beyond the variable $Z$ alone}

The quantity $Z$ is not a well-behaved measure of distance in the limit $H\to 0$. To consider this limit of the Fourier transforms, we work instead with the Fourier transform with respect to $L^2 = \frac{(\eta-\eta')^2-\Delta x^2}{H^2\eta\eta'}$, which we will denote by
\begin{equation}
\overline{G}(\eta,\eta';K) = \int\limits_{-\infty}^{\infty} \mathrm{d} (L^2)\ G(\eta,\eta';L^2) e^{-i K L^2}. 
\end{equation}

This is related in a simple manner to $\widetilde{G}(\eta,\eta';Q)$:
\begin{equation}
\widetilde{G}(\eta,\eta'; Q) = \frac{H^2e^{-iQ}}{2}\ \overline{G}\left(\eta,\eta';\frac{H^2Q}{2}\right).
\end{equation}

Substituting this in \eq{Gtilde_eqn} gives, with $Q = \tfrac{2K}{H^2}$
\begin{equation}
\frac{H^2}{2}K^2 \frac{\mathrm{d}^2}{\mathrm{d} K^2}\left(e^{-\frac{2iK}{H^2}}\overline{G}\right) + \frac{H^2}{2}\left(\frac{m^2}{H^2} + \frac{4K^2}{H^4}-2\right)e^{-\frac{2iK}{H^2}}\overline{G} = 0.
\end{equation}

On simplifying, we get
\begin{equation}
H^2K^2\frac{\mathrm{d}^2 \overline{G}}{\mathrm{d} K^2} - 4iK^2\frac{\mathrm{d} \overline{G}}{\mathrm{d} K} + (m^2-2H^2)\overline{G} = 0.
\end{equation}
which is essentially the Fourier transform of the equation satisfied by the de Sitter two point functions
\begin{equation}
(4L^2 + H^2L^4)\frac{\mathrm{d}^2 G}{\mathrm{d} (L^2)^2} + (8+4H^2L^2)\frac{\mathrm{d} G}{\mathrm{d} (L^2)} + m^2 G = 0.
\end{equation}

The Minkowski limit is now trivial - $H\to 0$ gives, for the latter
\begin{equation}
\label{G_L2FT_H0}
4L^2\frac{\mathrm{d}^2 G}{\mathrm{d} (L^2)^2} + 8\frac{\mathrm{d} G}{\mathrm{d} (L^2)} + m^2 G = 0,
\end{equation}
with the general solution (we do not assume $m^2 > 0$ to allow for the de Sitter case which is equivalent to a scalar field in a massless power law background)
\begin{equation}
G(L^2) = g_+ \frac{\mathrm{K}_1(i\sqrt{m^2 L^2})}{\sqrt{m^2 L^2}} + g_- \frac{\mathrm{K}_1(-i\sqrt{m^2 L^2})}{\sqrt{m^2L^2}},
\end{equation}
and for the former,
\begin{equation}
\label{FTL2_H0}
- 4iK^2\frac{\mathrm{d} \overline{G}}{\mathrm{d} K} + m^2\overline{G} = 0,
\end{equation}
which has the general solution
\begin{equation}
\overline{G} = g_0 e^{\frac{im^2}{4K}},
\end{equation}
where $g_+, g_-, g_0$ are integration constants.

Whereas \eq{G_L2FT_H0} is a second order differential equation in derivatives with respect to $L^2$, its Fourier transform \eq{FTL2_H0} is first order in derivatives with respect to $K$, admitting only one solution rather than two linearly independent solutions as in the former case. This is because the Fourier transform only exists for one of the two independent solutions written above (therefore invalidating dropping boundary terms when transforming the equation for the other solution).

To see this, we consider the asymptotic behaviour of the modified Bessel function $\mathrm{K}_\nu(z)$ from 10.25.3  of \cite{DLMF}
\begin{equation}
\mathrm{K}_\nu(z) \sim \sqrt{\frac{\pi}{2z}} e^{-z},\ \lvert z\rvert \rightarrow \infty, \lvert\mathrm{arg}(z)\rvert < \tfrac{3\pi}{2}.
\end{equation}
This shows that $\mathrm{K}_1(iz)$ diverges as $\mathrm{Im}\ z \to \infty$. We take $\sqrt{z}$ to denote the square root of $z$ in the upper half plane ($\mathrm{arg}(\sqrt{z})\in [0,\pi)$). This means that while the solution $\mathrm{K}_1(-i\sqrt{m^2L^2})$ remains convergent (and in fact, approaches zero) as $m^2L^2 \to -\infty$ i.e. for large (spacelike or timelike, depending on the sign of $m^2$) separations, $\mathrm{K}_1(i\sqrt{m^2L^2})$ diverges, and therefore its Fourier transform does not exist (moreover, a two point correlation function that diverges for large separations is unphysical).

To see the effect of the boundary terms, we explicitly transform the equation \eq{G_L2FT_H0} to \eq{FTL2_H0} assuming the existence of the Fourier transform, by multiplying it by $e^{-iKL^2}$ and integrating over all real values of $L^2$:
\begin{equation}
\int\limits_{-\infty}^{\infty}\mathrm{d}(L^2)\ \left(4L^2\frac{\mathrm{d}^2 G}{\mathrm{d} (L^2)^2} + 8\frac{\mathrm{d} G}{\mathrm{d} (L^2)} + m^2 G\right)e^{-iKL^2} = 0.
\end{equation}

Integrating by parts to shift the derivatives to the exponential factor, we get
\begin{multline}
\left( 4L^2\frac{\mathrm{d} G(L^2)}{\mathrm{d} (L^2)} e^{-iKL^2} + 4G(L^2)(1+iKL^2)e^{-iKL^2}\right)_{L^2 = -\infty}^{L^2 = \infty} \\ + \int\limits_{-\infty}^{\infty}\mathrm{d} (L^2)\ G(L^2)\left(-4iK^2\frac{\mathrm{d}}{\mathrm{d} K} + m^2\right)e^{-iKL^2} = 0.
\end{multline}
If the boundary term vanishes, we may consider the integral alone, yielding \eq{FTL2_H0}. This corresponds to the solution $G(L^2) = g_- \frac{\mathrm{K}_1(-i\sqrt{m^2 L^2})}{\sqrt{m^2L^2}}$. But if the solution diverges near $L^2 \to \mp\infty$ (depending on the sign of $m^2$), as it does for $G(L^2) = g_+ \frac{\mathrm{K}_1(i\sqrt{m^2 L^2})}{\sqrt{m^2L^2}}$ the boundary terms do not vanish, and therefore \eq{FTL2_H0} does not follow for this solution.

\subsection{Derivation of results in Sec. \ref{sec:warmup}}\label{app:derh}
We consider a massless scalar field in Minkoski spacetime. The Wightman function for such a field is expressed as
\bea
G^{+}(L_\epsilon^2) = - \frac{1}{4 \pi^2 L_\epsilon^2},
\eea
where $ L_\epsilon^2 = (t-t' - i\epsilon )^2 -r^2$. The {\it Fourier} transform w.r.t. $L_\epsilon^2$ is obtained as
\bea
\tilde{G}^+(K) = \int_{-\infty -i\epsilon(t,t') }^{\infty -i\epsilon(t,t')} \mathrm{d}L_\epsilon^2 e^{-i K L_\epsilon^2} G^{+}(L_\epsilon^2) = -\frac{1}{4 \pi^2}\left[\int_{-\infty}^{\infty} \mathrm{d}L^2 \frac{e^{-i K L^2}}{L^2 -i \epsilon(t,t')} \right]e^{-2 K \epsilon(t,t')},
\eea
with $\epsilon(t,t')\equiv 2\epsilon (t-t')$. In the above expression, we have converted $L_\epsilon^2$ into a real $L^2 -\epsilon^2$ and imaginary 
$-i \epsilon(t,t')$ part with $L^2 =(t-t')^2 -r^2$ (and have neglected the ${\cal{O}}(\epsilon^2)$ term). The massless Wightman function is given as
\bea
G(x,x') \equiv G(t,t';{\bf x}, {\bf x'}) = -\frac{1}{4 \pi^2} \frac{1}{(t-t' - i\epsilon )^2 -r^2} = -\frac{1}{4 \pi^2} \frac{1}{L^2 -2i \epsilon(t-t')}, \label{FltWghtMn1}
\eea
whereas if we flip the time co-ordinates we will be getting
\bea
G(t',t;{\bf x}, {\bf x'}) = -\frac{1}{4 \pi^2} \frac{1}{(t'-t - i\epsilon )^2 -r^2} = -\frac{1}{4 \pi^2} \frac{1}{L^2 -2i \epsilon(t'-t)},\label{FltWghtMn2}
\eea
Combining \eq{FltWghtMn1} and \eq{FltWghtMn2} we will readily obtain
\bea
G(x,x') = - \frac{1}{4 \pi^2 L^2 -2i \epsilon(t-t')} \equiv G^{+}(L_\epsilon^2).
\eea

Taking this integral in the complex plane we see that if $(t-t')>0$ and $K>0$, we can close the contour with the lower semi-circle avoiding any poles and the integral vanishes as a consequence. However, for $K<0$ we need to close the contour with the upper semicircle which invokes a simple pole at $z = i\epsilon(t,t') $, whereas for $ (t-t')<0$ the pole lies in the upper half plane and the residue survives for $K>0$. Thus, from the residue theorem we get as a result
\bea
\tilde{G}^+(K) = \frac{\text{sgn}(t-t')}{4\pi^2  i} \theta(-\text{sgn}(t-t')K). \label{MinkoskiGK}
\eea
With a given Wightman function we obtain the power spectrum (amplitude) as outlined before in Sec.\ref{sec:altpow}. The power spectrum per unit logarithmic integral will be obtained from the power spectrum amplitude through

\bea
\tilde{P} _k  =\Omega_q  k^q \int \frac{ \mathrm{d}^q (x-y)}{(2\pi)^q} e^{-i\sum_{j}^q k_j(x^j-y^j)}G(x^{\mu}, y^{\mu}),
\eea
 where $q$ is the number of the symmetric axes and $\Omega_q$ is the total solid angle of a $q-$ dimensional Killing space.
\subsubsection{Inertial Power Spectrum}
First we evaluate the power spectrum through the timelike Killing direction of the Minkowski spacetime. We will evaluate the power spectrum $P_+(\omega)$  with the realization that the  power spectrum corresponding to other convention  (Fourier transform w.r.t. $e^{-i \omega t}$) is trivially obtained as 
$P_+(\omega) =P_-(-\omega)$.
With the $\tilde{G}^+(K)$ obtained in previous appendix, we can obtain the power spectrum as
\bea
P_{\text{inertial}}(\omega) &=& \int_{-\infty}^{\infty} \frac{ \mathrm{d} t}{2 \pi} \left[\int_{-\infty}^{\infty} \mathrm{d} K   s_t\frac{\theta(- s_tK)}{4\pi^2  i} e^{i (K L_\epsilon^2 + \omega t)}\right], \\
&=& \lim_{r \rightarrow 0}\frac{1}{2r} \left[ \int_{-\infty}^{\infty} \frac{ \mathrm{d} t}{8 \pi^3} \left(\frac{1}{t -i \epsilon -r} - \frac{1}{t -i \epsilon  +r} \right)e^{i \omega t}\right],\\
&=& \int_{-\infty}^{\infty} \frac{ \mathrm{d} t}{8 \pi^3}\frac{1}{(t -i \epsilon)^2}e^{i \omega t},
\eea
with a redefinition $t = t-t'$ and $s_t = \text{sgn}(t-t')$. Again, as before, the integral is carried into the complex plane and the contour in the upper plane (i.e. $\omega >0$) only is able to survive, leading to the expression of the Power spectrum
\bea
P_{\text{inertial}}(\omega) = 2 \pi \frac{\omega e^{-\omega \epsilon}}{8 \pi^3} \theta(\omega)\big|_{\epsilon = 0^+} = \frac{\omega}{4\pi^2 }\theta(\omega).
\eea

\subsubsection{Through spatial Killing directions}
We again write the Green's function as
\bea
 G(t,t';{\bf x}, {\bf x'}) = -\frac{1}{4 \pi^2} \frac{1}{(t-t' - i\epsilon )^2 -r^2}.
\eea
For a spacelike separation, we can go to a frame putting $t=t'$ (which we will end up doing ultimately in the power spectrum calculation) at the end of the calculation. The power spectrum will then be defined as
\bea
P_{\text{inertial}}(k) = \int\frac{ \mathrm{d}^3{\bf r} }{(2 \pi)^3} \left[\int_{-\infty}^{\infty} \mathrm{d} K   s_t\frac{\theta(- s_tK)}{4\pi^2  i} e^{i (K L_\epsilon^2 + {\bf k} \cdot{\bf r})}\right]_{t=t'}, \\
= \frac{4 \pi}{(2 \pi)^5} \int \frac{ \mathrm{d} r}{k}r \sin{kr} \frac{1}{Z -2 i \epsilon (t-t') -\epsilon^2}\big|_{t=t'}.
\eea
In this integral $ k$ is the magnitude of  the vector ${\bf k}$ and is positive semi-definite. This integral can be evaluated to yield
\bea
P_{\text{inertial}}(k) &=& \frac{4 \pi}{(2 \pi)^5} \int \frac{ \mathrm{d} r}{k}r \sin{kr} \frac{1}{Z -2 i \epsilon (t-t') -\epsilon^2}\big|_{t=t'}, \nonumber\\
&=& \frac{4 \pi}{(2 \pi)^3} \int_0^{\infty} \frac{ \mathrm{d} r}{k}r \sin{kr} \frac{1}{4 \pi^2 (r^2 + \epsilon^2)}, \nonumber\\
&=& \frac{2 \pi}{(2 \pi)^3} \int_{-\infty}^{\infty} \frac{ \mathrm{d} r}{k}r \exp{ikr} \frac{1}{4 i \pi^2 (r^2 + \epsilon^2)} =\frac{1}{2k(2 \pi)^3}. \label{SpatialPS}
\eea
The standard logarithmic interval power spectrum will be obtained by again multiplying $\Omega_3k^3 = 4 \pi k^3$
\bea
\tilde{P}_{\text{inertial}}(k) = \frac{k^2}{4 \pi^2} = \frac{\omega^2}{4 \pi^2},
\eea
where $k$ or $\omega$ is positive semidefinite. The same result could have been obtained by setting $s_t =1$ in \eq{MinkoskiGK} \footnote{or equivalently, $s_t =-1$ as well for that matter, just like we argued for the spacelike surface case in de Sitter.}.
\subsubsection{Rindler co-ordinates, Minkowski spacetime}
We will now try to evaluate the power spectrum of the Minkowski vacuum for a Rindler observer in Minkowski spacetime. We introduce the Rindler co-ordinates $(\tau, \xi, \theta, \varphi)$ in terms of Minkowski spherical polar co-ordinates $(t,r, \theta, \varphi)$ (with the same angular co-ordinates $\theta, \varphi$):
\begin{align}
t &= a^{-1}e^{a\xi}\sinh(a\tau), \\
r &= a^{-1}e^{a\xi}\cosh(a\tau).
\end{align}
The geodesic distance $L^2$ between two points along the trajectory of a Rindler observer, $(\xi,\theta,\varphi) = \text{const.}$ is then given by:
\begin{equation}
L^2 = \frac{2e^{2a\xi}}{a^2}(\cosh(a\Delta \tau)-1) = \bar{A}(\cosh(a\Delta \tau)-1),
\end{equation}
where $\Delta \tau = \tau - \tau'$.
The power spectrum of the inertial vacuum as seen by the Rindler observer is
\begin{eqnarray}
P_{\text{Rindler}}(\omega) = \int\limits_{-\infty}^{\infty} \frac{\mathrm{d} \Delta\tau}{2 \pi}\ G^+(L^2)e^{i\omega \Delta\tau}.
\end{eqnarray}
As before, we write the power spectrum in terms of the Fourier transform $\tilde{G}^+(K)$ with respect to $L^2$ as follows
\begin{align}
P_{\text{Rindler}}(\omega) &= \frac{e^{a\xi}}{2\pi}\int\limits_{-\infty}^{\infty}\mathrm{d} K\  \tilde{G}^+(K) \int\limits_{-\infty}^{\infty}\mathrm{d}\frac{\Delta\tau}{2 \pi}\ e^{iKL^2+i\omega\Delta\tau}, \\
&= \frac{e^{a\xi}}{2\pi}\int\limits_{-\infty}^{\infty}\mathrm{d} K\  \tilde{G}^+(K) e^{-iK \bar{A}} \int\limits_{-\infty}^{\infty}\mathrm{d}\frac{\Delta\tau}{2 \pi}\ e^{iK\bar{A}\cosh(a\Delta\tau)+i\omega\Delta\tau}.
\end{align}
Using the known form of the Fourier transform of the Green's function:
\begin{equation}
\tilde{G}^+(K) = s_t\frac{e^{\frac{im^2}{4K}}}{4\pi^2 i}\theta(-s_tK),
\end{equation}
and setting $m = 0$ for a massless field gives\footnote{Again, the separation $\Delta \tau$ is supposed to hide $i\epsilon$ in it for convergence.}
\begin{align}
P_{\text{Rindler}}(\omega) &=\frac{1}{4 \pi^2 i} \int\limits_{-\infty}^{\infty} \mathrm{d}\Delta \tau \frac{e^{i \omega \Delta \tau}}{i  \bar{A}(\cosh{a \Delta \tau}-1)},\\
                           &= -\frac{1}{16 \pi^2}\int\limits_{-\infty}^{\infty} \mathrm{d}\Delta \tau \frac{e^{i \omega \Delta \tau}}{  \bar{A}\sinh^2{\left(\frac{a \Delta \tau}{2}\right)}}.
\end{align}
Thus we obtain,
\begin{eqnarray}
P_{\text{Rindler}}(\omega) 
= \frac{\omega}{4\pi^2}\left(\frac{e^{\frac{2\pi\omega}{a}}}{e^{\frac{2\pi\omega}{a}}-1} \right) = \frac{\omega}{4\pi^2}(1+ n_{\omega}).
\end{eqnarray}
reproducing \eq{rinps}. This invokes contribution from the usual Rindler thermal spectrum (with temperature $T = a/2\pi e^{a\xi} k_B$), over the Minkowski inertial power spectrum. Thus changing the observer supplements the background inertial power spectrum.
%\newpage

\subsection{Derivation of results in Sec. \ref{sec:bdpsdS}}\label{app:derj}
In this appendix as well, we will derive the power spectrum w.r.t. $e^{+ i \omega \tau}$. The result for the other convention  $e^{- i \omega \tau}$ can again be obtained trivially from here.
As we discussed previously, in the static co-ordinate system, the de Sitter metric has a timelike Killing vector and therefore, the natural Fourier transform to define the power spectrum w.r.t. is $\Delta \tau$, where $(\tau,R)$ is the static co-ordinate system,
\bea
P_{\text{static/BD}}=\int_{-\infty}^{\infty} \frac{d \Delta \tau}{2 \pi} e^{i  \omega \Delta \tau}G^{+}(Z_{\epsilon}),
\eea
where we are evaluating the $i\epsilon$ corrected geodesic distance $Z_{\epsilon}$ for spatially coincident points at $R=0$, for which
\bea
\eta &=& -\frac{1}{H}e^{-H\tau},\\
Z    &=& 1+\frac{(\Delta \eta)^2}{2 \eta \eta'} = \cosh{H \Delta \tau },\\
Z_{\epsilon} &=& \cosh{H(\Delta \tau -i \epsilon)},
\eea
whereas using \eq{Gplusdef} we get
\bea
G^{+}(Z_{\epsilon}) &=& \frac{H^2}{16\pi^2 \sqrt{2}} \int_{-\infty}^{\infty} \mathrm{d}u\frac{e^{-\nu u}}{\left( \cosh{u} -\cosh{H(\Delta \tau -i \epsilon)}\right)^{\frac{3}{2}}}, \\
&=& \frac{H^2}{16\pi^2 \sqrt{2}}  \int_{-\infty}^{\infty} \mathrm{d}u\frac{e^{-\nu u}}{\left( 2 \sinh{\left(\frac{u + H \Delta \tau}{2} - \frac{i \epsilon}{2}\right)}\sinh{\left(\frac{u - H \Delta \tau}{2} + \frac{i \epsilon}{2}\right)}\right)^{\frac{3}{2}}}.
\eea
Therefore,
\bea
P_{\text{static/BD}}(\omega) &=& \int_{-\infty}^{\infty} \frac{d \Delta \tau}{2 \pi} e^{i \omega \Delta \tau}\frac{H^2}{16\pi^2 \sqrt{2}} \int_{-\infty}^{\infty} \mathrm{d}u\frac{e^{-\nu u}}{\left( \cosh{u} -\cosh{H(\Delta \tau -i \epsilon)}\right)^{\frac{3}{2}}}, \\
&=&\frac{H^2}{16\pi^2 \sqrt{2}}  \int_{-\infty}^{\infty} \frac{d \Delta \tau}{2 \pi}\int_{-\infty}^{\infty} \mathrm{d}u\frac{e^{-\nu u}}{\left( 2 \sinh{\left(\frac{u + H \Delta \tau}{2} - \frac{i \epsilon}{2}\right)}\sinh{\left(\frac{u - H \Delta \tau}{2} + \frac{i \epsilon}{2}\right)}\right)^{\frac{3}{2}}}.\nonumber\\ \label{StatPsinBD}
\eea
Now, we can go to another set of variables defined as $T_\pm =(u \pm H \Delta \tau)/2$ to convert \eq{StatPsinBD} into
\bea
P_{\text{static/BD}}(\omega) &=& \frac{H^2}{16\pi^2 \sqrt{2}} \frac{1}{2\pi}\frac{2}{H}\int_{-\infty}^{\infty} \int_{-\infty}^{\infty} \mathrm{d}T_+ \mathrm{d}T_-\frac{e^{-\nu(T_++T_-)}e^{\frac{i\omega}{H}(T_+-T_-)}}{\left[ 2^{\frac{3}{2}}\left( \sinh{\left(T_+ - \frac{i \epsilon}{2}\right)}\right)^{\frac{3}{2}}\left(\sinh{\left(T_- + \frac{i \epsilon}{2}\right)}\right)^{\frac{3}{2}}\right]} ,\nonumber\\
&=& \frac{H^2}{16\pi^2 \sqrt{2}} \frac{1}{2\pi}\frac{1}{\sqrt{2}H}\left|\int_{-\infty}^{\infty} \mathrm{d}T\frac{e^{-\nu T}e^{\frac{i\omega}{H}T}}{\left( \sinh{\left(T - \frac{i \epsilon}{2}\right)}\right)^{\frac{3}{2}}}\right|^2.\label{StatPsinBD2}
\eea
Thus, we have to carry out the integration in \eq{StatPsinBD2} to obtain the power spectrum. For that we employ an identity
\bea
\frac{1}{\left( \sinh{\left(T - \frac{i \epsilon}{2}\right)}\right)^{\frac{3}{2}}} = \frac{e^{\frac{3 i \pi}{4}}}{\Gamma\left(\frac{3}{2}\right)}\int_0^{\infty}\mathrm{d}s \ s^{\frac{1}{2}}e^{-is\left(\sinh{\left(T - \frac{i \epsilon}{2}\right)} \right)},
\eea
which yields
\bea
\int_{-\infty}^{\infty} \mathrm{d}T\frac{e^{-\nu T}e^{\frac{i\omega}{H}T}}{\left( \sinh{\left(T - \frac{i \epsilon}{2}\right)}\right)^{\frac{3}{2}}}, &=&
\frac{e^{\frac{3 i \pi}{4}}}{\Gamma\left(\frac{3}{2}\right)}\int_0^{\infty}\mathrm{d}s \ s^{\frac{1}{2}}\int_{-\infty}^{\infty}\mathrm{d}T e^{-\nu T}e^{\frac{i\omega}{H}T}
e^{-is\left(\sinh{\left(T - \frac{i \epsilon}{2}\right)} \right)},\\
&=&\frac{2e^{\frac{3 i \pi}{4}}}{\Gamma\left(\frac{3}{2}\right)}\left(e^{\frac{i \pi}{2}-\frac{i \epsilon}{2}}\right)^{\nu -\frac{i\omega}{H}}\int_0^{\infty}\mathrm{d}s \ s^{\frac{1}{2}}K_{\nu -\frac{i\omega}{H}}(s), \\
&=& \frac{\sqrt{2}e^{\frac{3 i \pi}{4}}}{\Gamma\left(\frac{3}{2}\right)}\left(e^{\frac{i \pi}{2}-\frac{i \epsilon}{2}}\right)^{\nu -\frac{i\omega}{H}}\Gamma\left[\frac{3}{4}-\frac{\nu}{2}+\frac{i \omega}{2H}\right]\Gamma\left[\frac{3}{4}+\frac{\nu}{2}-\frac{i \omega}{2H}\right],\nonumber\\
\eea
where $\mathrm{K}_{\alpha}(z)$ is the Bessel function of order $\alpha$. Therefore, the power spectrum is obtained (in the limit of vanishing $\epsilon$) as
\bea
P_{\text{static/BD}}(\omega)=\frac{H^2}{4 \pi^2} e^{\frac{\pi \omega}{H}}\frac{1}{2}\left|\Gamma\left[\frac{3}{4}-\frac{\nu}{2}+\frac{i \omega}{2H}\right]\Gamma\left[\frac{3}{4}+\frac{\nu}{2}-\frac{i \omega}{2H}\right]\right|^2.
\eea
In the limit of massless scalar field $(\nu \rightarrow 3/2)$, we get the power spectrum to  be
\begin{align}
\label{dS_Static_PS_unevenresult}
P_{\text{static/BD}}(\omega) &= \frac{H^2}{4\pi^2\omega}\left(1+\frac{\omega^2}{H^2}\right)\frac{e^{\pi\frac{\omega}{H}}}{2\sinh\left(\pi\frac{\omega}{H}\right)}, \\
&= \frac{H^2}{4\pi^2\omega}\left(1+\frac{\omega^2}{H^2}\right)\frac{e^{\frac{2\pi\omega}{H}}}{e^{\frac{2\pi\omega}{H}}-1}
\end{align}
which again shows that the Bunch-Davies vacuum appears as thermally populated and the vacuum noise is supplemented by a thermal factor times the vacuum noise of the de Sitter (evaluated in the static vacuum).
\bea
P_{\text{static/BD}}(\omega) =\frac{H^2}{4\pi^2\omega}\left(1+\frac{\omega^2}{H^2}\right)(1+ n_{\omega}) = P_{\text{static}}(1+ n_{\omega}).
\eea
In terms of the amplitude, we get
\bea
\tilde{P}_{\text{static/BD}}(\omega) =\frac{H^2}{4\pi^2}\left(1+\frac{\omega^2}{H^2}\right)(1+ n_{\omega}) = \tilde{P}_{dS}(\omega)(1+ n_{\omega}),
\eea
where $\tilde{P}_{dS} = H^2/4 \pi^2(1 + \omega^2/H^2)$.

%\newpage

\subsection{Derivation of \eq{ftwrtm}}\label{app:derB1}
Finally, we verify that the Wightman function in \eq{intreppoly} reduces to the right $H\to 0$ limit i.e., the Minkowski Wightman function. The Fourier transformed Wightman function is:
\begin{equation}
\widetilde{G}^{+}_{\mathrm{dS}}(Q) = -\frac{H^2}{4\sqrt{2\pi}}e^{\frac{i\pi}{2}(\frac{1}{2}-\nu)}Q^{\frac{1}{2}}\mathrm{H}_\nu^{(2)}(Q)s_\eta \theta\left(- Qs_\eta \right).
\end{equation}
This gives
\begin{equation}
\overline{G}^{+}_{\mathrm{dS}}(K) = -\frac{1}{2H\sqrt{\pi}}e^{\frac{i\pi}{2}(\frac{1}{2}-\nu)}e^{\frac{2iK}{H^2}}K^{\frac{1}{2}}\mathrm{H}_\nu^{(2)}\left(\frac{2K}{H^2}\right)s_\eta \theta\left(-K s_\eta \right).
\end{equation}
To take the $H\to 0$ limit, we note that $\nu \to i m /H$, and for $\lvert\mathrm{Re}\ \nu\rvert < 1/2$, we have the following integral representation for the Hankel function (from 10.9.11  of \cite{DLMF})
\begin{equation}
\mathrm{H}_\nu^{(2)}\left(\frac{2K}{H}\right) \sim \mathrm{H}_{\tfrac{im}{H}}^{(2)}\left(\frac{2K}{H}\right) = -\frac{e^{-\frac{\pi m}{2H}}}{i\pi}\int\limits_{-\infty}^{\infty}\mathrm{d} t\ e^{-i\frac{2K}{H^2}\cosh t}e^{-i \frac{m}{H} t}.
\end{equation}
Using a variable transformation $e^{t} =z$ we can convert the above integral into a Bessel function
\begin{eqnarray}
\mathrm{H}_{\tfrac{im}{H}}^{(2)}\left(\frac{2K}{H}\right) &=& -\frac{e^{-\frac{\pi m}{2H}}}{i\pi}\int\limits_{0}^{\infty}\mathrm{d} z\ z^{-\frac{i m}{H}-1} e^{-i\frac{K}{H^2}\left(z-\frac{1}{z} \right) }, \nonumber\\
&=& -2 \frac{e^{-\frac{\pi m}{2H}}}{i\pi} K_{\frac{i m}{H}}\left(\frac{2K}{H^2}  \right) =-2\frac{e^{-\frac{\pi m}{2H}}}{i\pi} \mathrm{K}_{\frac{i m}{H}}\left(
\frac{m}{H}\frac{2 K}{m H}  \right).
\end{eqnarray}
Using the asymptotic order parameter expansion $m/H \rightarrow \infty$ of the Bessel function \cite{dunster}, we get
\begin{eqnarray}
\mathrm{H}_\nu^{(2)}\left(\frac{2K}{H}\right) = -\frac{e^{-\frac{i\pi}{4}}e^{-\frac{\pi m}{2 H}} e^{-\frac{2iK}{H^2}}}{i\sqrt{\pi}K^{\frac{1}{2}}} H e^{\frac{im^2}{4K}}.
\end{eqnarray}
The Fourier transform of the Wightman function in the Minkowski limit therefore reduces to
\begin{equation}
\overline{G}^+(K) = \frac{s_t}{2\pi i}e^{\frac{im^2}{4K}}\theta(-Ks_t),
\end{equation}
where $s_t = \mathrm{sgn}(t-t')$.
We obtain the Wightman function from
\begin{eqnarray}
G^+(L^2) &=& \frac{1}{2\pi}\int\limits_{-\infty}^{\infty}\mathrm{d} K\ \overline{G}^+(K)e^{iKL^2}, \\
 &=& -\frac{1}{4\pi^2 i}\int\limits_0^{\infty}\mathrm{d} K\ e^{-\frac{i s_t m^2}{4K}} e^{-i s_t KL^2}.
\end{eqnarray}
(we have replaced $K \to -s_t K$ and used $s^2_t = 1$ in the intermediate steps).
The integral (from 3.324(1)  of \cite{gd}) evaluates to
\begin{eqnarray}
G^+(L^2) &=& \frac{i}{4\pi^2} \sqrt{\frac{m^2}{L^2}} \mathrm{K}_{1}\left(\sqrt{-m^2L^2}\right), \\
&=& \frac{i}{4\pi^2} \sqrt{\frac{m^2}{L^2}} \mathrm{K}_{1}\left(-i\sqrt{m^2L^2}\right).
\end{eqnarray}
where, in the second line, we have used the upper-half-plane convention for the square root in the argument of $\mathrm{K}_\nu(z)$, giving us the solution we have previously shown to be vanishingly small at large spacelike distances.
This is the Wightman function for a massive scalar field in Minkowski spacetime. We also verify it has the right massless limit, using 10.30.2 of \cite{DLMF}, $\mathrm{K}_\nu(z) \sim \tfrac{1}{2}\Gamma(\nu)\left(\tfrac{z}{2}\right)^{-\nu},\ z\to 0$, giving the familiar result for the massless Wightman function  
\begin{equation}
G^+(L^2)\rvert_{m = 0} = -\frac{1}{4\pi^2L^2}.
\end{equation}

The integral representation in  the penultimate line of \eq{ftwrtm} can be obtained directly from more standard expression for Feynman propagator either in Lorentzian or in the Euclidean sector. We briefly outline this derivation. If we use the signature $(+,-,-,-)$, the standard expression for the Feynman propagator is given by
 \begin{equation}
 %% eqn 1
 G_F(x) = \int \frac{\mathrm{d}^D p}{(2\pi)^D} \, \frac{ie^{-ipx}}{p^2-m^2 + i\epsilon}.
  \label{7apr1}
 \end{equation} 
 Using the  Schwinger trick of writing the denominator $H \equiv p^2 -m^2 + i\epsilon$ as an integral over $\lambda$ of $\exp(i \lambda H)$ and performing the momentum integrals, we obtain the integral representation
  \begin{equation}
 %% eqn 2
 G_F = i \int_0^\infty  \frac{\mathrm{d}\lambda}{(2\pi)^D} \frab{\pi}{i\lambda}^{\frac{D}{2}} \exp\left[ -i \lambda (m^2 - i \epsilon) - \frac{ix^2 }{ 4\lambda}\right].
  \label{7apr2}
 \end{equation}
 This is a well known expression; what seems to be not so well known is its Fourier transform with respect to $m$ treated purely as a parameter. Straightforward calculation (after assuming a small imaginary part in $\lambda \rightarrow \lambda -i \delta$) shows that 
  \begin{equation}
%%  eqn 3
\int_{-\infty}^\infty G_F(x) \, e^{im\sigma} \ \mathrm{d}m= \frac{i}{(2\pi )^{D}} \frab{\pi}{i}^{\frac{D}{2}} \int_0^\infty \frac{\mathrm{d}\lambda}{\lambda^{D/2}} \left[ \frac{\pi}{i(\lambda - i \delta)}\right]^{1/2} \exp\frac{i(\sigma^2-x^2)}{4(\lambda - i \delta)}.
  \label{7apr3}
 \end{equation}
 Performing the integral, without worrying too much about convergence issues, we get the result
 \begin{equation}
  \int_{-\infty}^\infty G_F\, e^{im\sigma} \ \mathrm{d}m= \frac{1}{2} \frac{\Gamma(k)}{(\pi)^k} \frac{1}{(\sigma^2 - x^2 +i \epsilon)^k}; \qquad k\equiv \frac{1}{2} (D-1).
  \label{tpkey}
 \end{equation} 
 The same calculation can also be performed with the Euclidean propagator which has better convergence properties. Here we start with the expression 
 \begin{equation}
 %%% eqn 4 = rhs of eqn 5
  G_F(x) = \int \frac{\mathrm{d}^D p}{(2\pi)^D} \frac{e^{-ipx}}{(p^2 + m^2)} = \int_0^\infty \frab{1}{4\pi \lambda}^{D/2} \exp\left(- m^2\lambda - \frac{x^2}{4\lambda}\right) \mathrm{d}\lambda.
  \label{7apr4}
 \end{equation}
 The Fourier transform now leads to
 \begin{equation}
  \int_{-\infty}^\infty G_F\, e^{im\sigma} \ \mathrm{d}m= \frac{1}{2} \frac{\Gamma(k)}{\pi^k} \frac{1}{(\sigma^2 + x^2 )^k}; \qquad k\equiv \frac{1}{2} (D-1).
  \label{tpkey1}
 \end{equation} 
 Combining the two results, we have in the Euclidean and Lorentzian sector, the result
 \begin{equation}
 %%% eqn 6
 \int_{-\infty}^\infty G_F\ e^{im\sigma}\mathrm{d}m= 
 \begin{cases}
   \frac{1}{2} \frac{\Gamma(k)}{\pi^k} \frac{1}{(\sigma^2 + x^2 )^k} &\text{(Euclidean)},\\
   \frac{1}{2} \frac{\Gamma(k)}{\pi^k} \frac{1}{(\sigma^2 - x^2 +i \epsilon)^k} &\text{(Lorentzian)},
 \end{cases}
  \label{7apr6}
 \end{equation}
 where $k= (1/2)(D-1)$. Fourier inversion now leads to the representation used in  the penultimate line of \eq{ftwrtm} with $D=4$. This representation has some interesting implications for particle production in external backgrounds, which will be explored in a different publication.

\newpage

\bibliography{Gravity_1_full,Gravity_2_partial,My_References}

\bibliographystyle{./utphys1}

\end{document}